\documentclass[a4paper,11pt]{article}
\pdfoutput=1 

\usepackage{jinstpub} 
\usepackage{xspace}
\usepackage[toc,page]{appendix}
\usepackage{array}
\usepackage{tabulary}
\newcolumntype{K}[1]{>{\centering\arraybackslash}p{#1}}

\newcommand{\gf}{\textsc{Geant}4\xspace}

\title{\boldmath \gf Parameter Tuning Using Professor}

\author[a]{V. Elvira,}
\author[a,1]{L. Fields\note{Corresponding author.},}
\author[a]{K.~L.~Genser,}
\author[a]{R. Hatcher,}
\author[b,c]{V.~Ivanchenko,}
\author[d]{M.~Kelsey,}
\author[d]{T.~Koi,}
\author[a]{G.~N.~Perdue,}
\author[b]{A.~Ribon,}
\author[b]{V.~Uzhinsky,}
\author[d]{D.~H.~Wright,}
\author[a]{J. Yarba,} 
\author[a]{S.~Y.~Jun}

\affiliation[a]{Fermi National Accelerator Laboratory,
Kirk Road and Pine Street, Batavia, IL, 60510-5011 U.S.A.}
\affiliation[b]{CERN, EP Department, 27210, CH-1211 Geneva, Switzerland}
\affiliation[c]{Tomsk State University, 634050 Tomsk, Russia}
\affiliation[d]{SLAC National Accelerator Laboratory, 2575 Sand Hill Rd, Menlo Park, CA, 94025, USA}

\emailAdd{ljf26@fnal.gov}

\abstract{The \gf toolkit is used extensively in high energy physics to simulate the passage of particles through matter and to predict effects such as detector efficiencies and smearing.  \gf uses many underlying models to predict particle interaction kinematics, and uncertainty in these models leads to uncertainty in high energy physics measurements.  The \gf collaboration recently made free parameters in some models accessible through partnership with \gf developers.  We present a study of the impact of varying parameters in three \gf hadronic physics models on agreement with thin target datasets and describe fits to these datasets using the Professor model tuning framework~\cite{professor}.  We find that varying parameters produces substantially better agreement with some datasets, but that more degrees of freedom are required for full agreement.  This work is a first step towards a common framework for propagating uncertainties in \gf models to high energy physics measurements, and we outline future work required to complete that goal. }


\arxivnumber{1910.06417} 

\begin{document}
\maketitle
\flushbottom

\section{Introduction}
\label{sec:intro}

\gf~\cite{Agostinelli:2002hh,Allison:2006ve,Allison:2016lfl} is a toolkit for simulating the passage of particle through matter that is used in a variety of fields.  In high energy physics, it is routinely used to construct detailed simulations of particle detectors that are used to estimate quantities such as detector smearing, acceptance and efficiency.  In neutrino physics, it is additionally used to simulate neutrino beamlines and to predict the neutrino flux through detectors. 

\gf typically accepts an input list of particles (with initial 4-positions and 4-momenta) and a geometry from the user.  Each particle is stepped through the geometry, with probabilities of a variety of different kinds of interactions to happen at each step.  Once an interaction occurs, the final state particles and their momenta are determined from a model of that process based on theory and tuned to data where possible.  Various options for both the interaction cross sections and the final state models are available.  

Many of these models, and in particular those describing interactions of hadrons on nuclei, carry significant uncertainties.  When \gf-based simulations are used to correct observed data distributions and to extract measurements of fundamental parameters, these models introduce uncertainty in the measurements.  These uncertainties have historically been difficult to quantify, with experimenters typically guessing at what aspects of the \gf simulation have the biggest impact on a measurement, and then attempting to quantify the uncertainty in those effects.  

With the release of \gf version 10.4, the \gf collaboration made it possible to access some underlying parameters in several models, and to vary these parameters.  This opens the possibility of fitting these parameters to datasets, to extract optimal values of the parameters and uncertainties in those parameters that could be propagated to physics measurements.  

One significant hurdle to this goal is the fact that the available \gf parameters cannot be varied via event weighting (as discussed in futher detail in Section~\ref{sec:future}.  In order to produce a simulation with a varied set of parameters, one must execute a completely new simulation; it is not currently possible to vary the models by applying weights to events.  With simulations of typical experimental setup taking hours, fits of many parameters to many datasets are computationally challenging.  Professor is a framework for parameter fitting that has been successfully used to tune non-reweightable parameters of models in hadron collider event generators such as Pythia~\cite{pythia1,pythia2}.  

This article describes a first attempt to apply the Professor framework to \gf hadronic model parameter tuning. Section~\ref{sec:parameters} describes the models and parameters considered; thin target datasets considered in this study are listed in Section~\ref{sec:datasets}; Section~\ref{sec:sensitivity} compares scans of these parameters to available datasets; Section~\ref{sec:fits} describes the results of Professor fits.  While this represents the first step towards a common framework for assessing \gf model parameter uncertainty in high energy physics measurement, a broad program of work would be needed to fully achieve that goal.  The major components of that program are outlined in Section~\ref{sec:future} 

\section{Available Geant4 Parameters}
\label{sec:parameters}

\begin{table}[htbp]
\centering
\scriptsize
\caption{\label{tab:parameters} Continuous parameters available in the Bertini, PreCompound, and FTF models.  A * following the parameter name indicates parameters that were included in the fits and the $\dagger$ following the limits indicates that ranges of parameters 
are not defined in Geant4, but varied typically between 50\% and 200\% of the default or in the given range.  The asymmetry quantifies how much the parameter varies Monte Carlo Predictions as described in Sec.~\ref{sec:sensitivity}}.
\smallskip
\begin{tabular}{|c|K{4.5cm}|c|c|c|c|}
\hline
Parameter & Description & Default & Limits & Unit & Asymmetry \\
\hline
{\bf Bertini} & & & &  &\\

crossSectionScale* & Multiplicative factor applied to the nuclear 
cross sections & 1.0 & 0.05-2.0  & & .0136 \\ 
nuclearRadiusScale* & Nuclear radius multiplicative factor &2.82 &1.0-2.82 & & 0.214\\
fermiScale* & Fermi momentum multiplicative factor  & 0.6852 & 0.3426-1.3704 & & 0.137 \\
shadowingRadius* & Local depletion radius of nuclear density in  intra-nuclear collisions & 0.0 & 0.0-2.0 & & 0.134 \\ 
gammaQuasiDeutScale & Intra-nuclear pion absorption cross section scale factor for photon-nuclear interactions & 1.0 & 0.5-2.0$\dagger$ & & 0.000 \\ 
piNAbsorption* & Energy threshold for pion absorption & 0.0 & 0.0-1.0$\dagger$& ~$GeV$ & 0.043\\ 
smallNucleusRadius & Fixed radius for light ions (A$<$4) & 8.0 & 4.0-16.0$\dagger$ & & 0.000  \\ 
alphaRadiusScale & Fraction of light-ion radius for alphas & 0.7 & 0.35-1.40$\dagger$ & & 0.000\\ 
cluster2DPmax* & Momentum cut for pn~$\rightarrow$~D & 0.09 & 0.045-0.18$\dagger$ & $GeV$ & 0.068 \\ 
cluster3DPmax* & Momentum cut for pnn~$\rightarrow$~T, ppn~$\rightarrow$~3He & 0.108 & 0.054-0.216$\dagger$ & $GeV$ & .046 \\ 
cluster4DPmax* & Momentum cut for ppnn~$\rightarrow$~alpha & 0.115 & 0.0575-0.230$\dagger$ & $GeV$ & 0.023 \\ \hline

{ \bf PreCompound} & & & & & \\
LevelDensity* & Excited states level density  & 0.1 & 0.05-0.2$\dagger$ & MeV & 0.117 \\ 
R0* & Nuclear radius & 1.5  & 0.5-2.5$\dagger$ & fm & 0.026 \\ 
TransitionsR0* & Nuclear radius for transitions  & 0.6 & 0.1-1.1$\dagger$ & fm & 0.027\\ 
FermiEnergy* & Fermi energy level & 35.0 & 5.0-75.0$\dagger$ & MeV & 0.017\\ 
PrecoLowEnergy*  &  Low-energy excitation per nucleon limit & 0.1 & 0.05-0.2$\dagger$ & MeV & 0.037 \\ 
PhenoFactor & Phenomenological factor &  1.0 & 0.5-2.0$\dagger$ & & 0.000\\ 
MinExcitation* & Min excitation energy &  10.0 & 5.0-20.0$\dagger$ & ev & 0.002\\ 
MaxLifeTime & Time limit for long lived isomeres & 1000.0 & 500.0-2000.0$\dagger$ & s & 0.000 \\ 
MinExPerNucleounForMF & Min energy per nucleon for Multifragmentation & 100.0 & 50.0-200.0$\dagger$ & GeV & 0.000\\ \hline

\bf{FTF} & & & & &  \\
BaryonDiffMProj & Projectile baryon threshold for excited string mass sampling in diffractive interactions & 1.16 GeV; & 1.16-3.0 & GeV & 0.008\\ 
BaryonNondiffMProj* & Projectile baryon threshold for excited string mass sampling in non-diffractive interactions & 1.16 & 1.16-3.0 &  GeV & 0.025 \\ 
BaryonDiffMTgt & Target hadron threshold for excited string mass sampling in diffractive interactions & 
1.16 & 1.16-3.0 & GeV & 0.019\\
BaryonNonDiffMTgt* & Target hadron threshold for excited string mass sampling in non-diffractive interactions & 1.16 & 1.16-3.0 & GeV & 0.028\\ 
BaryonAvgPt* & Average transverse momentum squared in the excitation process & 0.15 & 0.08-1.0 & GeV$^2$ & 0.050\\ 
NucdestrP1Proj & $P_1$ in Eq.~\ref{cnd} for the projectile & 1.0 & 0.0-1.0 &  & 0.000\\ 
NucdestrP1Tgt* & $P_1$ in Eq.~\ref{cnd} for the target & 1.0 &  0.0-1.0 & & 0.025 \\ 
NucdestrP2Tgt* & $P_2$ in Eq.~\ref{cnd} for the target & 4.0 & 2.0-16.0 & & 0.001 \\ 
NucdestrP3Tgt* & $P_3$ in Eq.~\ref{cnd} for the target & 2.1 & 0.0-4.0 & & 0.002\\ 
Pt2NucdestrP1* & $C_1$ in Eq.~\ref{ptsq} & 0.035& 0.0-0.25 & & 0.013\\ 
Pt2NucdestrP2* & $C_2$ in Eq.~\ref{ptsq} & 0.004 & 0.0-0.25 & & 0.013\\
Pt2NucdestrP3 & $C_3$ in Eq.~\ref{ptsq} & 4.0 & 2.0-16.0 & & 0.001\\
Pt2NucdestrP4 & $C_4$ in Eq.~\ref{ptsq} & 2.5 & 0.0-4.0 & & 0.001\\ 
BaryonNucdestrR2* & $R_c^2$ in Eq.~\ref{prob} & 1.5 & 0.5-2.0 & fm$^2$  & 0.024 \\ 
BaryonExciEPerWndnucln* & Excitation energy per wounded nucleon & 40.0 MeV & 0.-100.0 & & 0.044\\ 
BaryonNucdestrDof & Dispersion parameter of the momentum distribution of the nucleons in the cascade & 0.3 & 0.1-0.4 & & .017 \\ \hline
\end{tabular}
\end{table}

As of release series 10.4, \gf provides
a configuration interface for the hadronic models Bertini Cascade, PreCompound, and Fritiof (FTF).  A configuration interface is also available to control and/or tweak modeling of the electromagnetic (EM) processes, but those are not considered here.  The configuration interfaces differ from model to model, and guidance for varying the models is available in Geant4 user documentation~\cite{ftfdoc,g4usermanual}.   The models also include a number of switches and parameters with discrete value options; the data comparisons and fits described here use the default value of the switches and discrete parameters.  

The Fritiof (FTF) model in GEANT4 simulates hadron-nucleus, nucleus-nucleus and antibaryon-nucleus interactions based on diffractive and non-diffractive quark-gluon strings 
reactions and the LUND-string fragmentation model.  The
valid energy range of the model is between 3 GeV and 10 TeV  per hadron or nucleon.
In Geant4 release 10.4, FTF offers 16 configurable parameters governing baryon projectiles only.  These parameters, their default values in Version 10.4, and the limits set by model developers are listed in Table~\ref{tab:parameters}.  Several of the parameters are associated with modeling nuclear destruction in baryon-nucleus interactions.  In these interactions, the probability of the nucleons to be involved in a reggeon cascade is given by  
\begin{equation}\label{prob}
P( | s_i - s_j | ) = C_{nd} e^{ -( s_i - s_j )^2 / R_c^2},
\end{equation} 
where $s_i$ and $s_j$ are projections of the radii of i-th and j-th nucleons on the impact parameter plane and the coefficient $C_{nd}$ is
\begin{equation}\label{cnd}
C_{nd} = P_1 e^{ P_2 (y-P_3) } / [ 1. + e^{ p_2 (y-P_3) } ].
\end{equation}
Other parameters involve modeling of momentum distributions of the nucleons involved in the cascade, which is described in greater detail in \cite{ftfdoc}; one key distribution is the average transverse momentum square of an ejected nucleon, which can be parameterized as
\begin{equation}\label{ptsq}
<P_T^2> = C_1 + C_2 \frac{ e^{C_3 (y_{lab} - C_4 )} }{ 1. + e^{C_ 3 (y_{lab} - C_4)} } [(GeV/c)^2],
\end{equation}
where $y_{lab}$ is the rapidity of the projectile nucleus in the rest frame of the target nucleus.

The Bertini intra-nuclear cascade model handles nuclear interactions initiated by
long-lived hadrons ($p, n, \pi, K, \Lambda, \Sigma, \Xi, \Omega$) and $\gamma$s
with energy up to 12 GeV where the de Broglie wavelength ($\lambda_B$) of the incident
particle is comparable to the average inter nucleon distance.  Below
200 MeV where the intra-nuclear model is no longer valid, the Bertini model
uses either its own pre-equilibrium model of the exciton by Griffin~\cite{Griffin} or
the Geant4 Pre-Compound model for nuclear de-excitation by setting a
``usePreCompound'' flag in the cascade model.
In the intra-nuclear model, the particle-nucleon cross sections and region-dependent
nucleon densities are used to sample path lengths of nucleons which follow
the Fermi gas momentum distribution.  The Fermi energy is calculated in a local
density approximation, $E_F = p_F^{2}(r)/2m_N$ where $p_{F}(r) = (3\pi^{2} \rho(r)/2)^{1/3}$ is the
radius-dependent Fermi momentum and
$m_N$ represents the nucleon mass.  The density of particles, $\rho(r)$ is different
for each incident particle and each region which is described by three concentric spheres
with radius,
\begin{equation}\label{fermiradius}
r_i(\alpha_i) = C_{1} + C_{2} \log (\frac{1+e^{-C_{1}/C_{2}}}{\alpha_i} -1)
\end{equation}
where $i = \{1, 2, 3\}$, $\alpha_i = \{0.01, 0.3, 0.7\}$, $C_1 = 3.3836 A^{1/3}$, and
$C_2 = 1.7234$ as an example for $A > 11$.  In Geant4, the internal cross sections,
the nuclear radius, and the Fermi momentum of bounded nucleons can be adjusted by
multiplicative factors, crossSectionScale, fermiScale, and nuclearRadiusScale,
respectively. As shown in Table~\ref{tab:parameters}, these are also
the most sensitive parameters of the Bertini model.

The Geant4 pre-compound model provides a transition from the kinetic stage of reaction
to the equilibrium stage of reaction described by the de-excitation models. At the
pre-equilibrium stage of reaction, the transition probability of the number of
excitions ($n$) are equiprobable for all three types of allowed transitions,
$\Delta n = 0, \pm 2$ and is characterized by the equilibrium number of excitons,
\begin{equation}\label{eq:neq}
n_{eq} = \sqrt{2gU}
\end{equation}
where $U$ is the excitation energy and $g$ is the density of the $n$-excition
that is approximated by the level density parameter $a$, i.e., $g \approx 0.0595 a A$.
The transition probabilities changing the exciton number by $\Delta n = \pm 2$
are assumed to be same as the probability for quasi-free scattering of a nucleon above
the Fermi level on a nucleon of the target nucleus,
\begin{equation}\label{eq:transition-prob}
\omega_{\Delta n =+2}(n,U) = \frac{\langle \sigma(v_{rel})\rangle \langle v_{rel} \rangle}{V_{int}}
\end{equation}
where $V_{int} = 4\pi (2r_{c} + \lambda_{B}/2\pi)^{3}/3$ is the interaction volume corresponding
to the relative velocity $\langle v_{rel} \rangle = \sqrt{2T_{rel}/m}$.  In Geant4,
the transition radius, $r_c$ and the Fermi energy ($E_F$) in the mean relative kinetic
energy, $2T_{rel} = 1.6 E_{F} + U/n$ along with the nuclear radius ($R_0$) and the level density $a$
can be varied within in the range shown in Table~\ref{tab:parameters}.
Even though the Bertini model uses the Pre-compound model in the  FTFP\_BERT physics
list by default, parameters of each model are tuned separately since the energy ranges
applicable to these two models are exclusive.
Of course, a combined tuning may be carried out once parameters of each model are
reasonably stabilized and ready to be optimized further.
 
\section{Datasets Considered}
~\label{sec:datasets}
There is a large body of data that could be used to tune hadronic models in \gf.  Based on advice from Geant4 model developer, we chose recent datasets with relatively large and precise results, including thin target datasets published by the HARP, IAEA, NA49, NA61, and ITEP771 collaborations, as detailed in table~\ref{tab:data}. The data clearly had to cover key energy range(s) relevant to the models and of interest to HEP experiments, so datasets used in this study
were chosen to span the range 0.8 - 158 GeV. Since
hadronic models are also largely dependent on the incident particle and colliding nucleus, data with all available beams and targets were included in the first round of tuning. Data from recent experiments with a large number of precise measurements was given priority.
These thin-target datasets used in this study were stored in Database of Scientific Simulation
and Experimental Results (DoSSiER)~\cite{DoSSiER} and made available through its programmatic interface.
\begin{table}[htbp]
\centering
\scriptsize
\caption{\label{tab:data} Datasets included in parameter fits, from the HARP~\cite{Apollonio:2009bu,Apollonio:2009re,Apollonio:2009en,Catanesi:2008aa}, IAEA~\cite{IAEA}, NA49~\cite{Alt:2006fr,Baatar:2012fua}, NA61~\cite{Abgrall:2015hmv}, and ITEP771~\cite{Bayukov:1986kf} collaborations.}
\smallskip
\begin{tabular}{|c|c|c|K{2cm}|c|K{3cm}|}
\hline
Model & Experiment & Projectile & Target & Final State & Distributions\\
\hline
Bertini & ITEP & 7.5 GeV protons & Be, C, Al, Ti, Fe, Cu, Nb, Sn, Ta, Pb, U & pX & $d^3\sigma/dp^3$  \\
 &  & 5 GeV protons & C, Pb & pX & $d^3\sigma/dp^3$  \\
 & HARP & 5 GeV protons & C, Pb & $\pi^+X$ & 
$d^2\sigma/dpd\Theta$ \\
& & & & $\pi^-X$ & $d^2\sigma/dpd\Theta$ \\
 &  & 5 GeV $\pi^-$ & B, C, Al, Cu, Ta, Pb & $\pi^-X$ & $d^2\sigma/dpd\Theta$ \\
 &  & 5 GeV $\pi^-$ & B, C, Al, Cu, Ta, Pb & $\pi^+X$ & $d^2\sigma/dpd\Theta$ \\
  &  & 5 GeV $\pi^+$ & B, C, Al, Cu, Ta, Pb & $\pi^-X$ & $d^2\sigma/dpd\Theta$ \\
 &  & 5 GeV $\pi^+$ & B, C, Al, Cu, Ta, Pb & $\pi^+X$ & $d^2\sigma/dpd\Theta$ \\
  &  & 8 GeV $\pi^-$ & B, C, Al, Cu, Ta, Pb & $\pi^-X$ & $d^2\sigma/dpd\Theta$ \\
 &  & 8 GeV $\pi^-$ & B, C, Al, Cu, Ta, Pb & $\pi^+X$ & $d^2\sigma/dpd\Theta$ \\
  &  & 8 GeV $\pi^+$ & B, C, Al, Cu, Ta, Pb & $\pi^-X$ & $d^2\sigma/dpd\Theta$ \\
 &  & 8 GeV $\pi^+$ & B, C, Al, Cu, Ta, Pb & $\pi^+X$ & $d^2\sigma/dpd\Theta$ \\
  & IAEA & 0.8 GeV protons & C, Al, Fe, Pb & nX & $d^2\sigma/dKEd\theta$  \\
  &  & 1.5 GeV protons & C, Al, Fe, In, Pb & nX & $d^2\sigma/dKEd\theta$  \\
&  & 3 GeV protons & C, Al, Fe, In, Pb & nX & $d^2\sigma/dKEd\theta$  \\

\hline
 PreCompound & IAEA & 0.8 GeV protons & C, Al, Fe, Pb & nX & $d^2\sigma/dKEd\theta$  \\
  &  & 1.5 GeV protons & C, Al, Fe, In, Pb & nX & $d^2\sigma/dKEd\theta$  \\
&  & 3 GeV protons & C, Al, Fe, In, Pb & nX & $d^2\sigma/dKEd\theta$  \\
 \hline
FTF & NA49 &  158 GeV protons & C & $\pi^+X$ & 
$dn/dx_F$ and $<p_T>$ vs $x_F$  \\
& & & & $\pi^-X$ & $dn/dx_F$ and $<p_T>$ vs $x_F$  \\ 
& & & & $pX$ & $dn/dx_F$ and $<p_T>$ vs $x_F$  \\
& & & & $\bar{p}X$ & $dn/dx_F$ and $<p_T>$ vs $x_F$  \\
& & & & $nX$ & $dn/dx_F$  \\
 & NA61 &  31 GeV protons & C & $\pi^+X$ &
$d^2\sigma/{dpd\theta}$  \\
& & & & $\pi^-X$ & $d^2\sigma/{dpd\theta}$  \\
& & & & $K^+X$ & $d^2\sigma/{dpd\theta}$  \\
& & & & $K^-X$ & $d^2\sigma/{dpd\theta}$  \\
& & & & $K^0X$ & $d^2\sigma/{dpd\theta}$ \\
& & & & $\Lambda X$ & $d^2\sigma/{dpd\theta}$  \\
& & & & $pX$ & $d^2\sigma/{dpd\theta}$  \\
 & ITEP771 & 5 GeV protons & C, Pb & $pX$ & 
$d\sigma/dx_F$ at 59.1, 89, 119 and 159 degrees  \\
& & & & $nX$ & $d\sigma/dx_F$ at 119 degrees  \\
& HARP & 5 GeV protons & C, Pb & $\pi^+X$ & 
$d^2\sigma/dpd\Theta$ \\
& & & & $\pi^-X$ & $d^2\sigma/dpd\Theta$ \\
& IAEA & 3 GeV protons &  C, Pb & nX & $d^2\sigma/d\phi dE$ \\ \hline
\end{tabular}
\end{table}

\section{Sensitivity to Parameters in Modeling Thin Target Data}
\label{sec:sensitivity}
Some of the available Geant4 parameters have a greater impact on predictions of the thin target datasets listed in Table~\ref{sec:datasets} than others.  Because the computing resources required for parameter fits grow with the number of parameters considered, it is useful to understand which parameters create the greatest changes in model predictions, and are therefore most important for inclusion in fits.  

\begin{figure}[htbp]
\centering 
\includegraphics[width=0.9\textwidth,trim=0 0 0 0,clip]{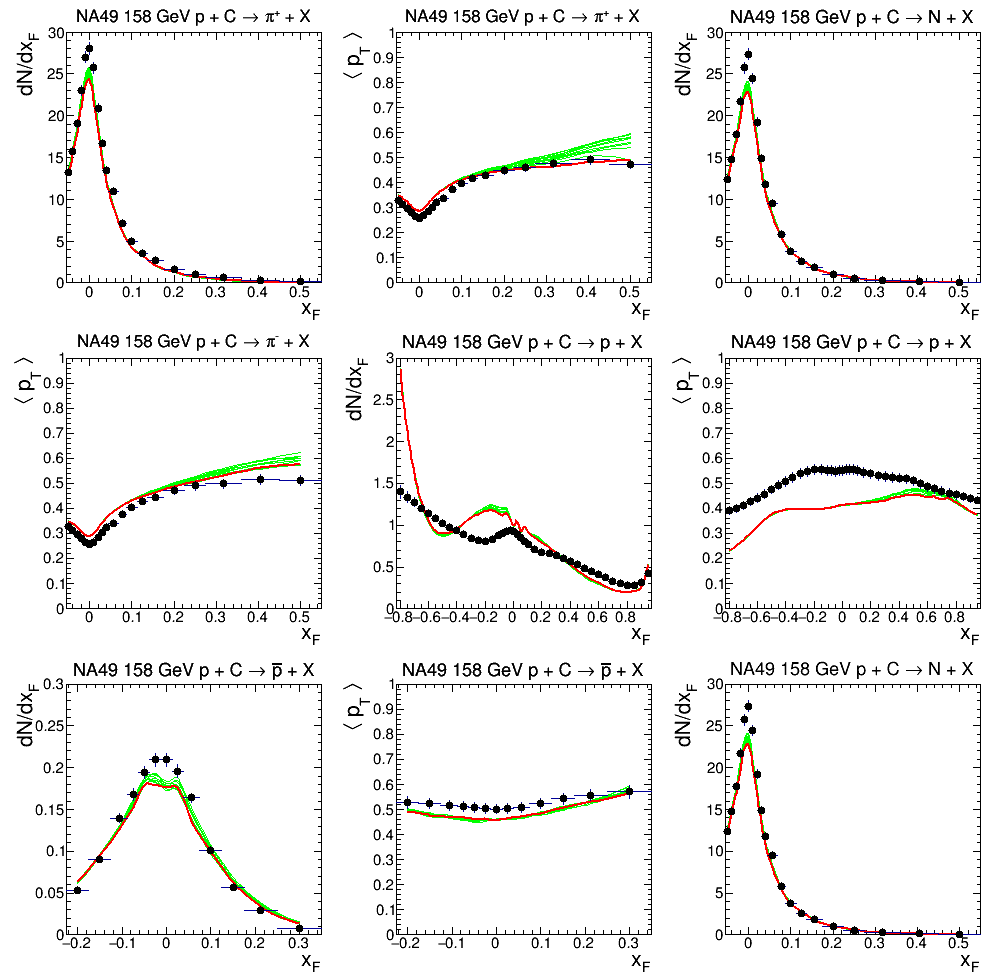}
\caption{\label{fig:scan_ftf} Predictions for 158 GeV proton interactions on Carbon in default Geant4 (red) and with the FTF parameter BARYON\_NONDIFF\_M\_PROJ varied within the its allowed limits (green); NA49 measurements~\cite{Alt:2006fr,Baatar:2012fua} of the same quantities are also shown (black).  }
\end{figure}

\begin{figure}[htbp]
\centering 
\includegraphics[width=0.9\textwidth,trim=0 0 0 0,clip]{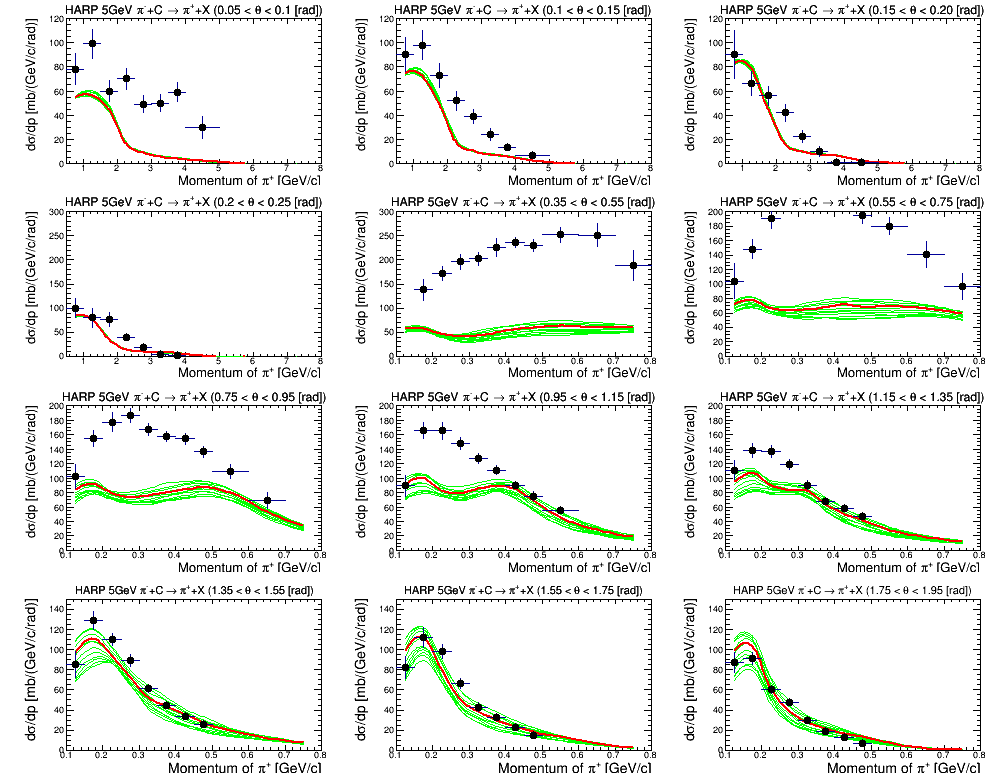}
\caption{\label{fig:scan_bertini} Predictions for 5 GeV $\pi^-$ interactions on Carbon using default Geant4 (red) and with the Bertini parameter RadiusScale varied within the its allowed limits (green).  HARP measurements~\cite{Alt:2006fr,Baatar:2012fua} of the same quantities are also shown (black). }
\end{figure}

\begin{figure}[htbp]
\centering 
\includegraphics[width=0.9\textwidth,trim=0 0 0 0,clip]{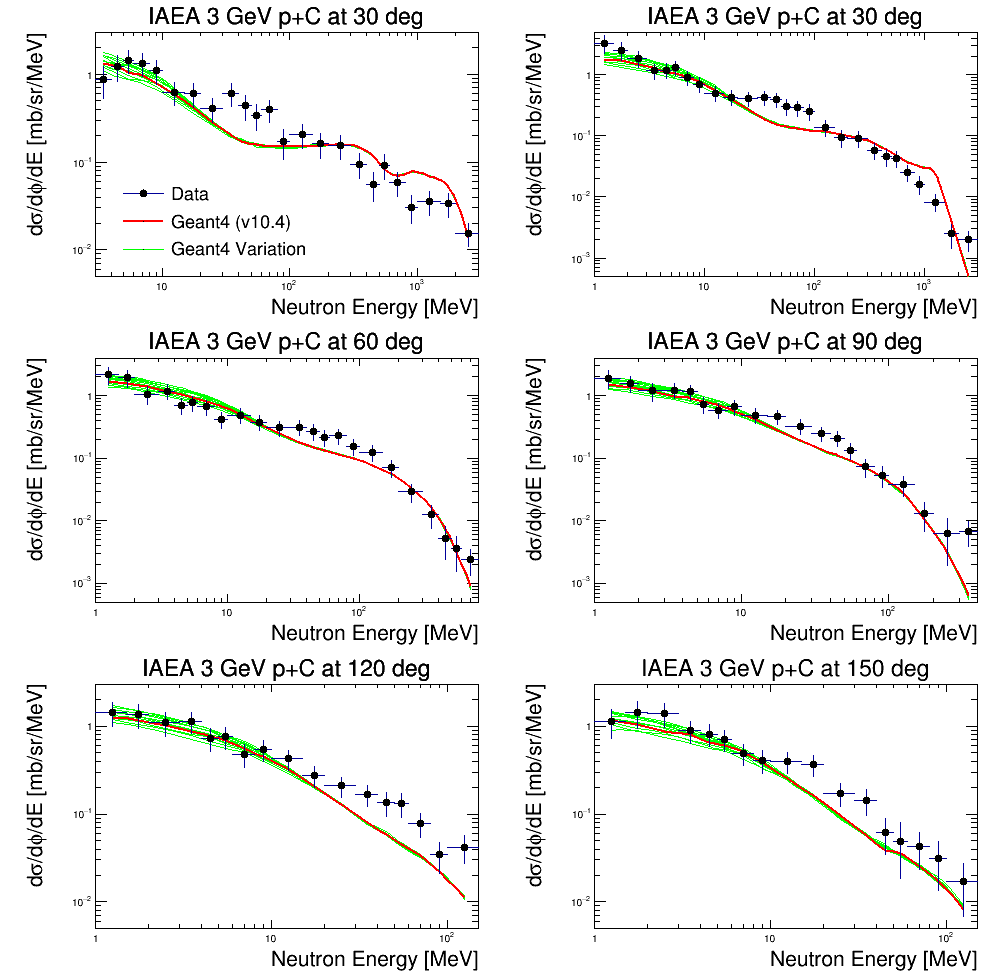}
\caption{\label{fig:scan_precompound} Predictions for 3 GeV proton interactions on Carbon using default Geant4 (red) and with the PreCompound parameter LevelDensity varied within the its allowed limits (green).  IAEA measurements~\cite{IAEA} of the same quantities are also shown (black). }
\end{figure}

To estimate the sensitivity of each parameter to thin target predictions, we produced simulations of several thin-target measurements where all other parameters are fixed to their default values and the parameter in question was varied within the limits specified by the model developers (or between 50\% and 200\% of the default value if limits are not specified).  An example of such a parameter scan is shown in Figure~\ref{fig:scan_ftf} for several quantities measured by NA49 for 158 GeV proton interactions on carbon and for the FTF parameter BARYON\_NONDIFF\_M\_PROJ.  Example parameter scans for parameters in the Bertini and PreCompound are shown in Figures ~\ref{fig:scan_bertini} and ~\ref{fig:scan_precompound}, respectively. 

For each distribution considered in the parameter scan, we calculate the asymmetry for a particular bin as $|y_{max}-y_{min}|/|y_{max}+y_{min}|$ where $y_{max}$ ($y_{min}$) is the maximum value of the predicted distributions in bin in question.  This is then averaged over all bins and all distributions to produce a total asymmetry, shown in Table~\ref{tab:parameters}.  Based on these sensitivities and other details of the parameter scans, the parameters noted with an * in Table~\ref{tab:parameters} were chosen for inclusion in fits.  

\section{Parameter Fits using Professor}

\label{sec:fits}

Generating a model prediction for all of the thin target datasets for a given set of parameters requires multiple CPU hours.  Traditional fitting methods wherein  many possible parameter values are scanned, simulations are performed, and chi-square differences between models and data are minimized, is not computationally feasible.  Instead, the Professor~\cite{professor} fitting framework was used.  Professor parameterizes the prediction for a given bin of an observable distribution using the $n$-degree polynomial of parameters.  The coefficients of the polynomial are analytically evaluated by the singular value decomposition using a simulated data ensemble in which the underlying parameters are thrown randomly within their {\it a priori} probability distributions (which are taken to be flat distributions within parameter limits for the Geant4 parameters discussed here).  The resulting polynomials can then be used to construct predictions of each dataset given any set of parameters, substantially reducing the time required to produce dataset predictions from hours to a fraction of a second, and allowing traditional chi-square minimization. 

\subsection{Parameter Fits}

\begin{figure}[htbp]
\centering 
\includegraphics[width=15cm, height=10.5cm]{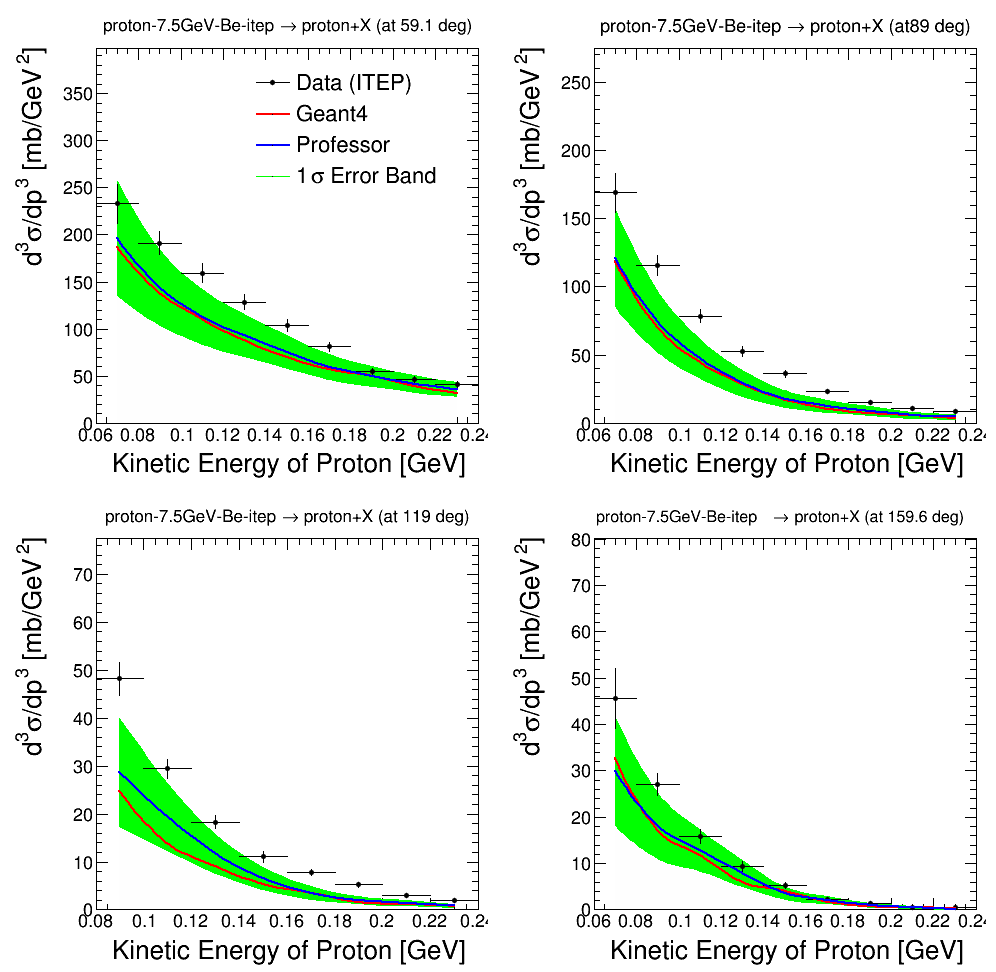}
\caption{\label{fig:fit_example_bertini} Results of the global Bertini parameter fit, compared to ITEP 7.5 GeV $pBe\rightarrow pX$ data in bins of final state proton angle.  Data points are shown in black; default Geant4 is red and the global fit result are blue; the green band shows uncertainties propagated from parameter uncertainties returned by the fit.   }
\end{figure}

\begin{figure}[htbp]
\centering 
\includegraphics[width=0.9\textwidth,trim=0 0 0 0,clip]{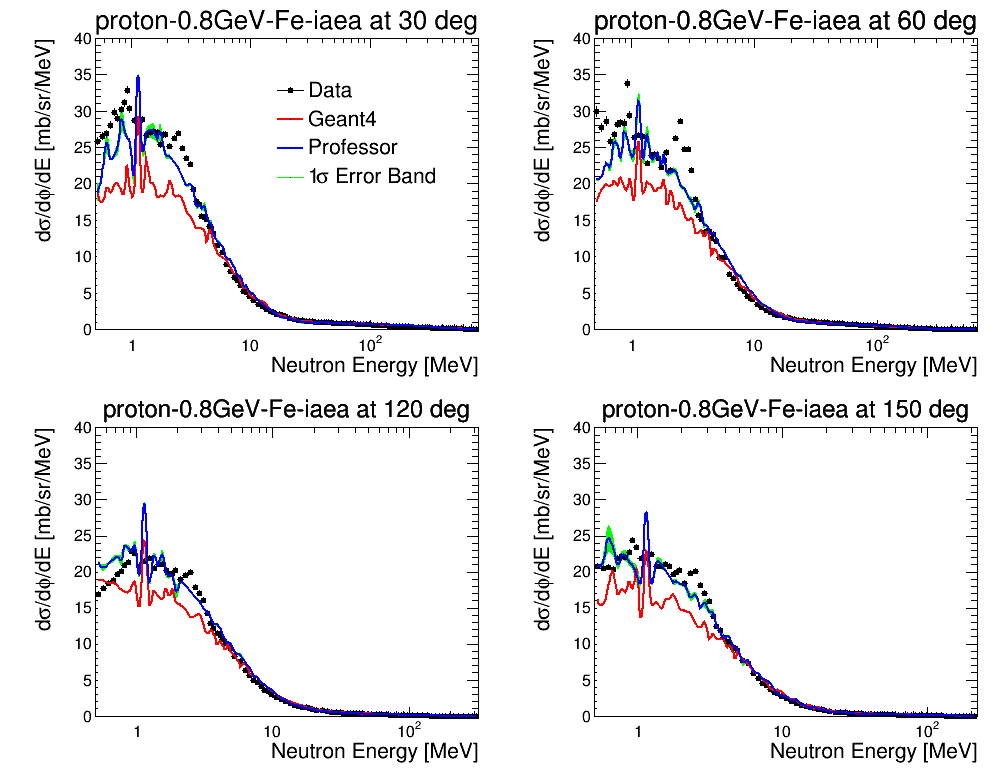}
\caption{\label{fig:fit_example_precompound} Results of the global PreCompound parameter fit, compared to IAEA 0.8 GeV $pFe\rightarrow n$ data.  Data points are shown in black; default Geant4 in red and Geant4 with best fit parameters in blue; the green band shows uncertainties propagated from parameter uncertainties returned by the fit.      }
\end{figure}

\begin{figure}[htbp]
\centering 
\includegraphics[width=0.9\textwidth,trim=0 0 0 0,clip]{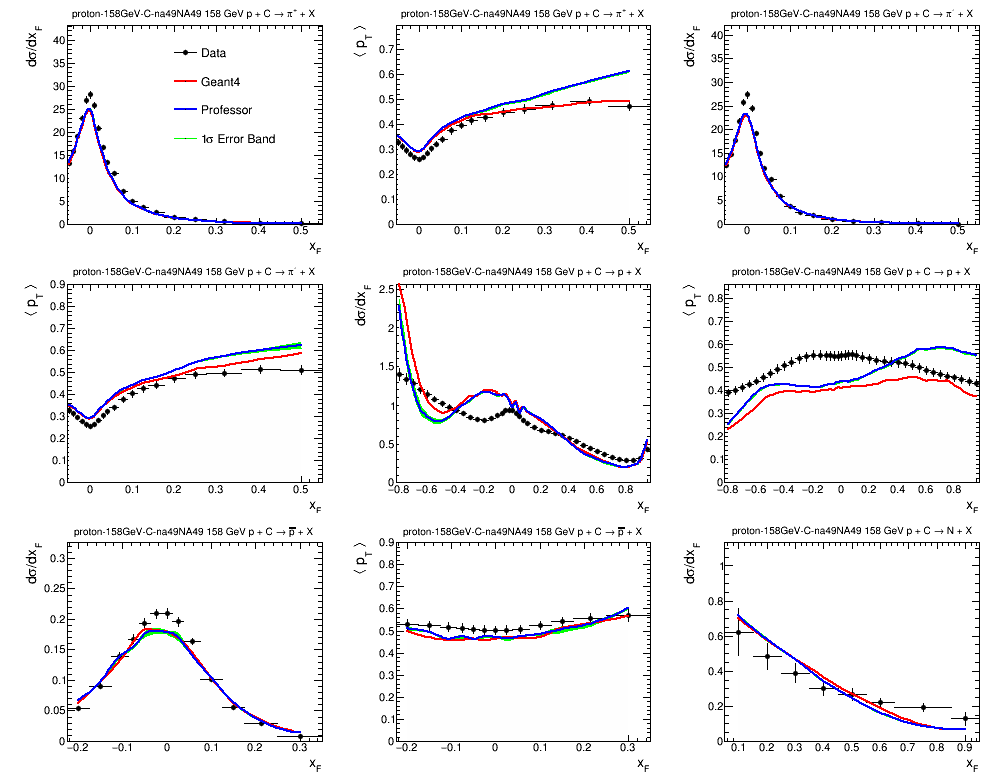}
\caption{\label{fig:fit_example_ftf} Results of the global FTF parameter fit, compared to NA49 31 GeV $pC\rightarrow K_S^0X$ and $pC\rightarrow K^-X$ data.  Data points are shown in black; default Geant4 in red and Geant4 with best fit parameters in blue; the green band shows uncertainties propagated from parameter uncertainties returned by the fit.      }
\end{figure}

Three "global fits" were performed, one for each of the three models considered here.  In each case, the parameters of the model in question were fit to all the datasets listed in Table~\ref{tab:data} simultaneously.  Example fit results are shown in Figures ~\ref{fig:fit_example_bertini}-\ref{fig:fit_example_ftf} for selected datasets.  Comparisons of additional datasets with these global fit results are available in Appendix~\ref{app:global_fits}.  The appendix includes results compared to all fitted datasets except those considered in the Bertini fit, where only ITEP 7.5 GeV proton data are shown for the sake of brevity.  In each figure, data points from the relevant datasets are compared with \gf predictions with both the default and best fit parameters.  The error band on the fit result is taken from parameter errors returned by the fit, accounting for correlations.  However, because the best fit chi-square per degree of freedom is much more than one, these error bands are not complete and do not include uncertainty associated with the models' inability to produce good agreement with data~\cite{PDG}.  A summary of the default and best-fit chi-squares is available in Tables~\ref{tab:chisquares3}-\ref{tab:chisquares1}.  Fits were also performed individually for several datasets; the chi-squares and best fit parameters for those fits are also available in Tables~\ref{tab:chisquares3}-\ref{tab:chisquares1}.  While improvement in fit quality is obtained by fitting each dataset separately, fit quality is still poor for most datasets, and the best fit parameters vary significantly between these fits.   

Geant4 predictions using the best fit parameters do improve agreement with the data in some areas.  For example, Bertini agreement with ITEP proton scattering data is much improved for many (but not all) nuclei.  In other cases, the fit causes data Monte Carlo agreement to be substantially worse, e.g. in comparisons with NA61 31 GeV $pC\rightarrow \pi-X$ data using FTF.  In other cases, e.g. in predictions of IAEA 1.5 GeV $pC\rightarrow nX$ data, the best fit is close to the default prediction.  

\begin{table}[htbp]
\centering
\scriptsize
\caption{\label{tab:chisquares3} Summary of fit chi-squares, number of degrees of freedom (NDOF), and best fit parameters for the global Bertini fit and Bertini fits to subsets of the data. The chi-squares are evaluated using data points of all observables with respect to the simulated result with the default parameter values of Geant4 (default) and the predicted curve from the Professor fit
(the global fit and the best fit of each dataset).}
\smallskip
\begin{tabular}{|c|c|c|c|c|c|}
\hline
 &  &  &                                  
   Proton  &  $\pi^+$   &   $\pi^-$   \\ 
 &   Default   &   Global &  Beam Data &  Beam Data  & Beam Data 
 \\ \hline
 $\chi^2$ & 521950   & 417510 &                  128554  &   104156 & 131006 \\ 
NDOF        &  15098         &   15090 &        4513    &    5269    &   5292 \\  \hline
FermiScale    &       0.69  &   0.58 &  0.50 &  0.61  & 0.55 \\ 
RadiusScale    &      2.82  &   2.98  & 2.62 &  2.67	& 2.92  \\ 
TrailingRadius   &    0.0   &     1.01 &  0.88  & 1.08 &	 1.38 \\ 
XSecScale   &         1.0  &  	2.00 &  2.00&   1.40&	 1.64 \\ 
cluster2DPmax  &     0.09      & 0.05 	 & 0.05  & 0.10  & 0.05	  \\ 
cluster3DPmax & 0.11         & 0.05 	 & 0.05 & 0.05  &	0.20  \\ 
 cluster4DPmax &     0.12     & 0.13 	 & 0.18 & 0.08  &	0.13  \\ 
  piNAbsorption & 0.0          &  0.67	 & 0.21 & 0.00  &	1.00  \\ \hline
 \end{tabular}
\end{table}

\begin{table}[htbp]
\centering
\scriptsize
\caption{\label{tab:chisquares2} Summary of fit chi-squares, number of degrees of freedom (NDOF), and best fit parameters for the global PreCompound fit and PreCompound fits to individual datasets.}
\smallskip
\begin{tabular}{|c|c|c|c|c|c|c|c|c|}
\hline
 &                       & &            IAEA  &     IAEA  &     IAEA &     IAEA     &    IAEA   &   IAEA \\
 &  Default   &   global &    3GeV-C  &   3GeV-Pb & 1.5GeV-C & 1.5GeV-Pb  &  0.8GeV-C  & 0.8GeV-Pb\\ \hline

$\chi^2$        & 45360  &                33806 &  246 &     266 &    198 &  128 &     5971 &   11875  \\
 NDOF &  2858   &                         2852   &     122    &    125   &     118  &      118      &   344    &    400  
   \\ \hline
LevelDensity     &       0.10  &   0.14 &   0.11 &   0.05 &   0.07 &   0.19 &    0.06 &   0.14\\
R0 [e-12] &               1.50  &   1.22 &   0.93 &  0.50 &	 0.50 &  1.17  &  0.50 &  1.24 \\ 
TransitionsR0 [e-12]  &  0.60  &   0.62 &  0.74 &  1.10 &	 1.10 &   1.10 &    0.28 &   0.82 \\
FermiEnergy &             35.0 &   	20.2 &   30.2 &   74.9 &	 60.1 &   44.7 &    31.4 &   6.9 \\
PrecoLowEnergy  &         0.10   &  0.09 &   0.20 &  0.20 &	 0.20 &   0.05 &    0.10 &   0.20 \\
MinExcitation [e-5] &    1.00 &     2.00 &   1.75 &  0.84 &	 0.59 & 2.00 &    0.77 &   2.00 \\ \hline


 &              IAEA   &    IAEA   &    IAEA    &   IAEA   &    IAEA  &     IAEA    &     IAEA   &   IAEA\\ 
 &      3GeV-Al  &   3GeV-Fe  &  3GeV-In & 1.5GeV-Al &  1.5GeV-Fe & 1.5GeV-In &   0.8GeV-Al & 0.8GeV-Fe \\ \hline

$\chi^2$  &              128     & 135 &     147 &     128 &     110 &      102 &     5025 &    5160 \\ 
NDOF & 126    &   127   &     127    &    123    &    123     &   122  &     399  &      400 \\ \hline
LevelDensity      &      0.06 &  0.06 &  0.06 & 0.09 & 0.09 & 0.11 &   0.17  &  0.20 \\
R0 [e-12]        &       0.70 &   0.50 &  1.98 & 2.50 & 0.50 & 0.50 &    2.03 &  1.08 \\
TransitionsR0 [e-12] &    0.10 &  0.10 &   0.42 &  1.08 &  1.03 &  1.10 &  1.10 &  1.10 \\ 
FermiEnergy      &       49.41 &   53.58 & 5.08 & 74.93 &   5.08 &   42.61 &    65.14 &   61.98 \\ 
PrecoLowEnergy        &  0.16 &   0.20 &   0.05 &  0.20 &  0.16 &  0.20 &    0.14 &   0.15 \\ 
MinExcitation [e-5]   &  1.79 &   0.50 &  2.00 &  2.00 & 0.50 &  2.00  &  0.65   & 0.70 \\ \hline
%

 \end{tabular}
\end{table}

\begin{table}[htbp]
\centering
\scriptsize
\caption{\label{tab:chisquares1} Summary of fit chi-squares, number of degrees of freedom (NDOF), and best fit parameters for the global FTF fit and FTF fits to individual  datasets.}
\smallskip
\begin{tabular}{|c|c|c|c|c|c|}
\hline

 & & & NA49  & NA61 & ITEP771  \\    
  &      Default & global & 158GeV-C & 31GeV-C & 5GeV-C   \\ \hline
 $\chi^2$                  & 39981         &  24619 & 3656  & 10895 &   319 \\ 
 NDOF & 2319    &                        2309 & 184 &   1355&  34 \\ \hline
BaryonNondiffMProj  &   1.16 &  1.75 & 1.16 &  1.16 &  2.64 \\
BaryonNondiffMTgt   &    1.16  & 1.16 & 1.16 &  1.16 & 2.19 \\ 
BaryonAvrgPt2 &      0.15  &  0.34 &  0.23 &  0.39 &  0.11 \\
NucdestrP1Tgt & 1.00 & 0.39 &  0.35 &  0.00 &  0.00 \\
NucdestrP2Tgt & 4.00 &    10.52 &  2.01 &  11.51 &  14.92 \\
NucdestrP3Tgt &      2.10 &    3.43 &  0.05 &   1.99 &  2.50 \\
Pt2NucdestrP1 &             0.04  & 0.07 &  0.06 &  0.13 &  0.02 \\
Pt2NucdestrP2 &
            0.04  & 0.12 & 0.24 &  0.19 &  0.00 \\
BaryonNucdestrR2 [e-24] &              1.5 &   0.51 & 1.07 &  1.11 &  1.09 \\ 
BaryonExciEPerWndNucln & 40.0 &   29.6 & 0.1 &  8.1 &  0.1 \\
\hline
 & ITEP771 & HARP &      HARP  & 	IAEA &     IAEA \\    
 &  5GeV-Pb  & 5GeV-C   & 5GeV-Pb  & 3GeV-C &   3GeV-Pb  \\ \hline
 $\chi^2$                  &      348 & 1089 &  1219 & 292 & 145 \\ 
 NDOF & 40 & 189 & 198 & 118 &  121 \\ \hline
BaryonNondiffMProj  &  1.54 &  1.16 & 1.62 &  1.41 &  1.94 \\
BaryonNondiffMTgt   &   1.56 & 1.98 & 1.96 & 2.90 & 1.68 \\ 
BaryonAvrgPt2 &  0.08 &  0.85 &  1.00 &  0.28 &  1.00 \\
NucdestrP1Tgt &  0.00 & 0.00 &  0.00 &  0.00 &  0.00 \\
NucdestrP2Tgt & 5.43 &  9.69 &  15.99 &  2.27 &  15.99 \\
NucdestrP3Tgt &  3.61 &  3.23 &  3.42 &  0.65 &  2.59 \\
Pt2NucdestrP1 &  0.06 &  0.00 &  0.08 &  0.11 &  0.05 \\
Pt2NucdestrP2 &
 0.03 &  0.19 &  0.00 &  0.00 &  0.02 \\
BaryonNucdestrR2 [e-24] &         0.72 &  0.50 &  2.00 &  0.50 &  0.50 \\ 
BaryonExciEPerWndNucln & 57.9 &  0.1 &  38.9 &  11.7 &  38.9 \\
\hline
\end{tabular}
\end{table}

\subsection{A Dependence of Fit Results}

\begin{figure}[htbp]
\centering 
\includegraphics[width=0.9\textwidth,trim=0 0 0 0,clip]{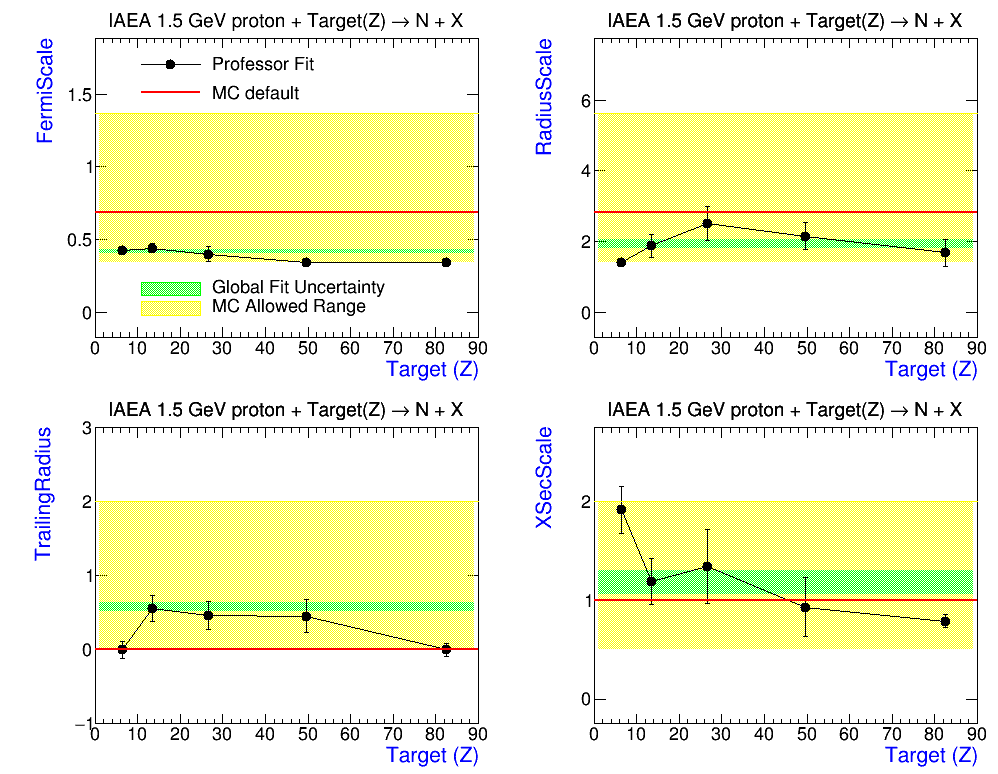}
\caption{\label{fig:adependence} Best fit parameters obtained from Bertini parameter fits to IAEA 1.5 GeV proton on various nuclear targets (black).  The default Geant4 value is shown by the red line.  The yellow band shows the allowed parameter range of the values while the green band is the uncertainty of the global Professor fit with all target data.  
}
\end{figure}

It is clear from the fits described above that expanded Geant4 parameter space will be required to achieve good agreement with many datasets.  One natural expansion is to allow different parameter values for different nuclear targets.  To explore this possibility, separate Bertini model parameter fits were performed to IAEA 1.5 GeV proton datasets on several nuclear targets.  The best fit parameters versus nuclear mass number (A) is shown in Figure ~\ref{fig:adependence}.  In some cases, there are clear trends versus nuclear mass.  The preferred value of the FermiScale parameter decreases as nuclear mass increases, whereas RadiusScale is nearly flat (perhaps because the fitter pushes that parameter to the lower limit of the allowed range for that parameter).  It appears that better parameter fits could be obtained if Geant4 offered the possibility of setting different parameters for different nuclear targets.  

\subsection{Comparison of Professor Fit Results with \gf Simulations}

As discussed above, the Professor fits do not actually run Geant4 simulations as part of the fitting algorithm.  Rather, interpolation between many simulations with randomly chosen parameters is used to estimate the result one would obtain with a true Geant4 simulation.  

\begin{figure}[htbp]
\centering 
\includegraphics[width=0.9\textwidth,trim=0 0 0 0,clip]{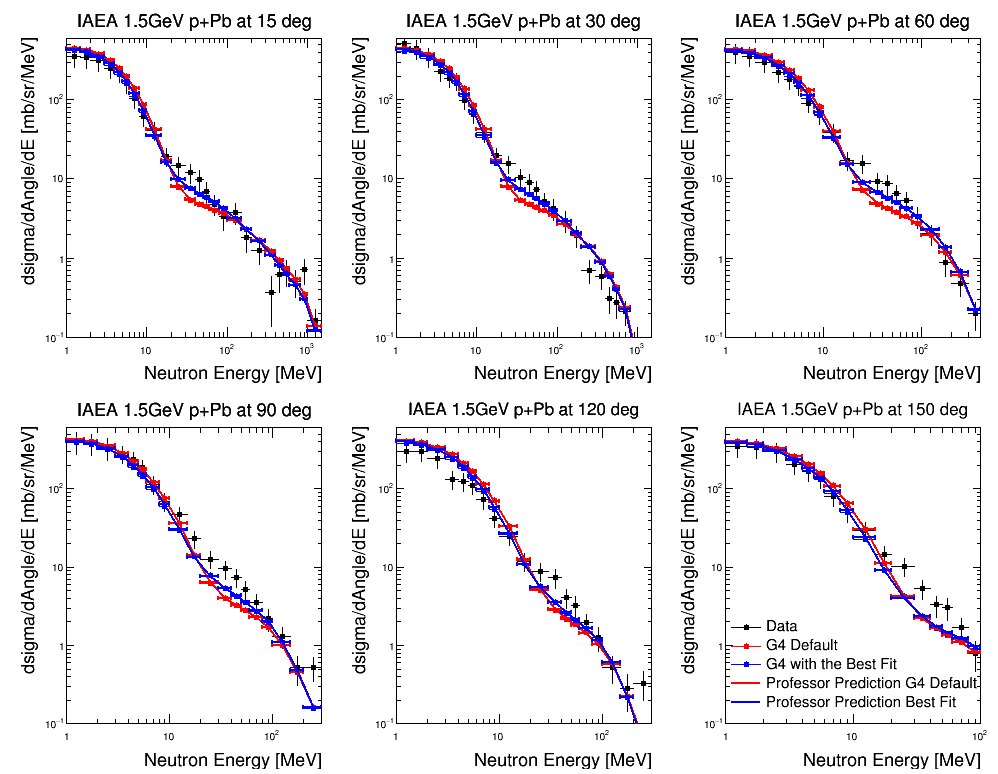}
\caption{\label{fig:professorcomp} Results of the best-fit Professor prediction (blue line) from the Bertini Global fit, compared to IAEA 1.5 GeV $pPb\rightarrow nX$ data (black points) and to a Geant4 simulation run with the best fit parameters extracted from the Professor fit.  In general, the professor prediction agrees with a full simulation, but the Professor prediction fails in some areas of phase space.}
\end{figure}

To study the efficacy of the interpolation, we compared the best-fit predictions returned by Professor with full Geant4 simulations run with the best-fit parameters.  In general, Professor does a reasonably good job of predicting the full simulation, but breaks down in some areas of phase space, as shown in Figure~\ref{fig:professorcomp}.  We find this to be particularly likely when parameters are pushed to the edges of the allowed limits. 
By default, the Professor uses the third order polynomial expansion with all 
parameters involved in the fit, which seems to be sufficient for most of cases when parameters are well behaved within the allowed range. Nonetheless,
increasing the order of the polynomial fit function used by Professor may improve this behavior.  

\section{Future Work}
\label{sec:future}
This work illustrates that recently available variable parameters in the Bertini, PreCompound and FTF models in Geant4 provide a powerful mechanism for tuning Geant4 predictions to data.  The work was motivated by a desire for a framework for propagating Geant4 model uncertainties to measurements, similar to that provided by the GENIE neutrino event generator~\cite{genie1,genie2}.  Many steps remain before that could be a reality, including
\begin{itemize}
\item {\bf Additional datasets and degrees of freedom in fits:}
While the fits described in Section~\ref{sec:fits} do improve data/simulation agreement, it is clear that the parameters currently available in Geant4 are not sufficient to bring predictions into agreement with the data.  Fits to individual datasets indicate that simple extensions of parameters, e.g. allowing different parameters for different projectiles or target nuclei would improve agreement, but that these simple extensions would need to be supplemented by additional new parameters, and/or significantly expanded ranges of existing parameters.  Also, there are clear limitations of the current parameters; FTF provides parameters that alter models only for baryon projectiles, for example.  There is also additional thin target data that could be included in additional fits, that may require yet more parameters beyond what is indicated here.    
\item {\bf Inclusion of cross sections}
All of the parameters discussed here modify the final states simulated by Geant4 given that some type of interaction has occurred.  They do not modify the probability that an interaction or particular type of interaction will occur.  Any complete assessment of Geant4 uncertainties would need to include uncertainties in the cross sections themselves, not just final state models.  In fact, the interaction cross sections are among the most important sources of uncertainty in some high energy physics measurements, and experiments have developed event weighting mechanisms to assess their impact~\cite{minerva_flux,t2kflux}.  
\item {\bf Expansion to more \gf models}
The models discussed here are the main hadronic models used by the Geant FTFP\_BERT physics list to simulate the passage through matter of hadronic particles with energy between 0 and 100 TeV.  They were chosen because they are important models for a variety of planned and future high energy physics experiments.  But there are a number of interaction types that are not covered by the models or parameters discussed here.  Electromagnetic models, for example, although fairly well understood, can lead to uncertainties in modeling of quantities such as shower shape in detectors. Other Geant4 hadronic models that could be included in future fits are INCL++ (Inter-nuclear Cascade), QGS (Quark Gluon String) and BIC (Binary Light Ion Cascade).
\item {\bf Treatment of correlated errors in datasets}
All of the fits described above assume that there are no bin-to-bin correlations in the thin-target data.  This approach was taken because the experiments reporting thin target data have generally not provided covariance matrices.  However, it is likely that some of the uncertainties in these datasets are correlated.  For example, the NA49 data included a 3.8\% systematic uncertainty which may be correlated across bins.  Future efforts at fitting \gf parameters should consider the possibility of bin-to-bin correlations in the datasets, which may substantially alter the fit results.  
\item {\bf Reweightable parameters}
A major hurdle to using the parameter framework currently available in Geant4 for propagation of model-related systematics to physics measurements is that the parameters are not reweightable.  That is, a simulation with varied parameters cannot be produced by applying event weights to some nominal simulation; when a parameter is changed, the user must run an entirely new simulation with the parameter change.  Thus, to propagate \gf systematics to measurements, experiments would have to produce many different varied simulations.  Currently, the computational resources required to generate a single simulation make this prohibitively expensive for most high energy physics collaborations.  Although modification of models and assessment of systematic uncertainties via event weighting has drawbacks (e.g. weighting can never create events in phase space that was not generated in the first place), it is the only way many experiments can realistically propagate model uncertainties at present.  Implementation of a reweighting engine for the Geant4 parameters would therefore dramatically improve the feasibility of using these parameters for error propagation.  One method of reweighting that effectively modifies Geant4 models is reweighting of double or triple differential cross sections~\cite{minerva_flux,t2kflux}.  The advent of high performance computing may make generation of alternative samples more feasible in the future.  
\end{itemize}

\section{Conclusion}
We have studied the hadronic model parameters recently made available by developers of the Bertini,PreCompound, and FTF models in Geant4.  These parameters facilitate variation of the final state content of hadronic interactions in detector and beam simulations.  We have varied each of the parameters within the ranges set by developers and compared the resulting simulations to an array of thin target datasets.  We have identified the parameters that create the largest variations in predictions, and have tuned those parameters to the data using the Professor fitting framework.  Although agreement with data is improved, model predictions are still quite far from data in many areas of phase space.    Steps for further work that would develop this parameter infrastructure into a framework for error propagation have also been outlined.  The toolkit and fitting framework used for these studies is available for use through correspondence with the authors.

\acknowledgments

This manuscript has been authored by Fermi Research Alliance, LLC under Contract No. DE-AC02-07CH11359 with the U.S. Department of Energy, Office of Science, Office of High Energy Physics.

We gratefully acknowledge the efforts of the Geant4 collaboration, and the authors of the Bertini, PreCompound, and FTF models in particular, for their efforts to make available model parameters.    

We also gratefully acknowledge the GENIE collaboration, whose framework for model parameter variation was an impetus for this work.  

The authors would also like to thank Holger Schulz for his advice on using the Professor tool.

\appendix
\section{Appendix 1: Global Fits}
\label{app:global_fits}

\begin{figure}[htbp]
\centering 
\includegraphics[width=15cm, height=10.5cm]{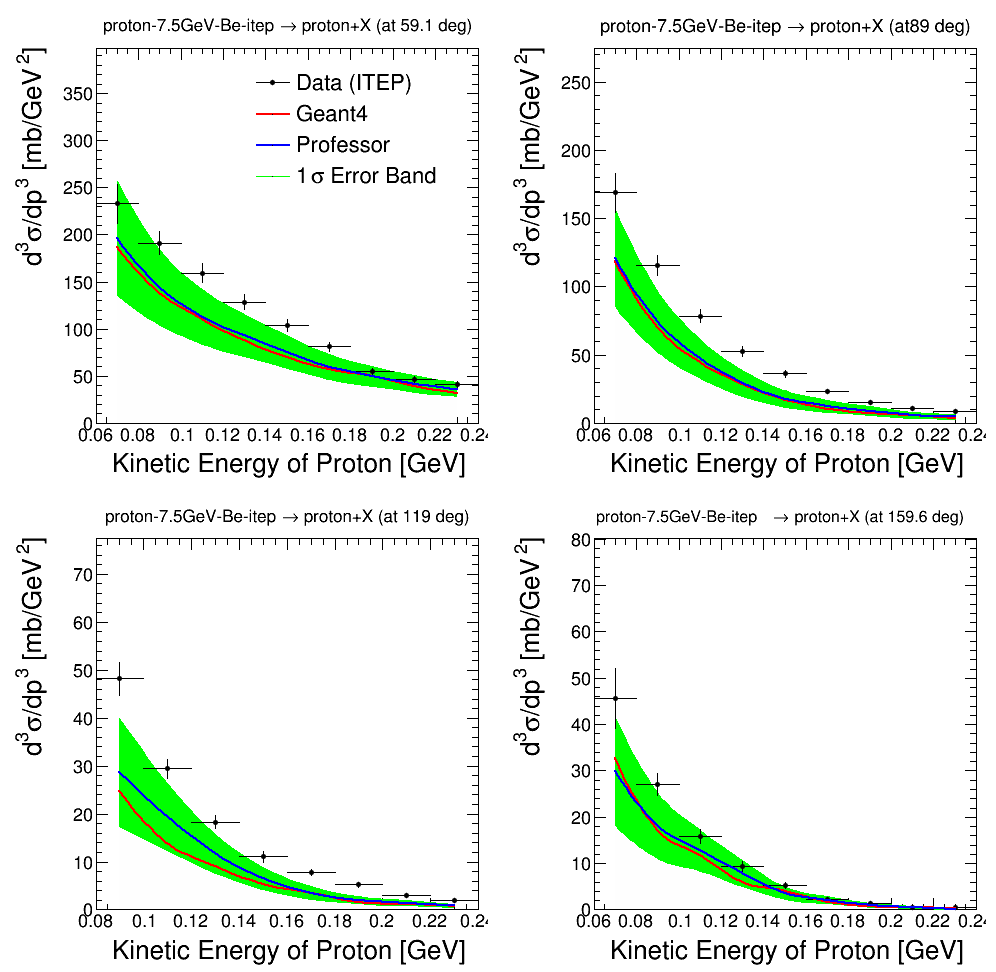}
\caption{\label{fig:fit_bertini1} Results of the global Bertini parameter fit, compared to ITEP 7.5 GeV $pBe\rightarrow pX$  data in bins of final state proton angle.  Data points are shown in black; default Geant4 is red and the global fit result are blue; the green band shows uncertainties propagated from parameter uncertainties returned by the fit.   }
\end{figure}

\begin{figure}[htbp]
\centering 
\includegraphics[width=15cm, height=10.5cm]{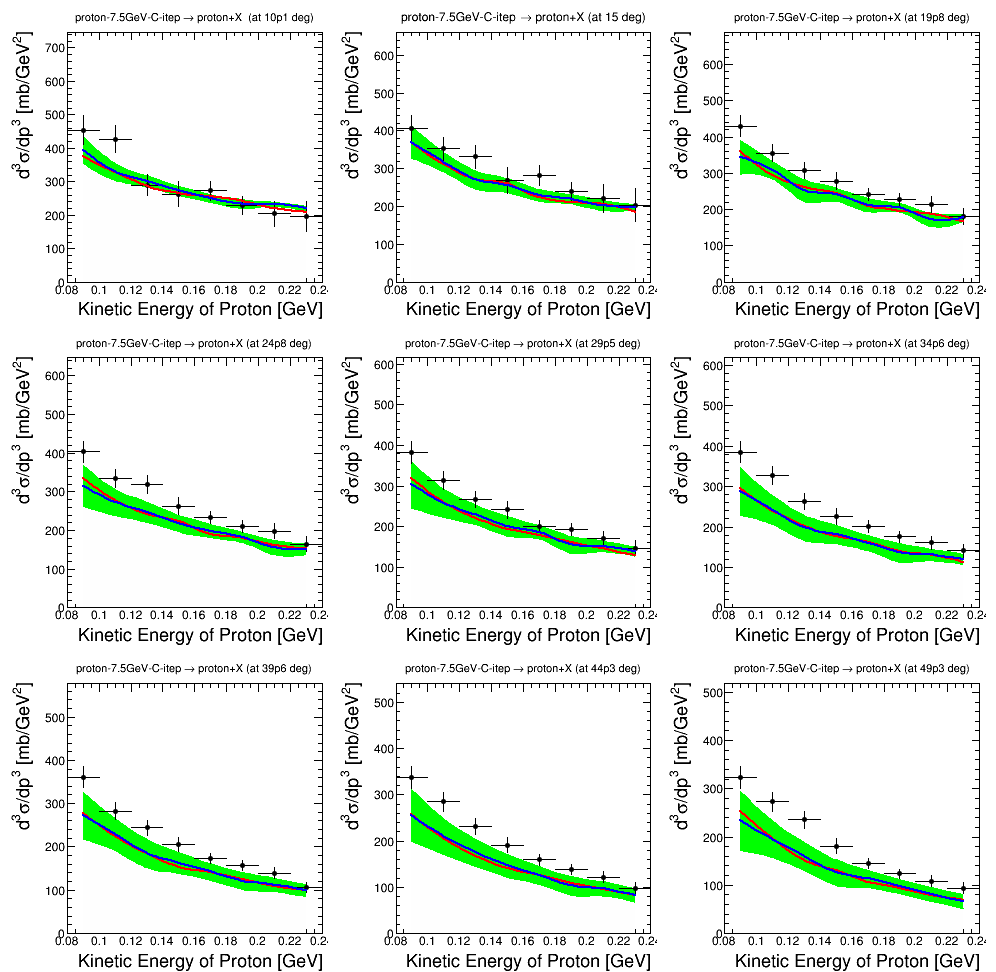}
\includegraphics[width=15cm, height=10.5cm]{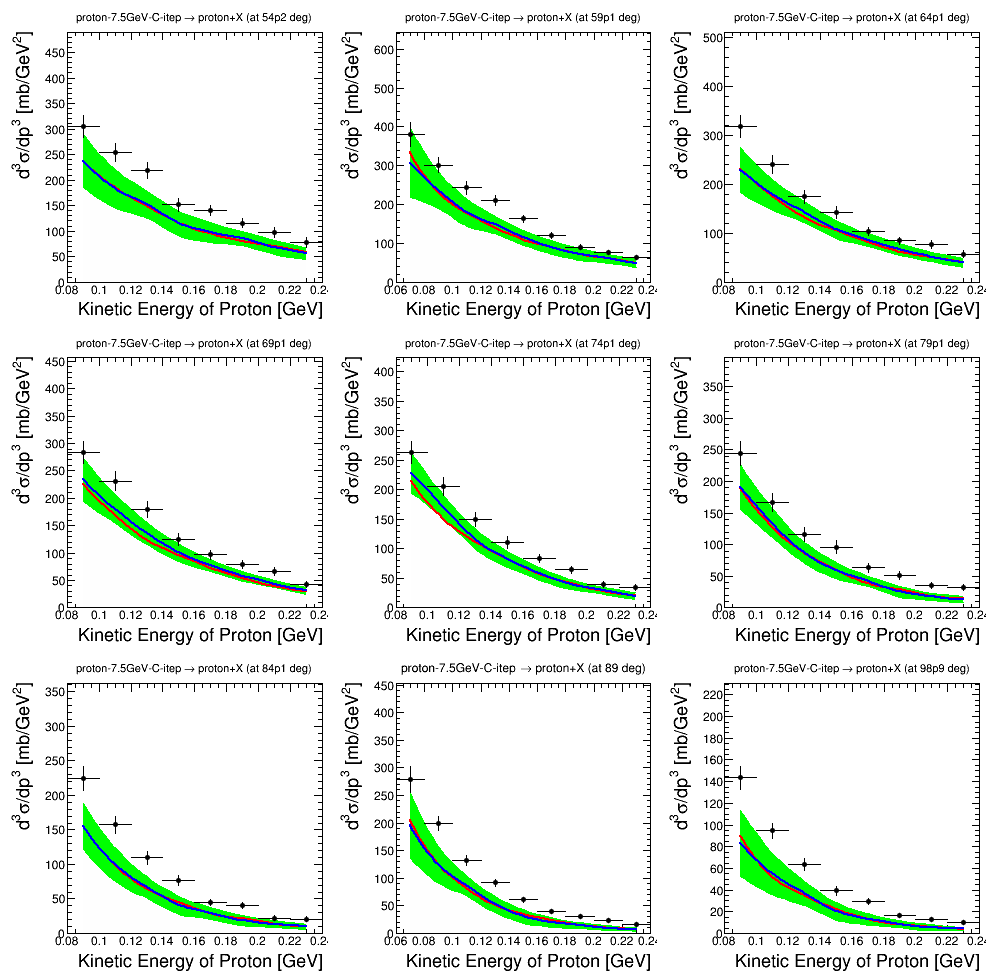}
\caption{\label{fig:fit_bertini2} Results of the global Bertini parameter fit, compared to ITEP 7.5 GeV $pC\rightarrow pX$ data in bins of final state proton angle.  Data points are shown in black; default Geant4 is red and the global fit result are blue; the green band shows uncertainties propagated from parameter uncertainties returned by the fit.   }
\end{figure}

\begin{figure}[htbp]
\centering 
\includegraphics[width=15cm, height=10.5cm]{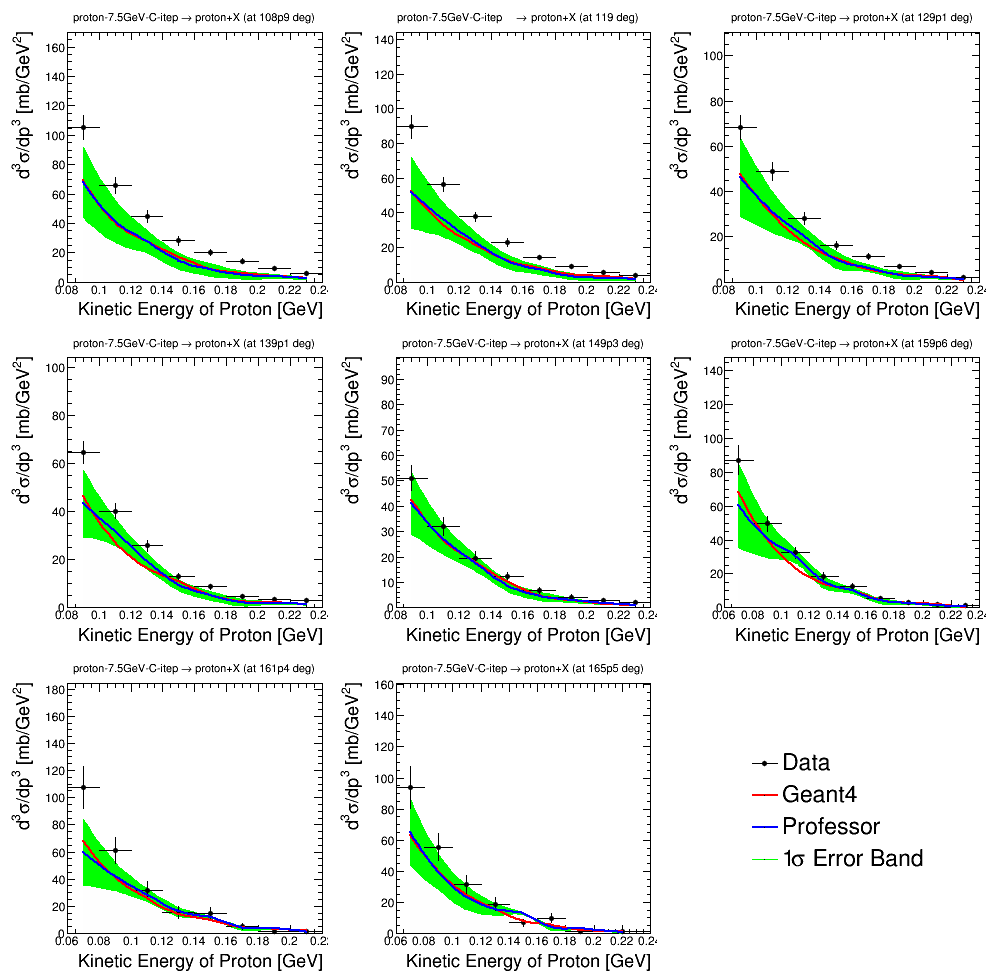}
\caption{\label{fig:fit_bertini3} Results of the global Bertini parameter fit, compared to ITEP 7.5 GeV $pC\rightarrow pX$ data in bins of final state proton angle.  Data points are shown in black; default Geant4 is red and the global fit result are blue; the green band shows uncertainties propagated from parameter uncertainties returned by the fit.   }
\end{figure}

\begin{figure}[htbp]
\centering 
\includegraphics[width=15cm, height=7.3cm]{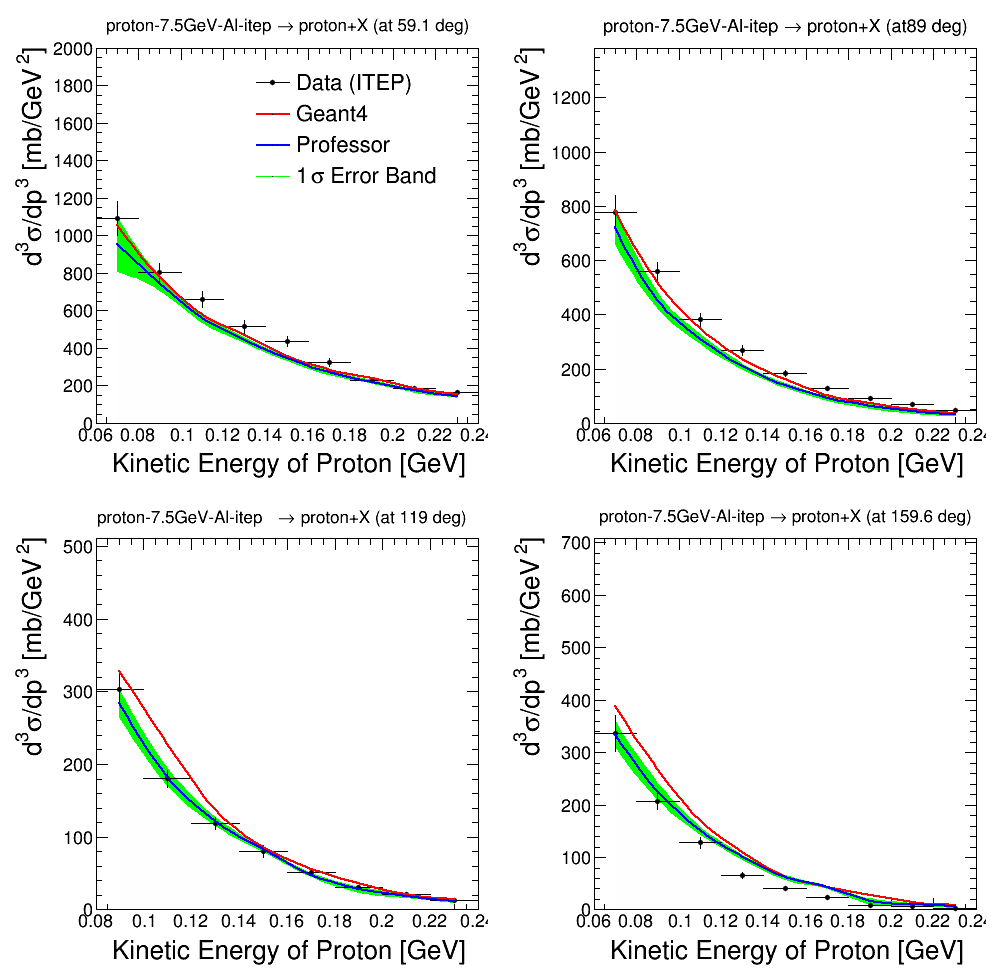}
\includegraphics[width=15cm, height=7.3cm]{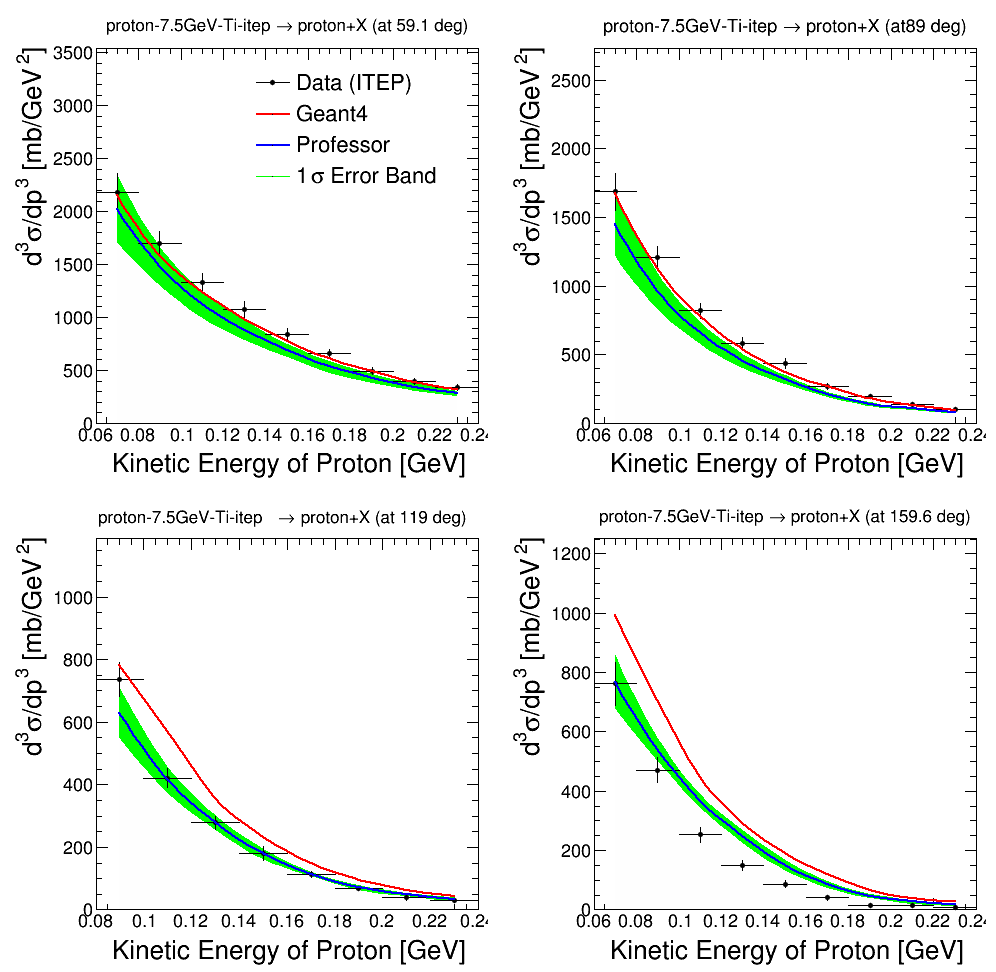}
\includegraphics[width=15cm, height=7.3cm]{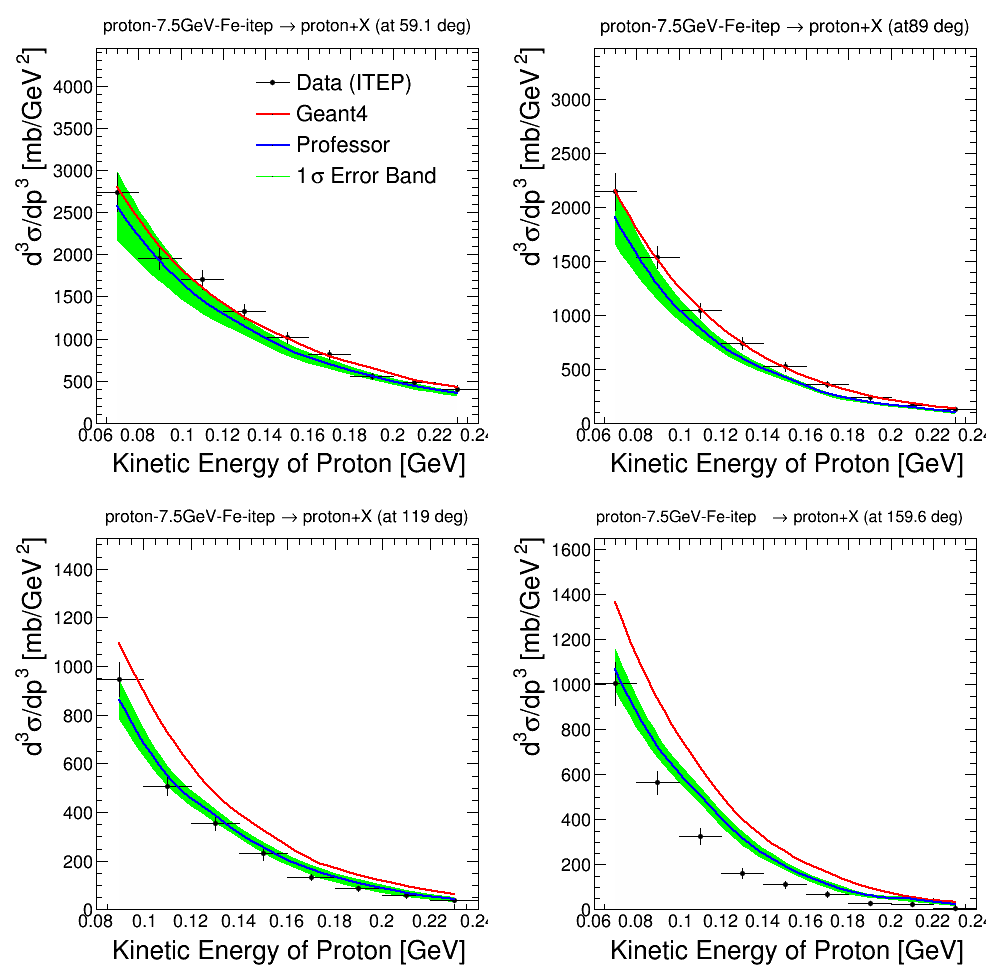}
\caption{\label{fig:fit_bertini4} Results of the global Bertini parameter fit, compared to ITEP 7.5 GeV $pAl\rightarrow pX$, $pTi\rightarrow pX$, and $pFe\rightarrow pX$ data in bins of final state proton angle.  Data points are shown in black; default Geant4 is red and the global fit result are blue; the green band shows uncertainties propagated from parameter uncertainties returned by the fit.   }
\end{figure}

\begin{figure}[htbp]
\centering 
\includegraphics[width=15cm, height=10.5cm]{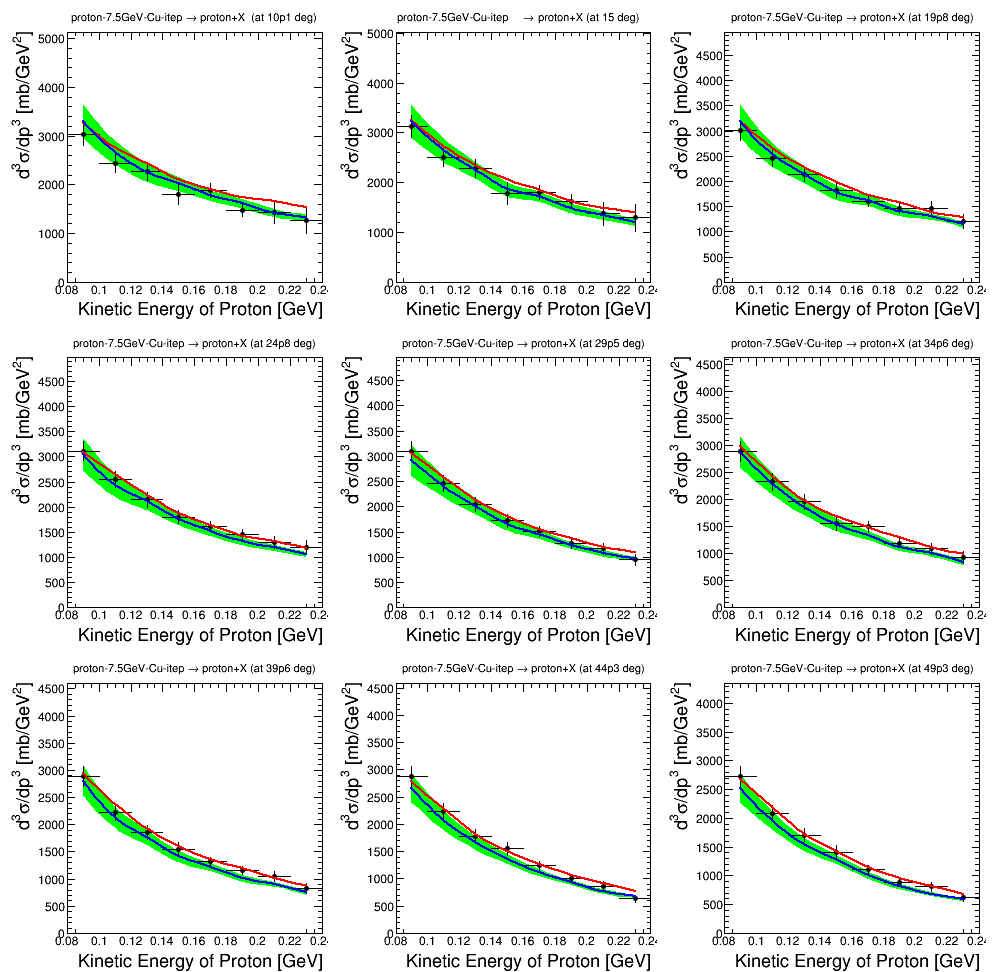}
\includegraphics[width=15cm, height=10.5cm]{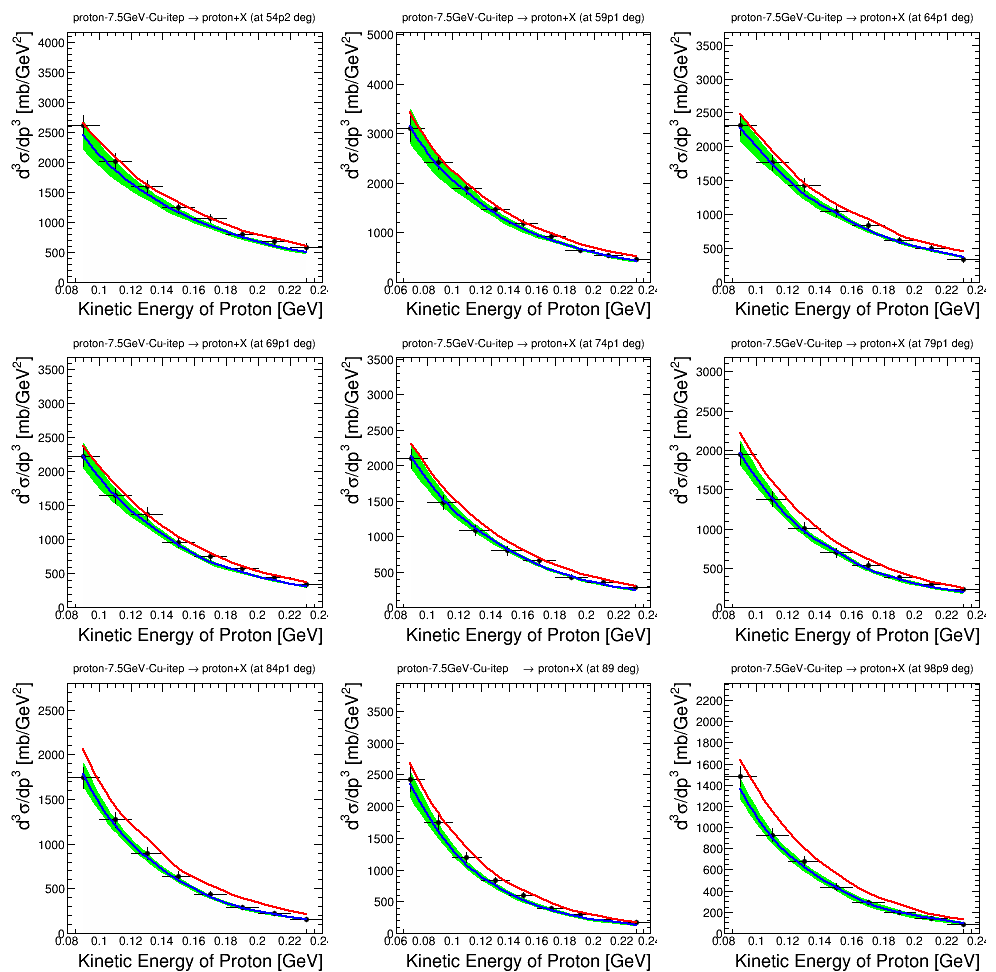}
\caption{\label{fig:fit_bertini5} Results of the global Bertini parameter fit, compared to ITEP 7.5 GeV $pCu\rightarrow pX$ data in bins of final state proton angle.  Data points are shown in black; default Geant4 is red and the global fit result are blue; the green band shows uncertainties propagated from parameter uncertainties returned by the fit.   }
\end{figure}

\begin{figure}[htbp]
\centering 
\includegraphics[width=15cm, height=10.5cm]{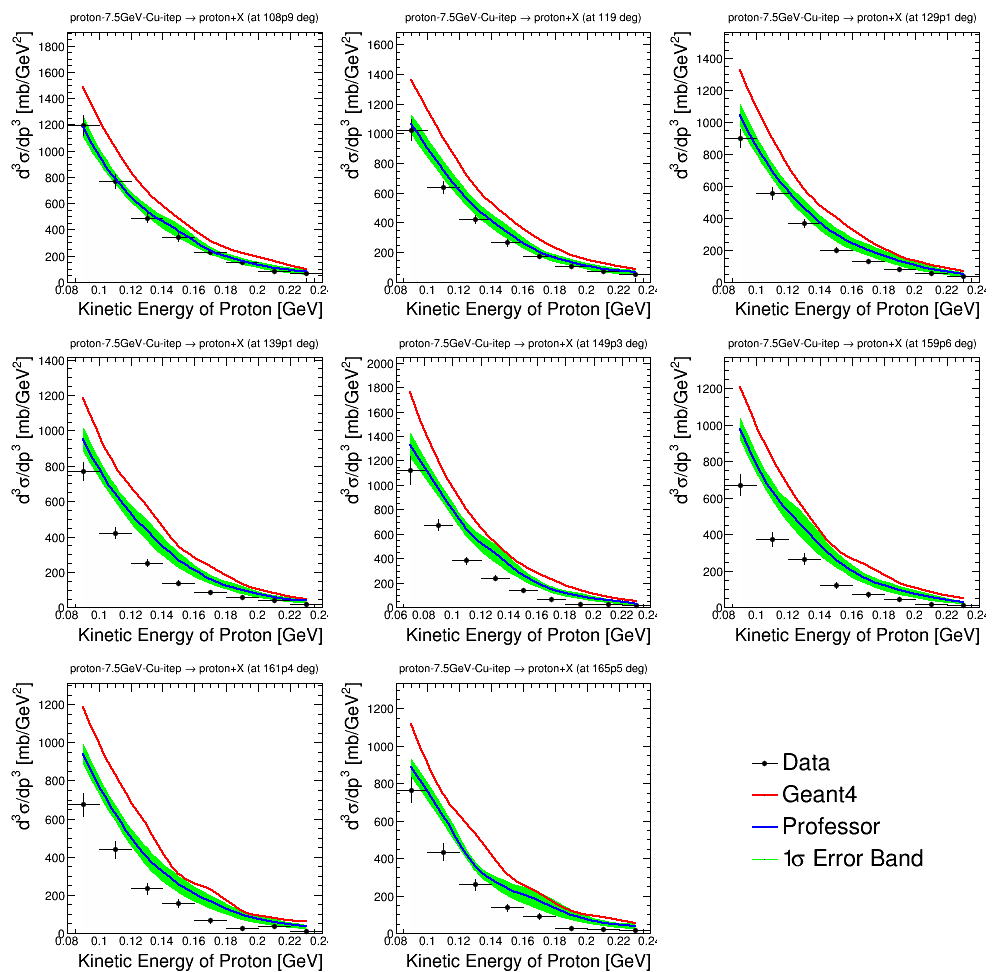}
\caption{\label{fig:fit_bertini6} Results of the global Bertini parameter fit, compared to ITEP 7.5 GeV $pCu\rightarrow pX$ data in bins of final state proton angle.  Data points are shown in black; default Geant4 is red and the global fit result are blue; the green band shows uncertainties propagated from parameter uncertainties returned by the fit.   }
\end{figure}

\begin{figure}[htbp]
\centering 
\includegraphics[width=15cm, height=7.3cm]{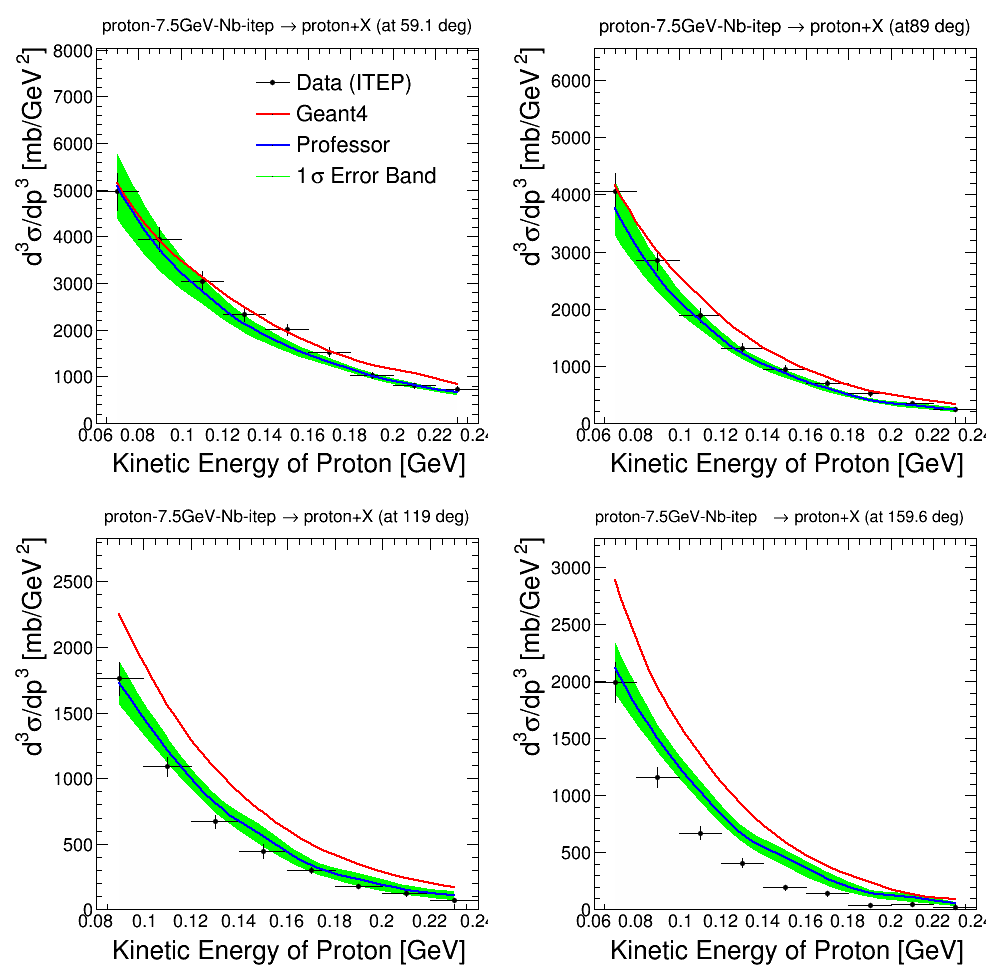}
\includegraphics[width=15cm, height=7.3cm]{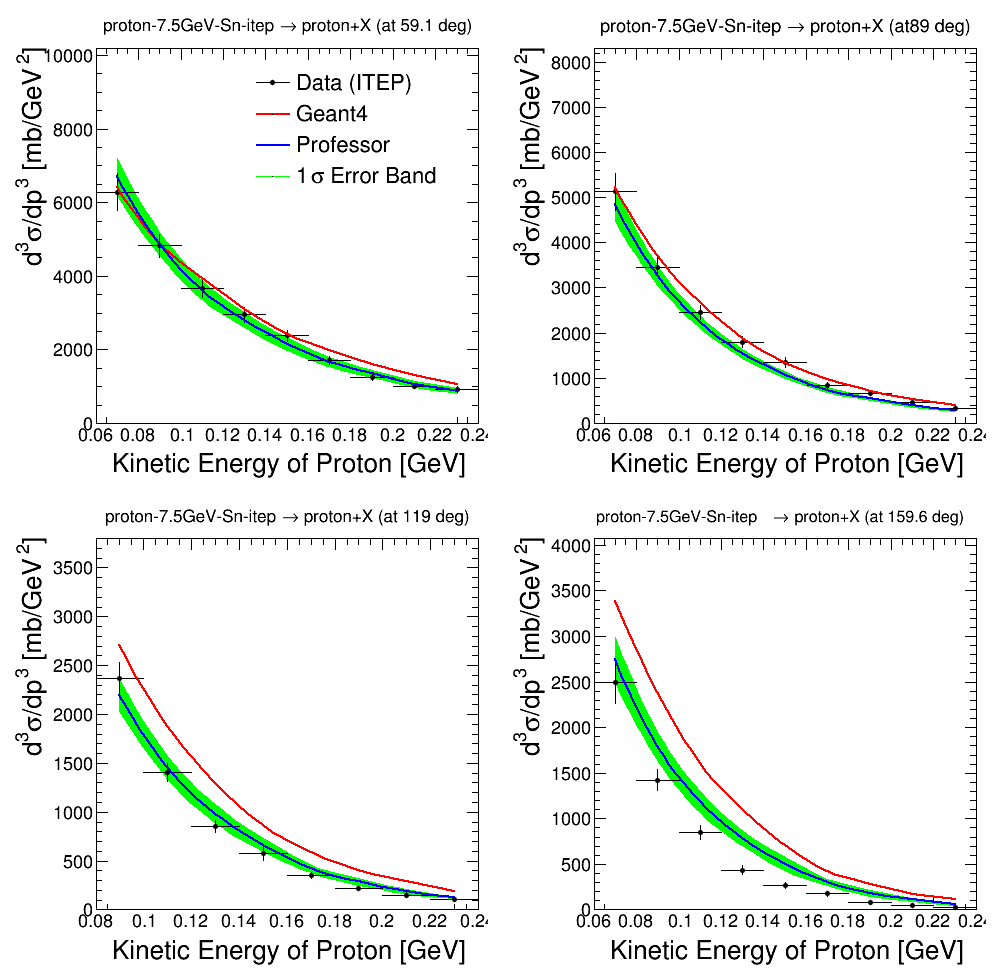}
\includegraphics[width=15cm, height=7.3cm]{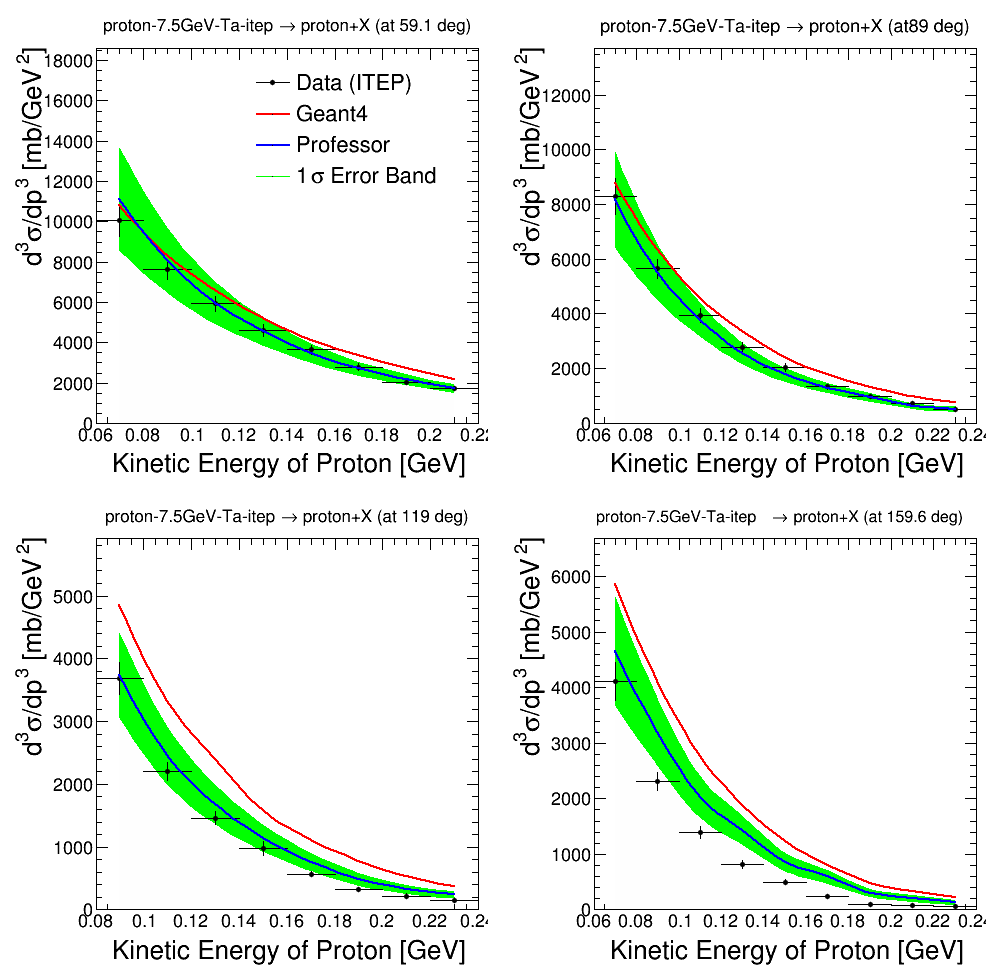}
\caption{\label{fig:fit_bertini7} Results of the global Bertini parameter fit, compared to ITEP 7.5 GeV $pNb\rightarrow pX$, $pSn\rightarrow pX$, and $pTa\rightarrow pX$ data in bins of final state proton angle.  Data points are shown in black; default Geant4 is red and the global fit result are blue; the green band shows uncertainties propagated from parameter uncertainties returned by the fit.   }
\end{figure}

\begin{figure}[htbp]
\centering 
\includegraphics[width=15cm, height=10.5cm]{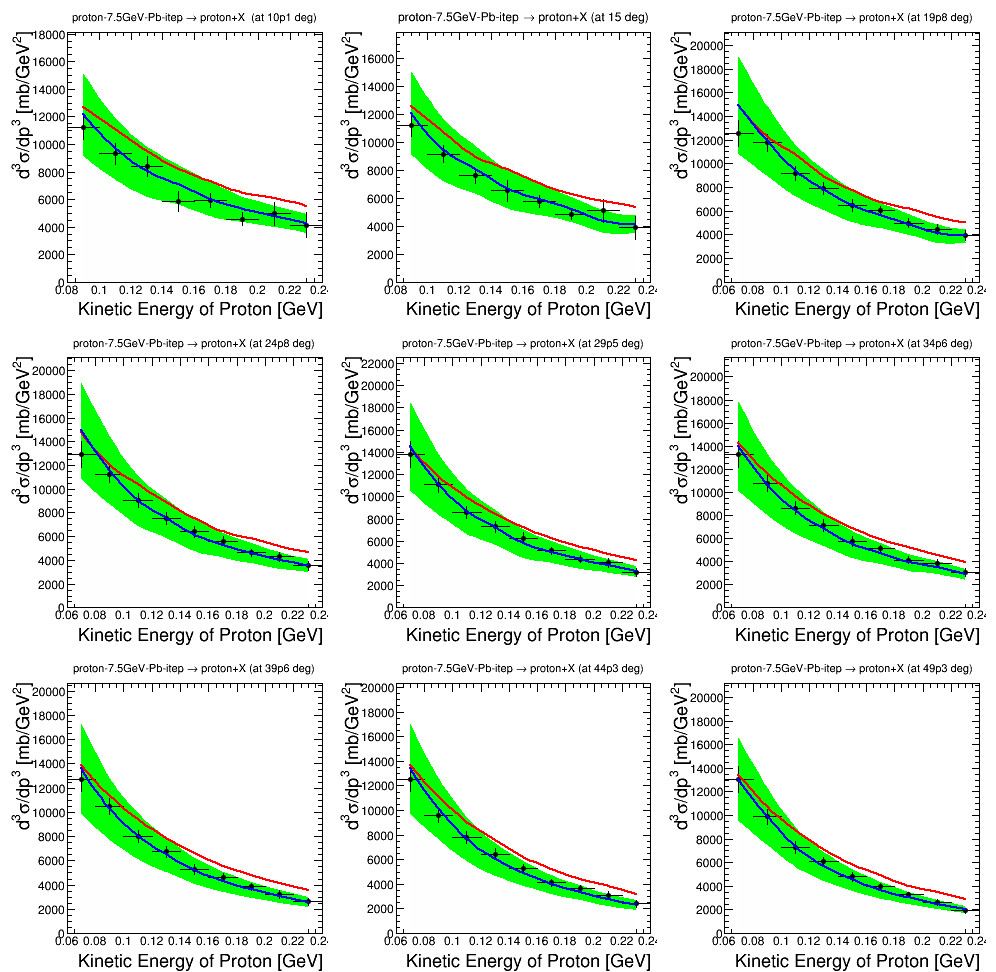}
\includegraphics[width=15cm, height=10.5cm]{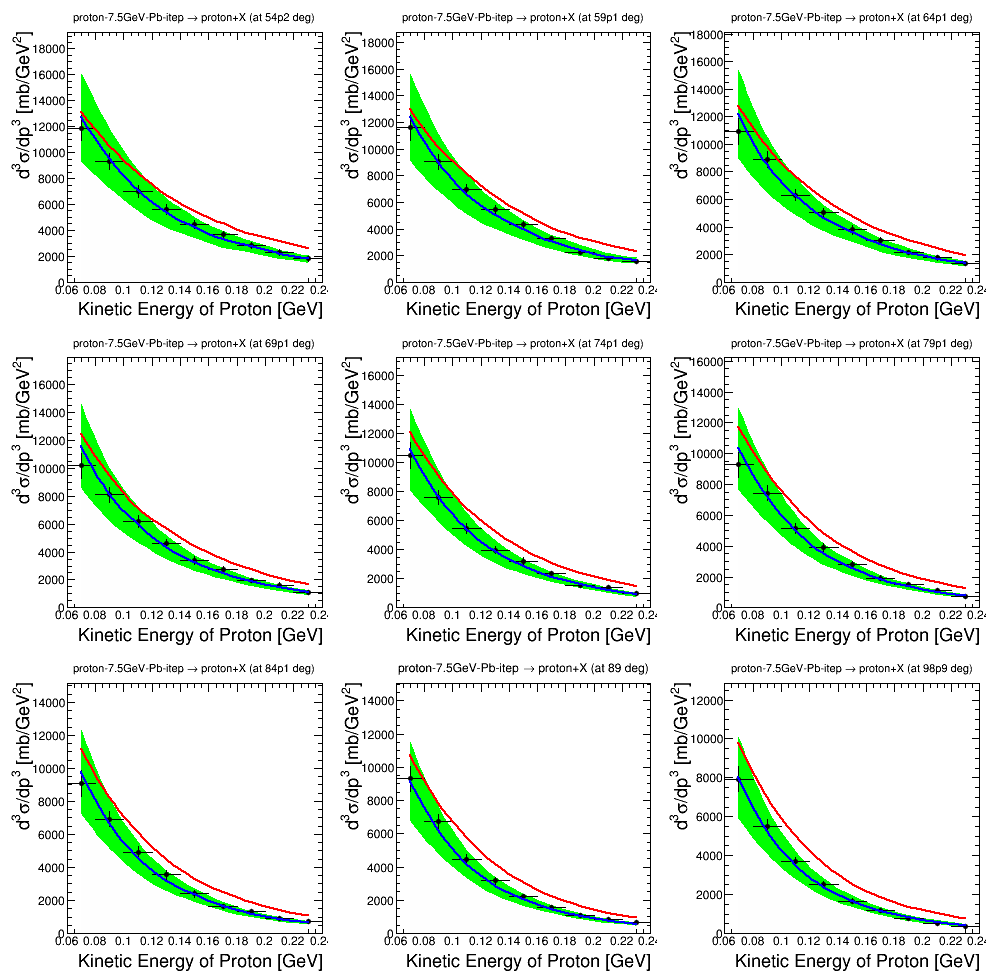}
\caption{\label{fig:fit_bertini8} Results of the global Bertini parameter fit, compared to ITEP 7.5 GeV $pPb\rightarrow pX$ data in bins of final state proton angle.  Data points are shown in black; default Geant4 is red and the global fit result are blue; the green band shows uncertainties propagated from parameter uncertainties returned by the fit.   }
\end{figure}

\begin{figure}[htbp]
\centering 
\includegraphics[width=15cm, height=10.5cm]{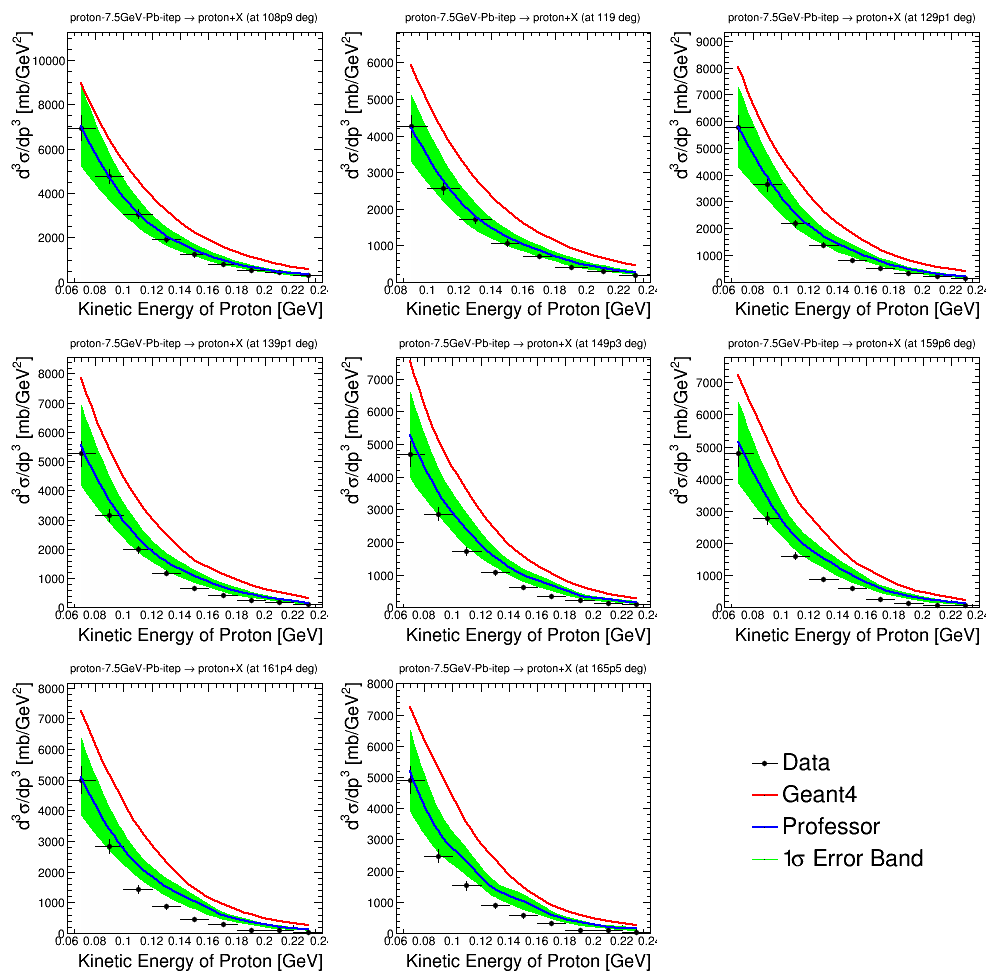}
\caption{\label{fig:fit_bertini9} Results of the global Bertini parameter fit, compared to ITEP 7.5 GeV $pPb\rightarrow pX$ data in bins of final state proton angle.  Data points are shown in black; default Geant4 is red and the global fit result are blue; the green band shows uncertainties propagated from parameter uncertainties returned by the fit.   }
\end{figure}

\begin{figure}[htbp]
\centering 
\includegraphics[width=15cm, height=10.5cm]{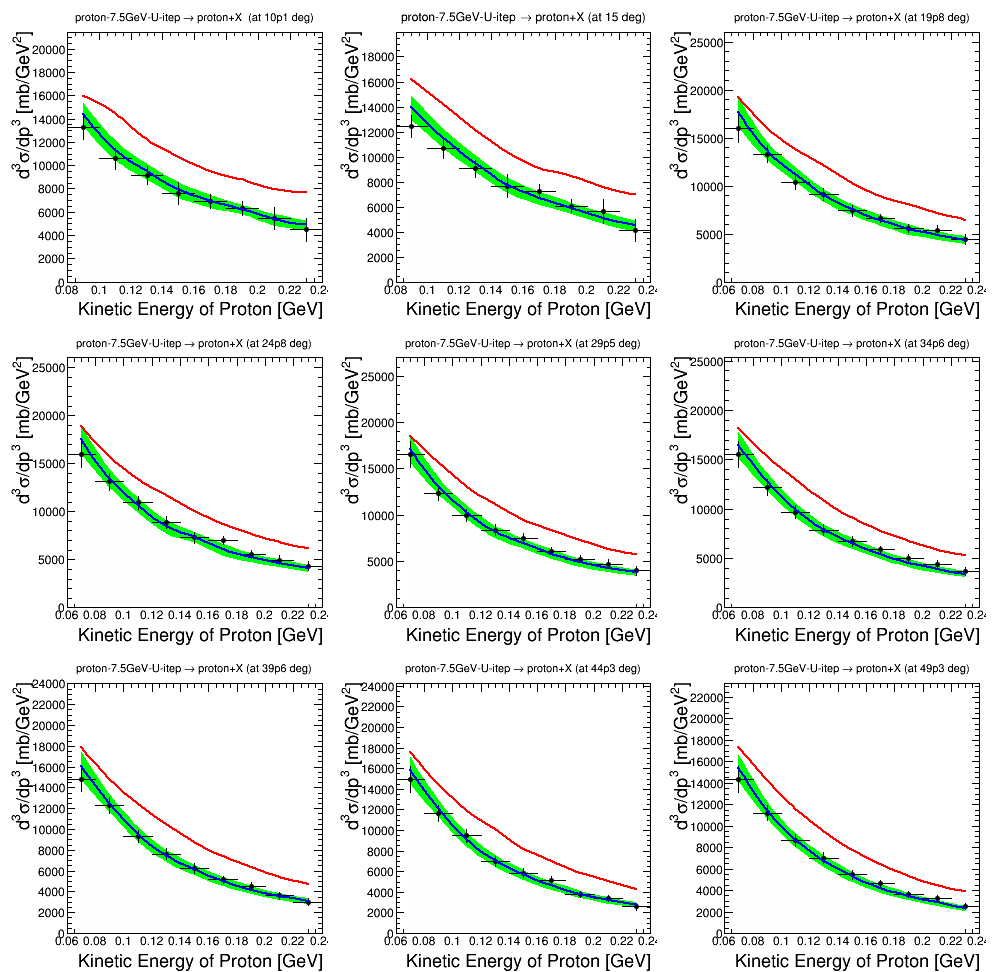}
\includegraphics[width=15cm, height=10.5cm]{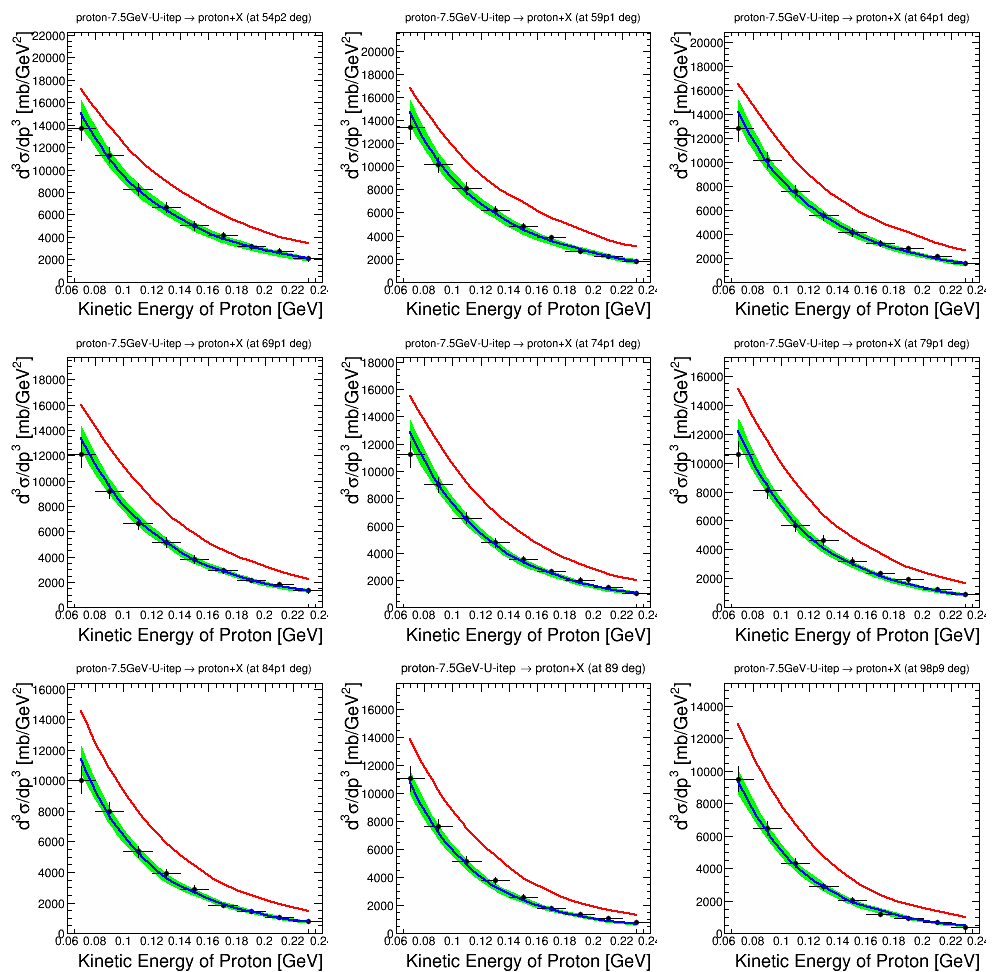}
\caption{\label{fig:fit_bertini10} Results of the global Bertini parameter fit, compared to ITEP 7.5 GeV $pCu\rightarrow pX$ data in bins of final state proton angle.  Data points are shown in black; default Geant4 is red and the global fit result are blue; the green band shows uncertainties propagated from parameter uncertainties returned by the fit.   }
\end{figure}

\begin{figure}[htbp]
\centering 
\includegraphics[width=15cm, height=10.5cm]{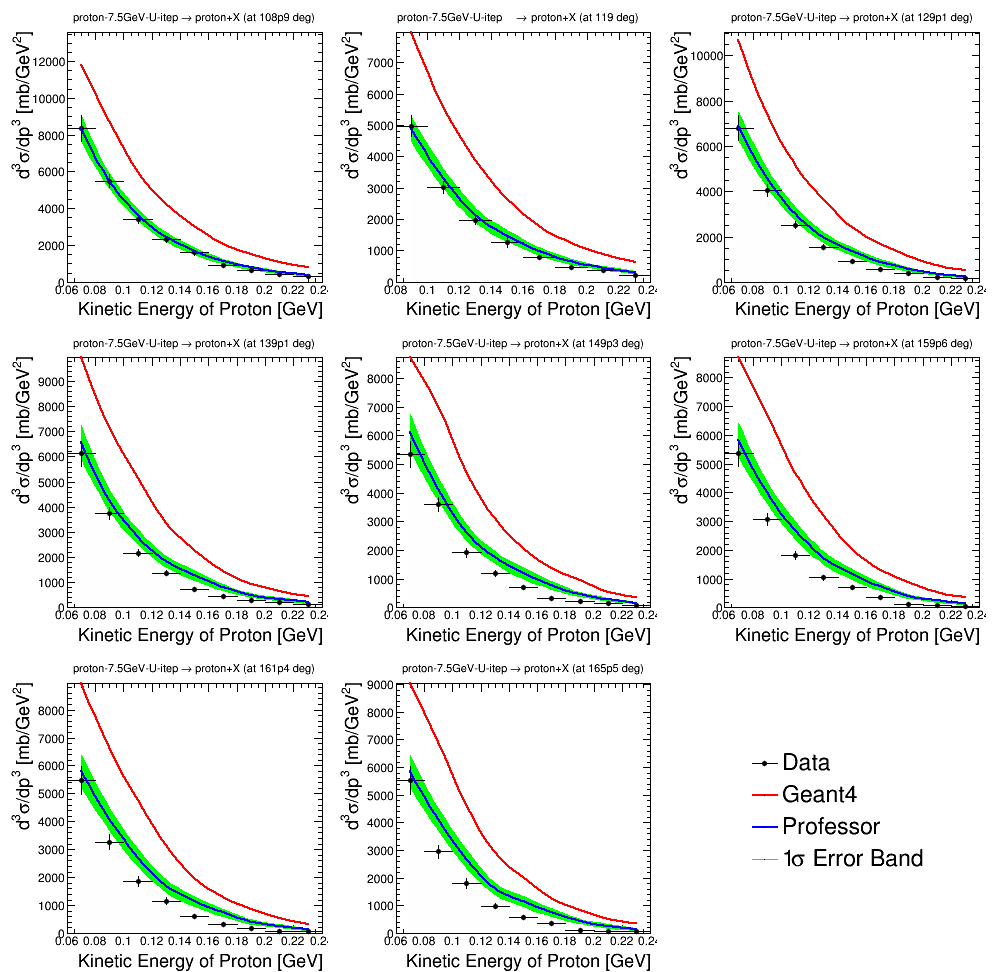}
\caption{\label{fig:fit_bertini11} Results of the global Bertini parameter fit, compared to ITEP 7.5 GeV $pU\rightarrow pX$ data in bins of final state proton angle.  Data points are shown in black; default Geant4 is red and the global fit result are blue; the green band shows uncertainties propagated from parameter uncertainties returned by the fit.   }
\end{figure}

\begin{figure}[htbp]
\centering 
\includegraphics[width=15cm, height=10.5cm]{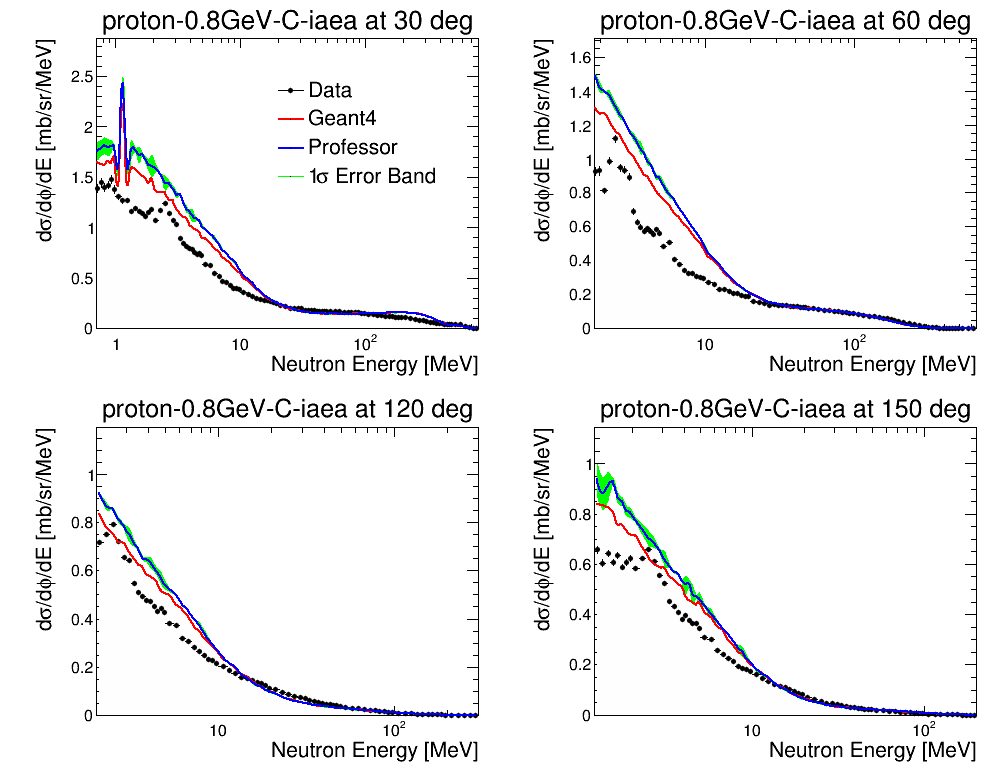}
\includegraphics[width=15cm, height=10.5cm]{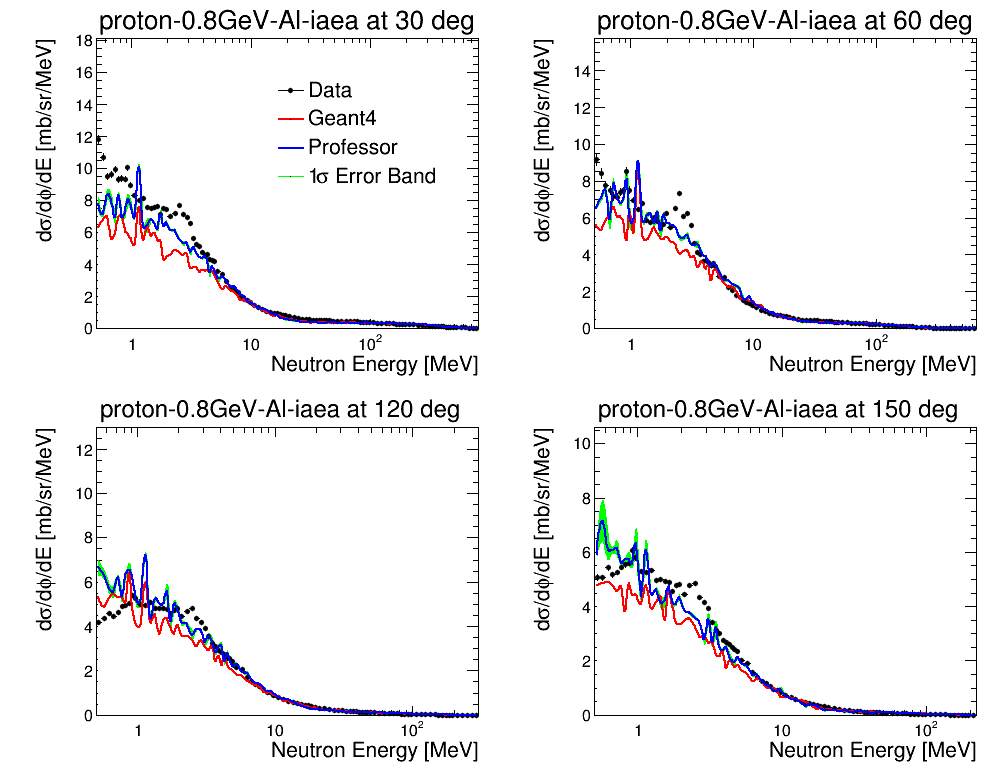}
\caption{\label{fig:precompound2} Results of the global Precompound parameter fit, compared to IAEA 0.8 GeV $pC\rightarrow nX$ and $pAl\rightarrow nX$ data in bins of final state neutron angle.  Data points are shown in black; default Geant4 is red and the global fit result in blue; the green band shows uncertainties propagated from parameter uncertainties returned by the fit.   }
\end{figure}

\begin{figure}[htbp]
\centering 
\includegraphics[width=15cm, height=10.5cm]{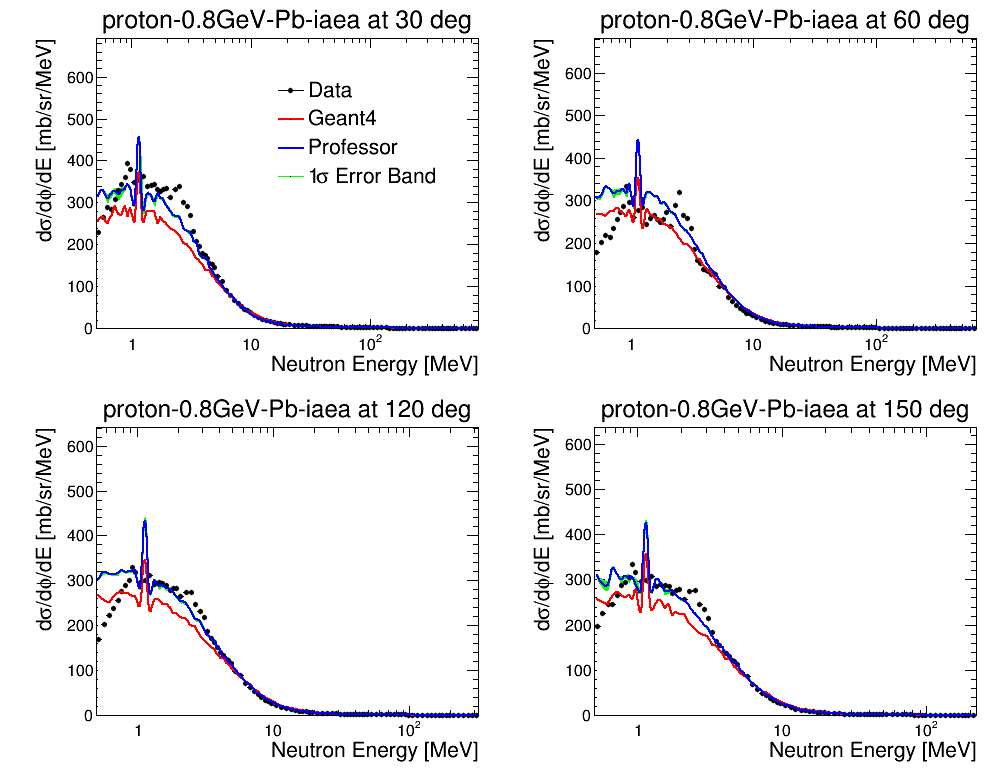}
\caption{\label{fig:fit_precompound3} Results of the global Precompound parameter fit, compared to IAEA 0.8 GeV $pPb\rightarrow nX$ data in bins of final state neutron angle.  Data points are shown in black; default Geant4 is red and the global fit result in blue; the green band shows uncertainties propagated from parameter uncertainties returned by the fit.    }
\end{figure}

\begin{figure}[htbp]
\centering 
\includegraphics[width=15cm, height=10.5cm]{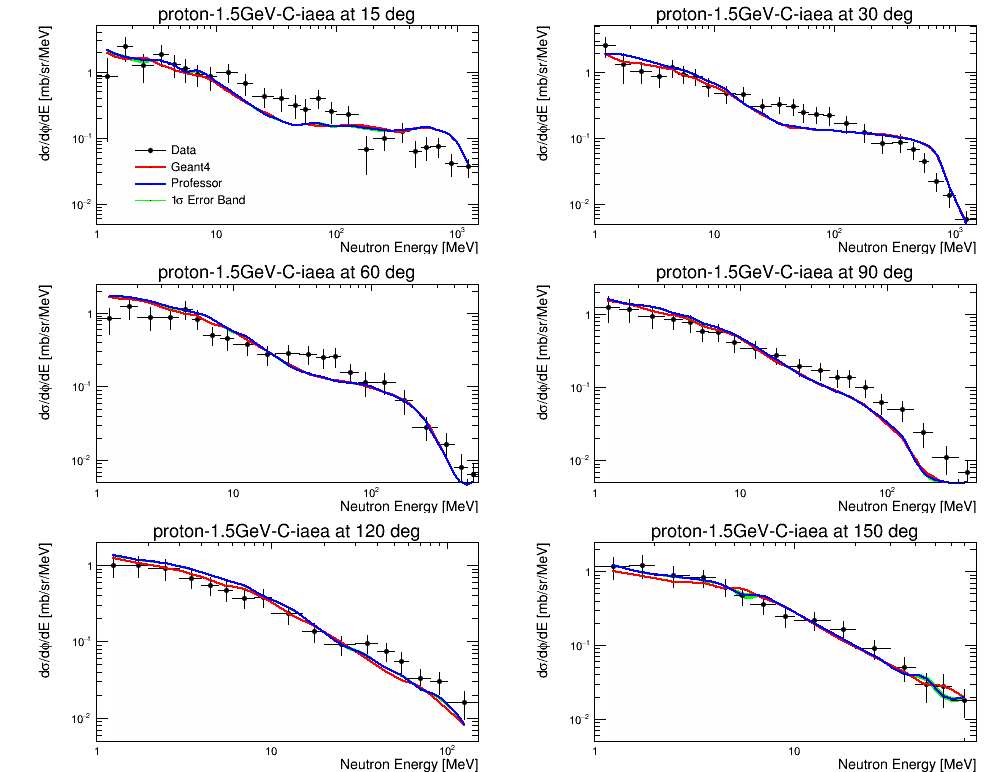}
\includegraphics[width=15cm, height=10.5cm]{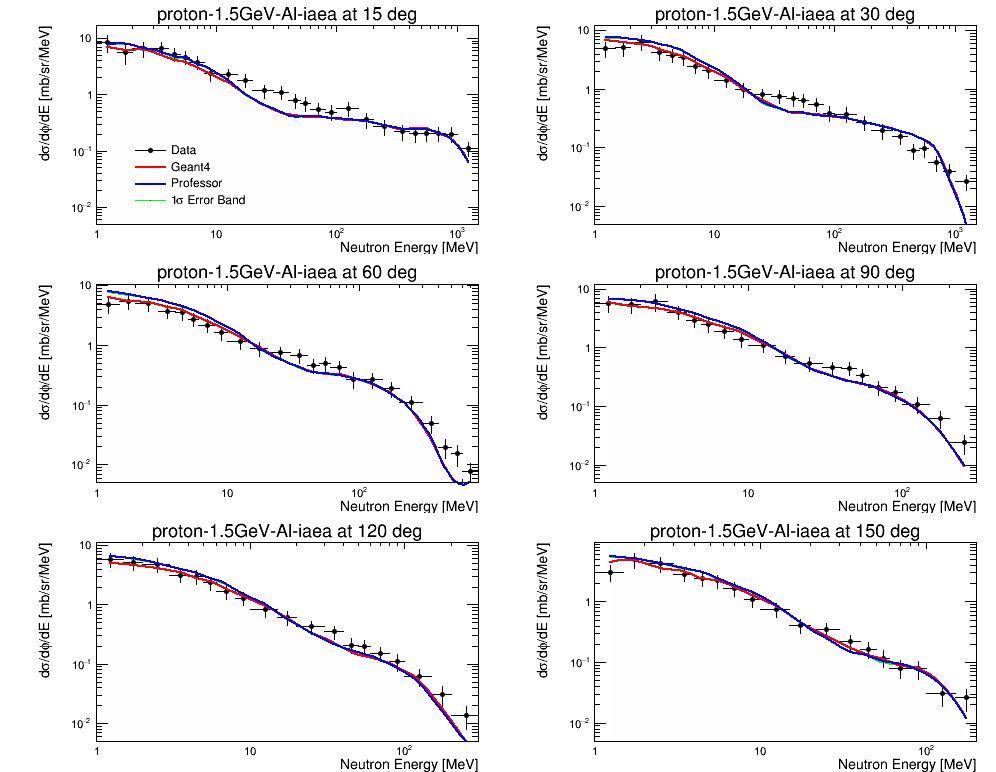}
\caption{\label{fig:fit_precompound4} Results of the global Precompound parameter fit, compared to IAEA 1.5 GeV $pC\rightarrow nX$ and $pAl\rightarrow nX$ data in bins of final state neutron angle.  Data points are shown in black; default Geant4 is red and the global fit result in blue; the green band shows uncertainties propagated from parameter uncertainties returned by the fit.    }
\end{figure}

\begin{figure}[htbp]
\centering 
\includegraphics[width=15cm, height=10.5cm]{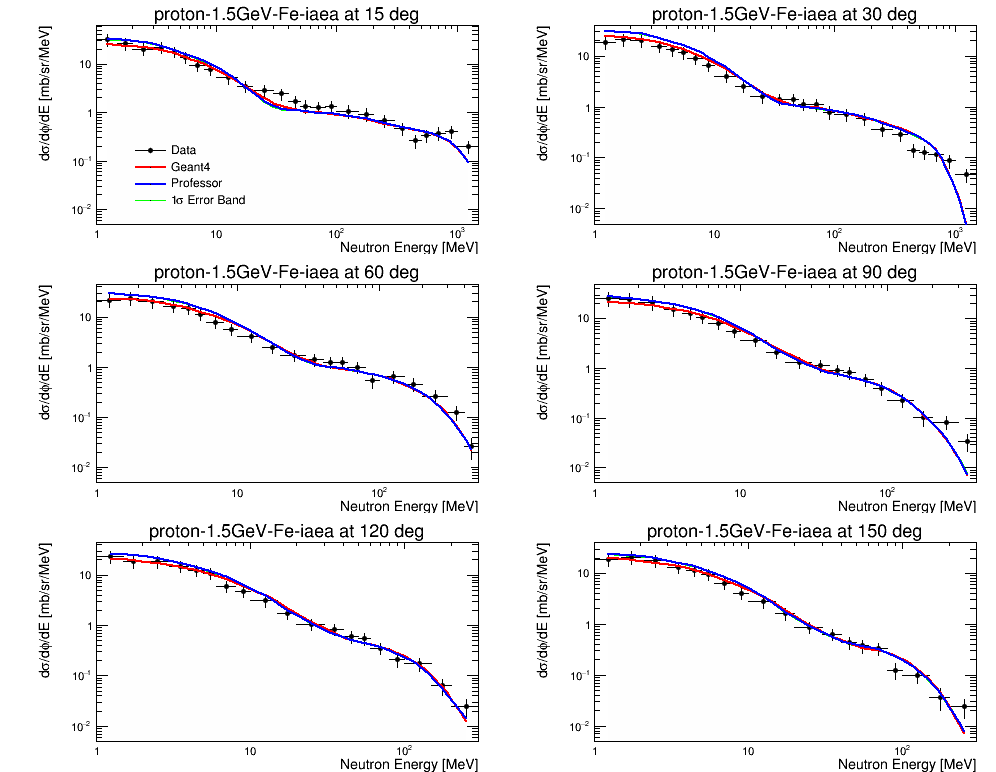}
\includegraphics[width=15cm, height=10.5cm]{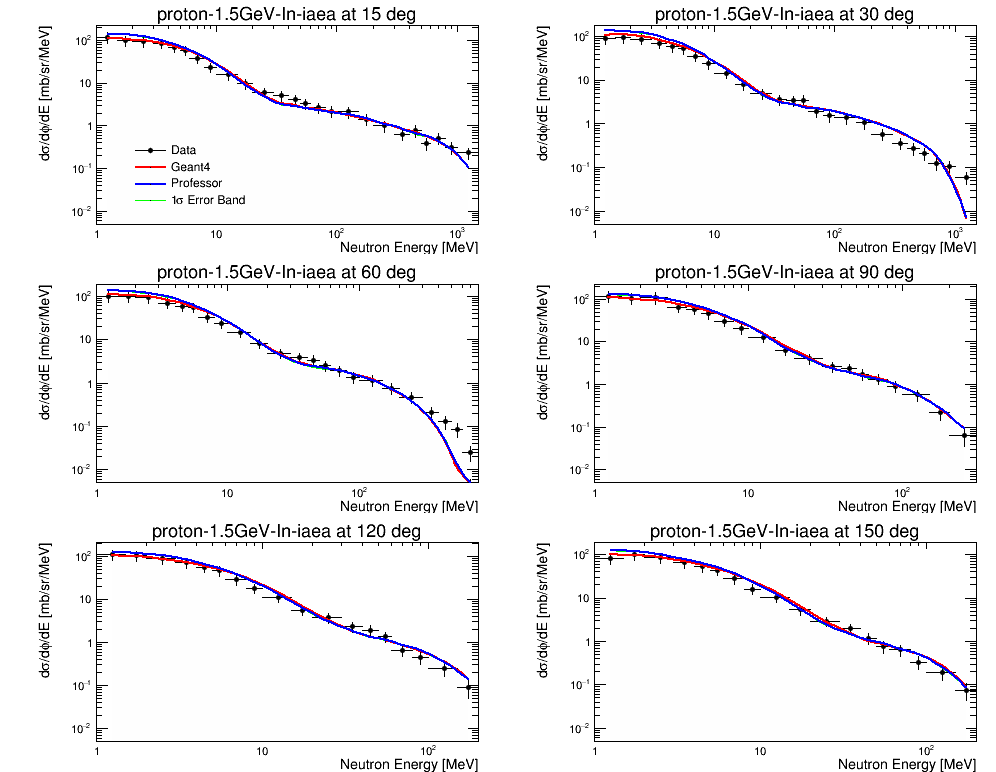}
\caption{\label{fig:fit_precompound5} Results of the global Precompound parameter fit, compared to IAEA 1.5 GeV $pFe\rightarrow nX$ and $pIn\rightarrow nX$ data in bins of final state neutron angle.  Data points are shown in black; default Geant4 is red and the global fit result in blue; the green band shows uncertainties propagated from parameter uncertainties returned by the fit.   }
\end{figure}

\begin{figure}[htbp]
\centering 
\includegraphics[width=15cm, height=10.5cm]{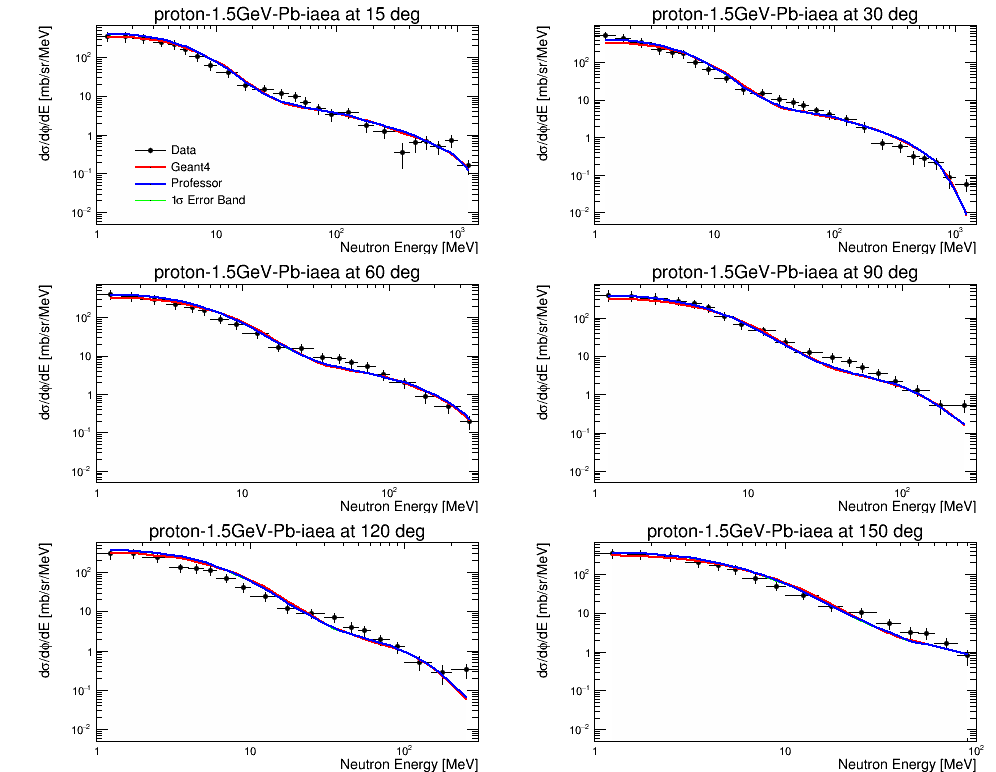}
\caption{\label{fig:fit_precompound7} Results of the global Precompound parameter fit, compared to IAEA 1.5 GeV $pPb\rightarrow nX$ data in bins of final state neutron angle.  Data points are shown in black; default Geant4 is red and the global fit result in blue; the green band shows uncertainties propagated from parameter uncertainties returned by the fit.    }
\end{figure}

\begin{figure}[htbp]
\centering 
\includegraphics[width=15cm, height=10.5cm]{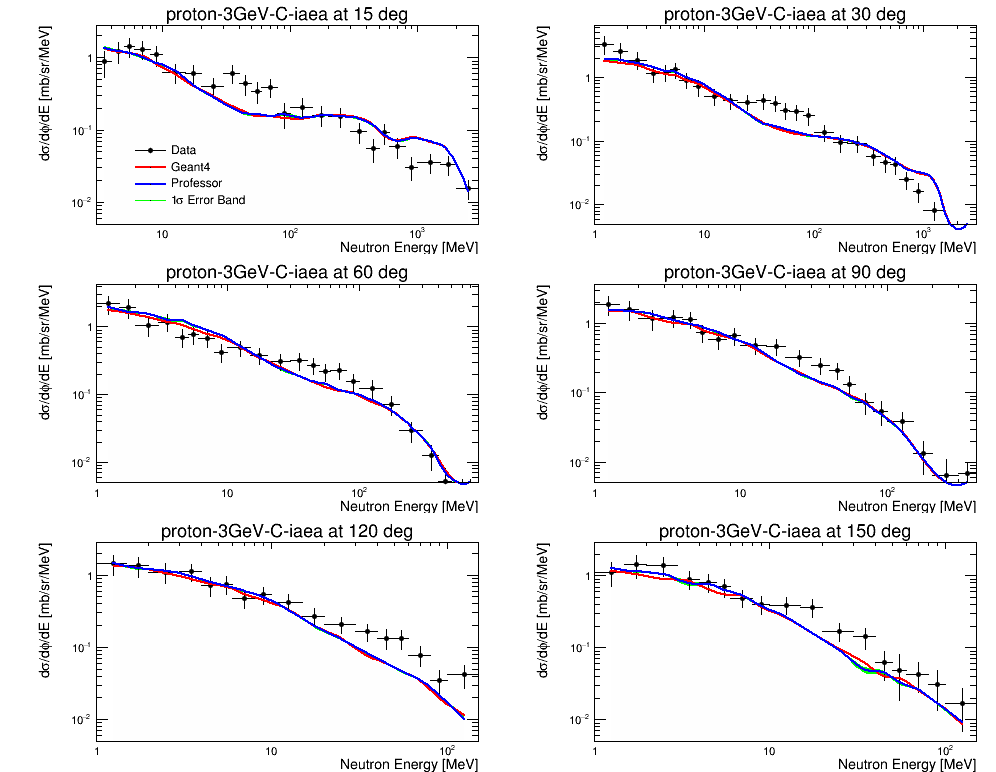}
\includegraphics[width=15cm, height=10.5cm]{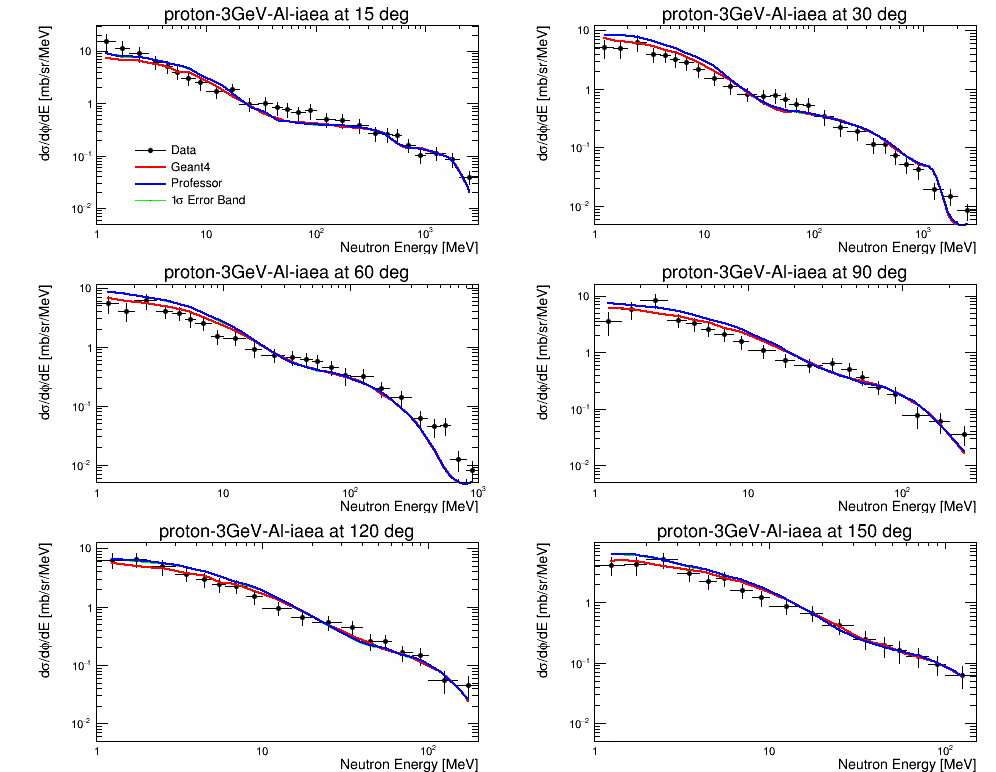}
\caption{\label{fig:fit_precompound8} Results of the global Precompound parameter fit, compared to IAEA 3 GeV $pC\rightarrow nX$ and $pAl\rightarrow nX$ data in bins of final state neutron angle.  Data points are shown in black; default Geant4 is red and the global fit result in blue; the green band shows uncertainties propagated from parameter uncertainties returned by the fit.    }
\end{figure}

\begin{figure}[htbp]
\centering 
\includegraphics[width=15cm, height=10.5cm]{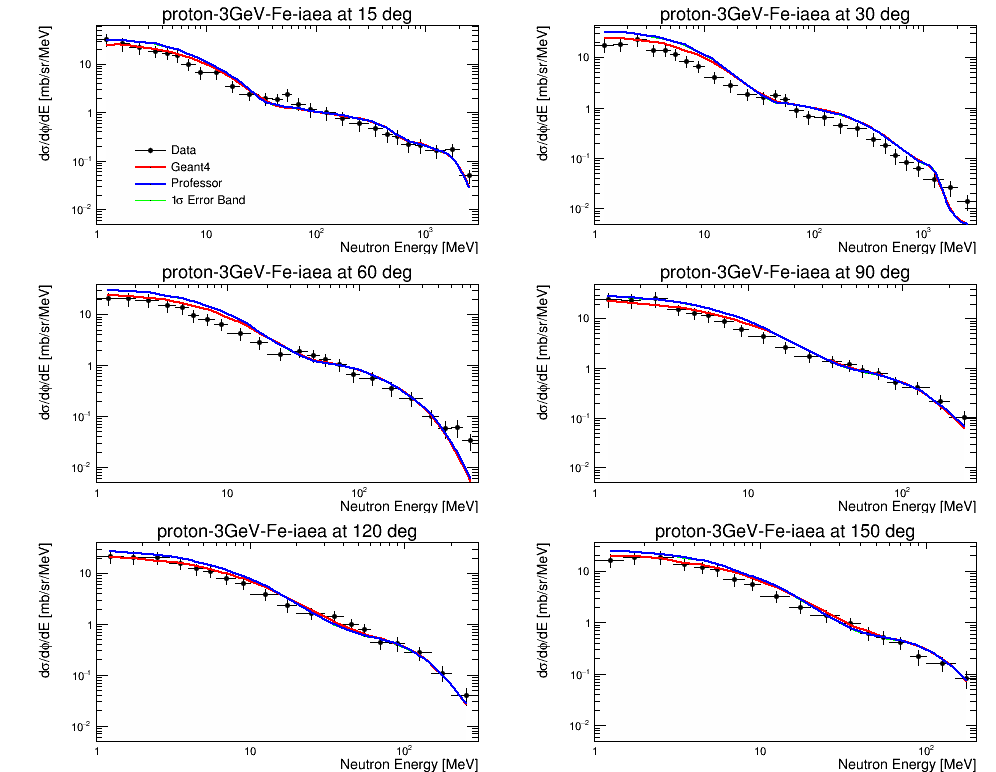}
\includegraphics[width=15cm, height=10.5cm]{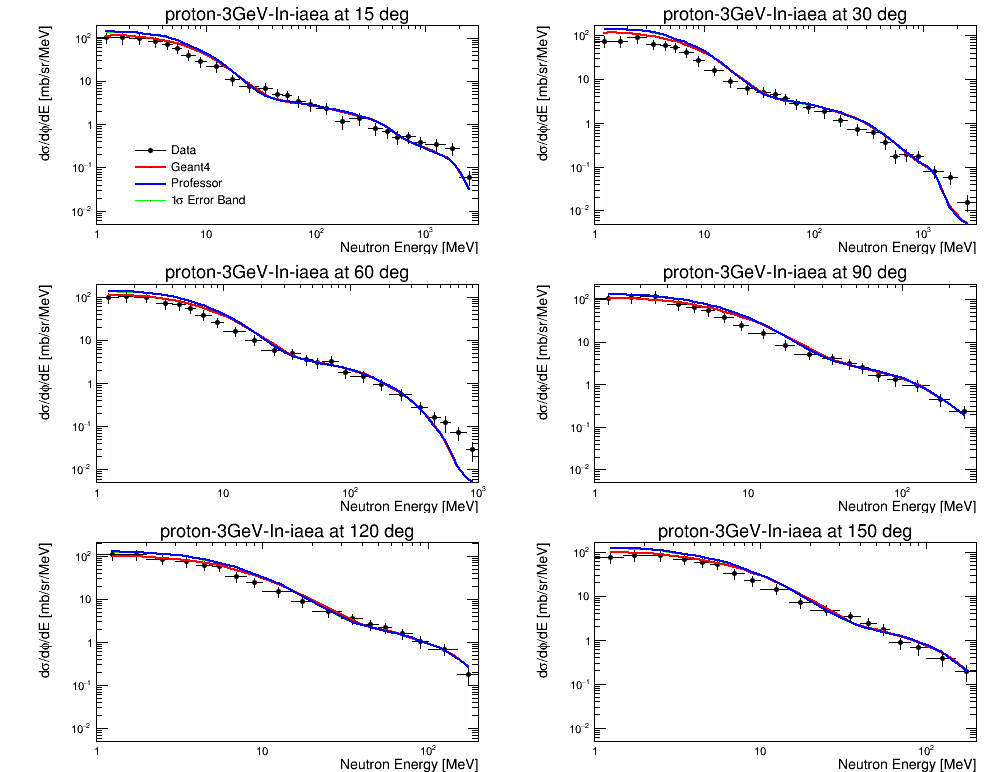}
\caption{\label{fig:fit_precompound9} Results of the global Precompound parameter fit, compared to IAEA 3 GeV $pFe\rightarrow nX$ and $pIn\rightarrow nX$ data in bins of final state neutron angle.  Data points are shown in black; default Geant4 is red and the global fit result in blue; the green band shows uncertainties propagated from parameter uncertainties returned by the fit.    }
\end{figure}

\begin{figure}[htbp]
\centering 
\includegraphics[width=15cm, height=10.5cm]{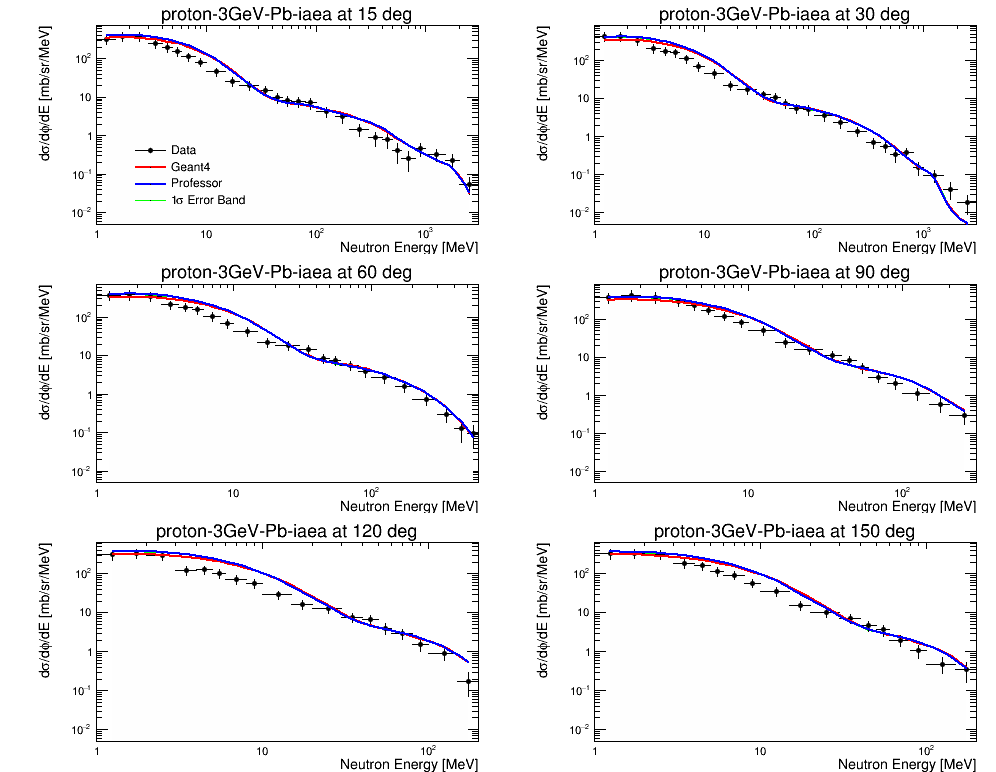}
\caption{\label{fig:fit_precompound10} Results of the global Precompound parameter fit, compared to IAEA 3 GeV $pPb\rightarrow nX$ data in bins of final state neutron angle.  Data points are shown in black; default Geant4 is red and the global fit result in blue; the green band shows uncertainties propagated from parameter uncertainties returned by the fit.   }
\end{figure}

\begin{figure}[htbp]
\centering 
\includegraphics[width=15cm, height=10.5cm]{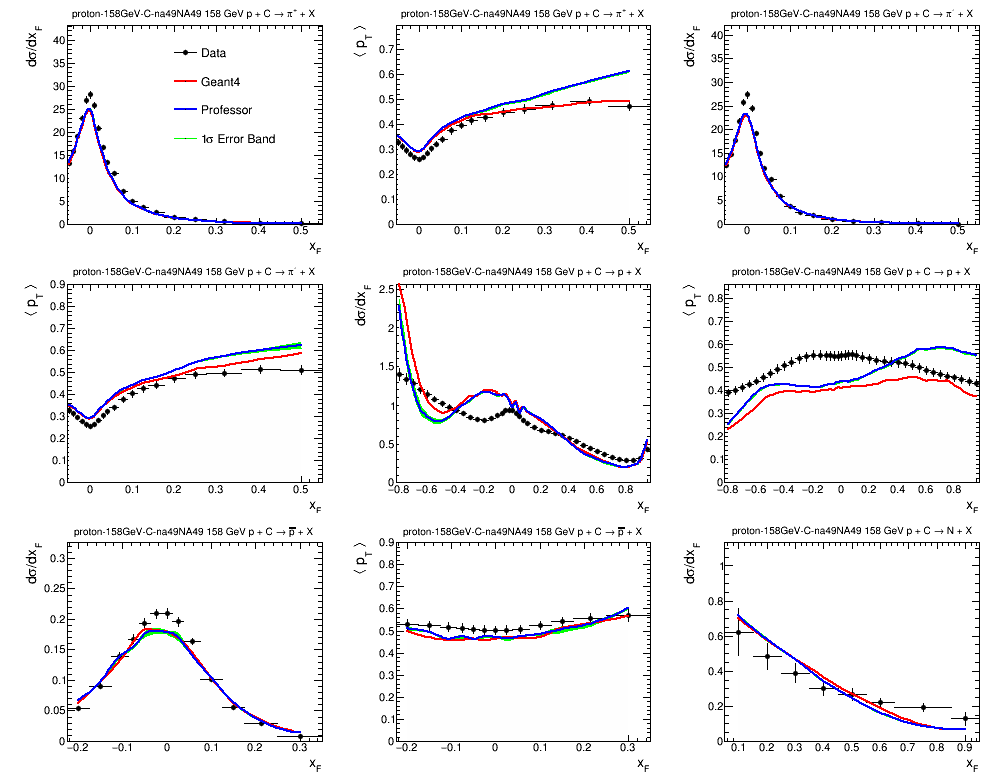}
\caption{\label{fig:fit_ftf1} Results of the global FTF parameter fit, compared to NA49 158 GeV $pC\rightarrow \pi^-X$, $pC\rightarrow \pi^+X$, and $pC\rightarrow pX$,
$pC\rightarrow \bar{p}X$ and $pC\rightarrow nX$ 
data.  Data points are shown in black; default Geant4 is red and Geant4 with best fit parameters in blue;
the green band shows uncertainties propagated from parameter uncertainties returned by the fit. }
\end{figure}

\begin{figure}[htbp]
\centering 
\includegraphics[width=15cm, height=10.5cm]{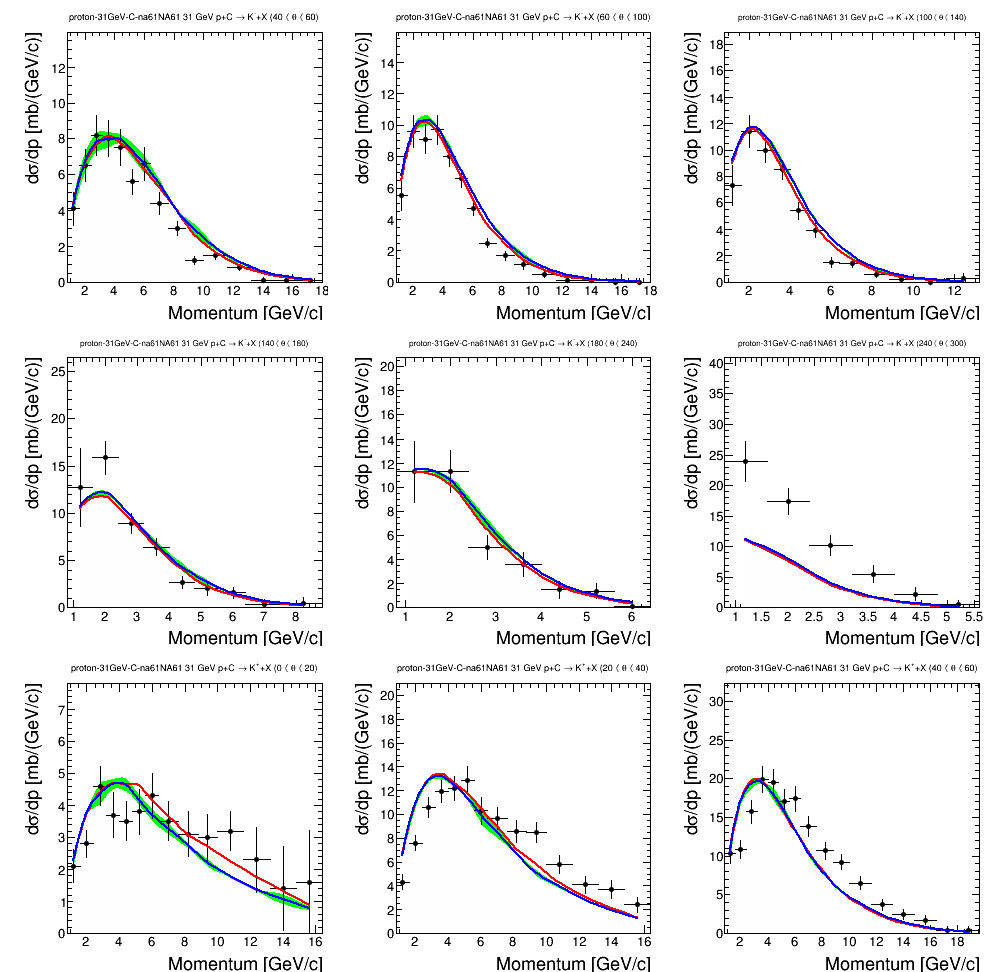}
\includegraphics[width=15cm, height=10.5cm]{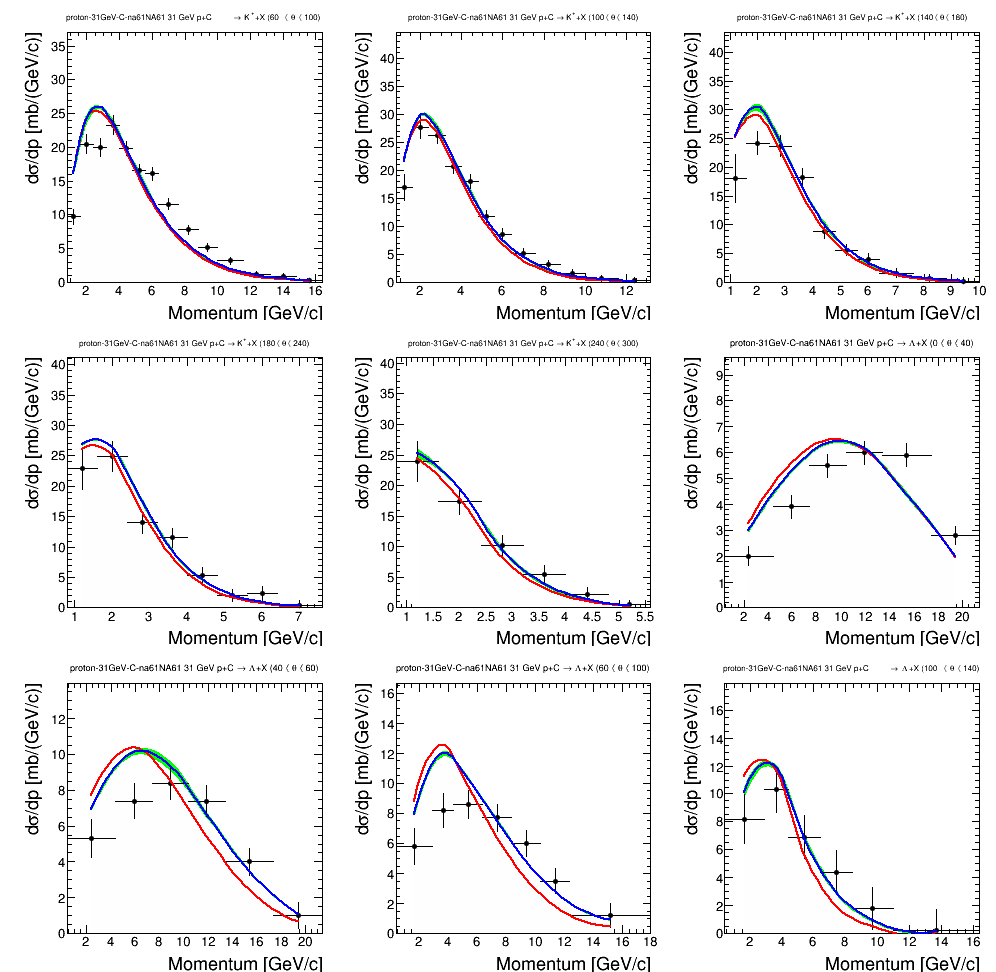}
\caption{\label{fig:fit_ftf2} Results of the global FTF parameter fit, compared to NA61 31 GeV $pC\rightarrow K^-X$, $pC\rightarrow K^+X$ and $pC\rightarrow \Lambda X$ data in bins of final state hadron angle.  Data points are shown in black; default Geant4 is red and Geant4 with best fit parameters in black; the green band shows uncertainties propagated from parameter uncertainties returned by the fit.   }
\end{figure}

\begin{figure}[htbp]
\centering 
\includegraphics[width=15cm, height=10.5cm]{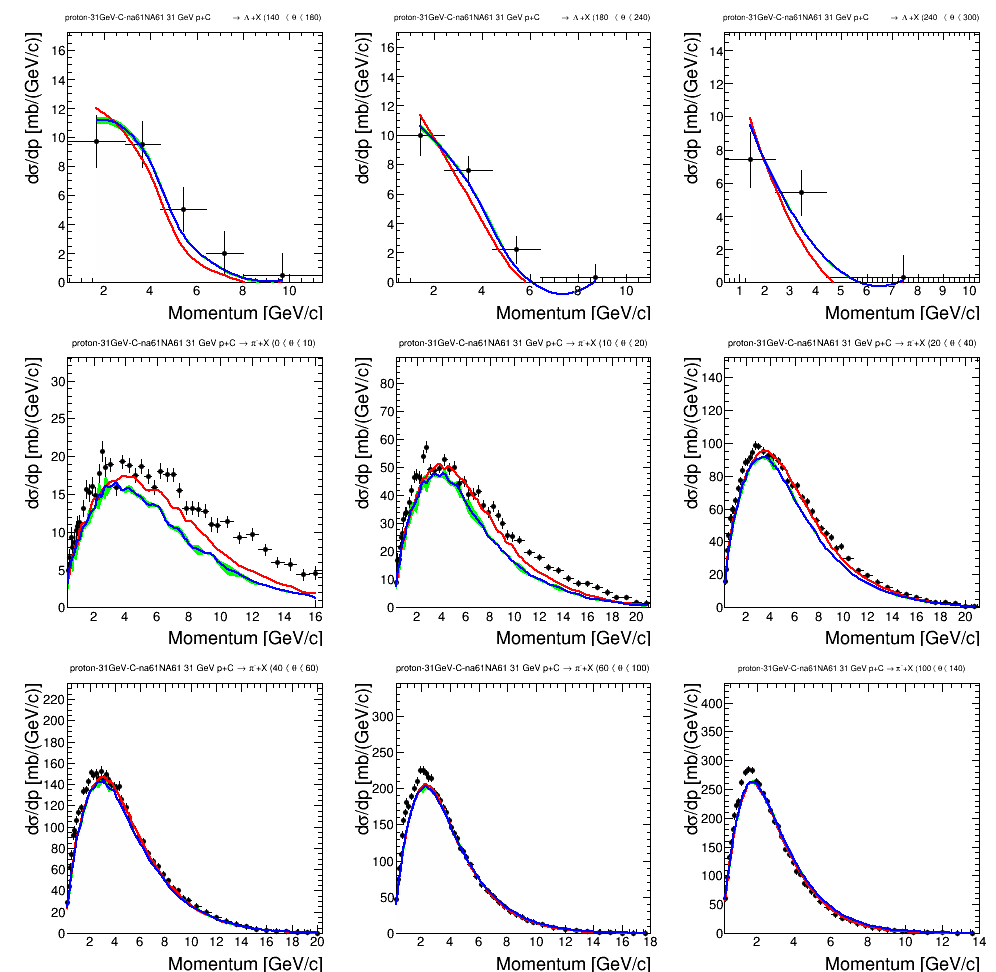}
\includegraphics[width=15cm, height=10.5cm]{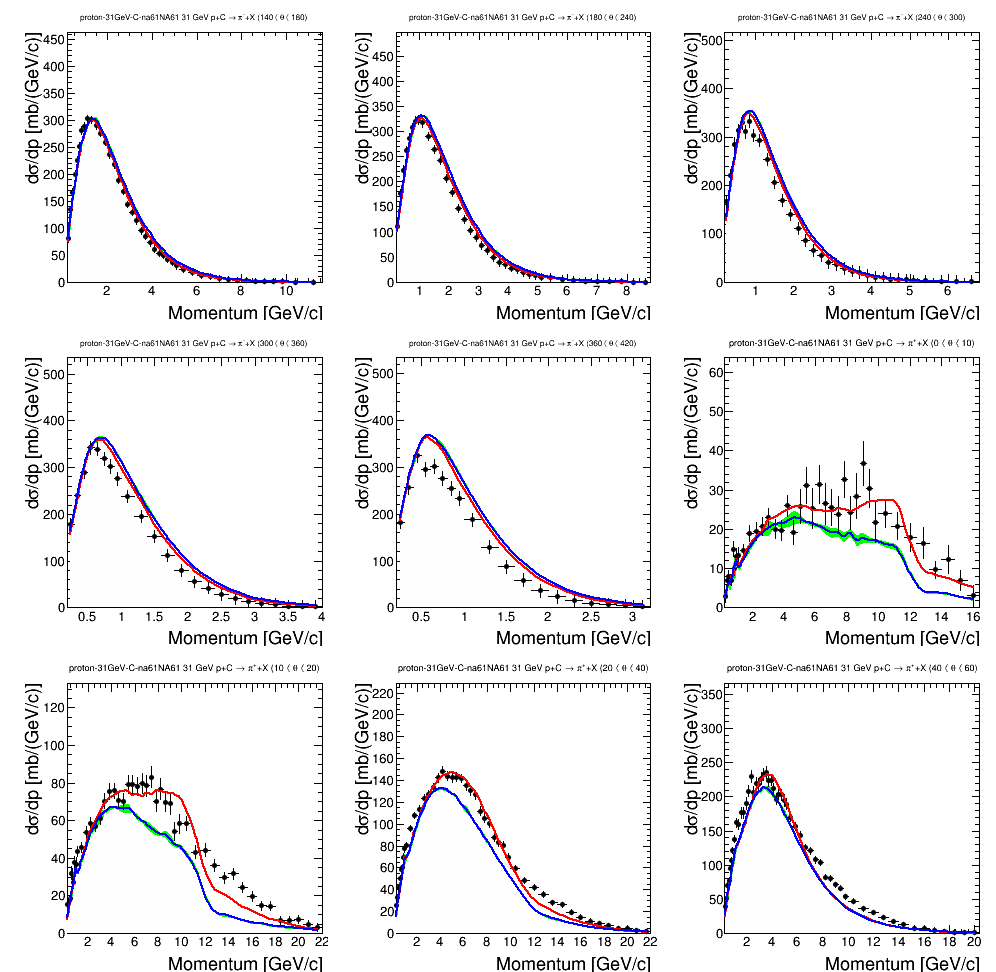}
\caption{\label{fig:fit_ftf3} Results of the global FTF parameter fit, compared to NA61 31 GeV $pC\rightarrow \Lambda X$, $pC\rightarrow \pi^-X$, and $pC\rightarrow \pi^+X$ data in bins of final state hadron angle.  Data points are shown in black; default Geant4 is red and Geant4 with best fit parameters in black; the green band shows uncertainties propagated from parameter uncertainties returned by the fit.   }
\end{figure}

\begin{figure}[htbp]
\centering 
\includegraphics[width=15cm, height=10.5cm]{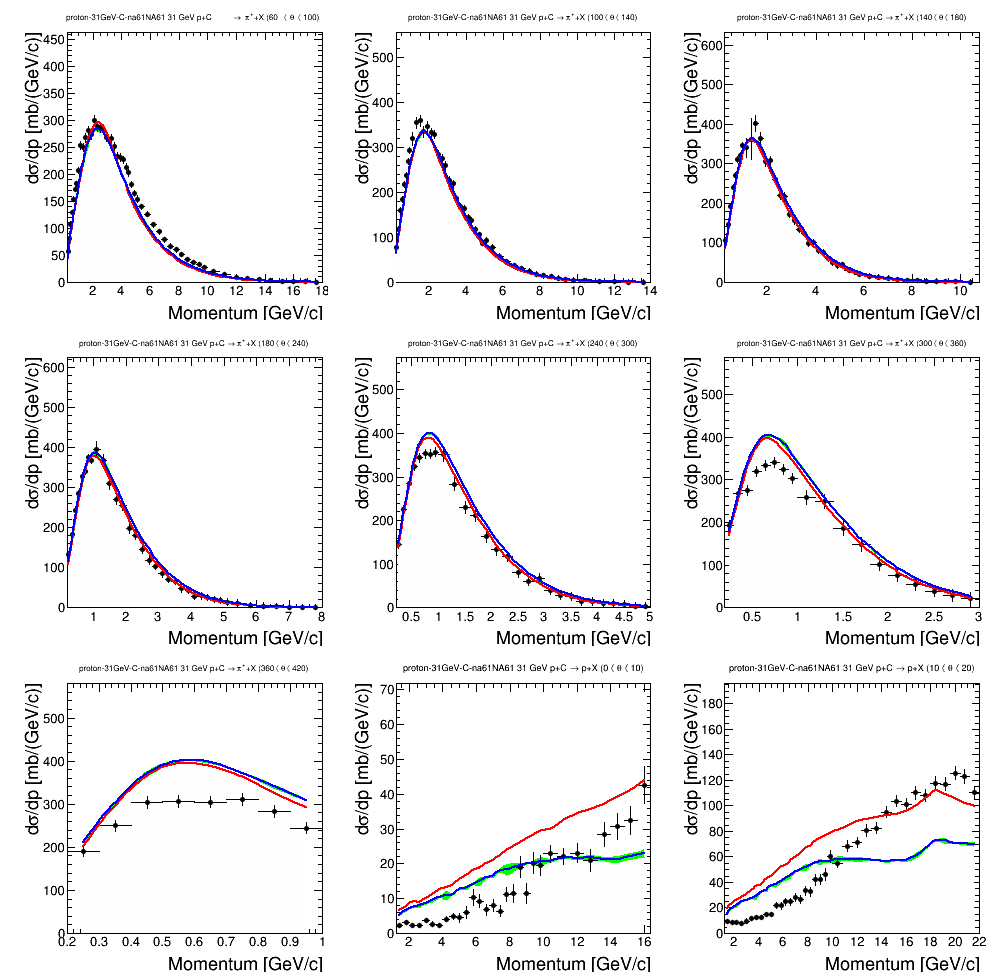}
\includegraphics[width=15cm, height=10.5cm]{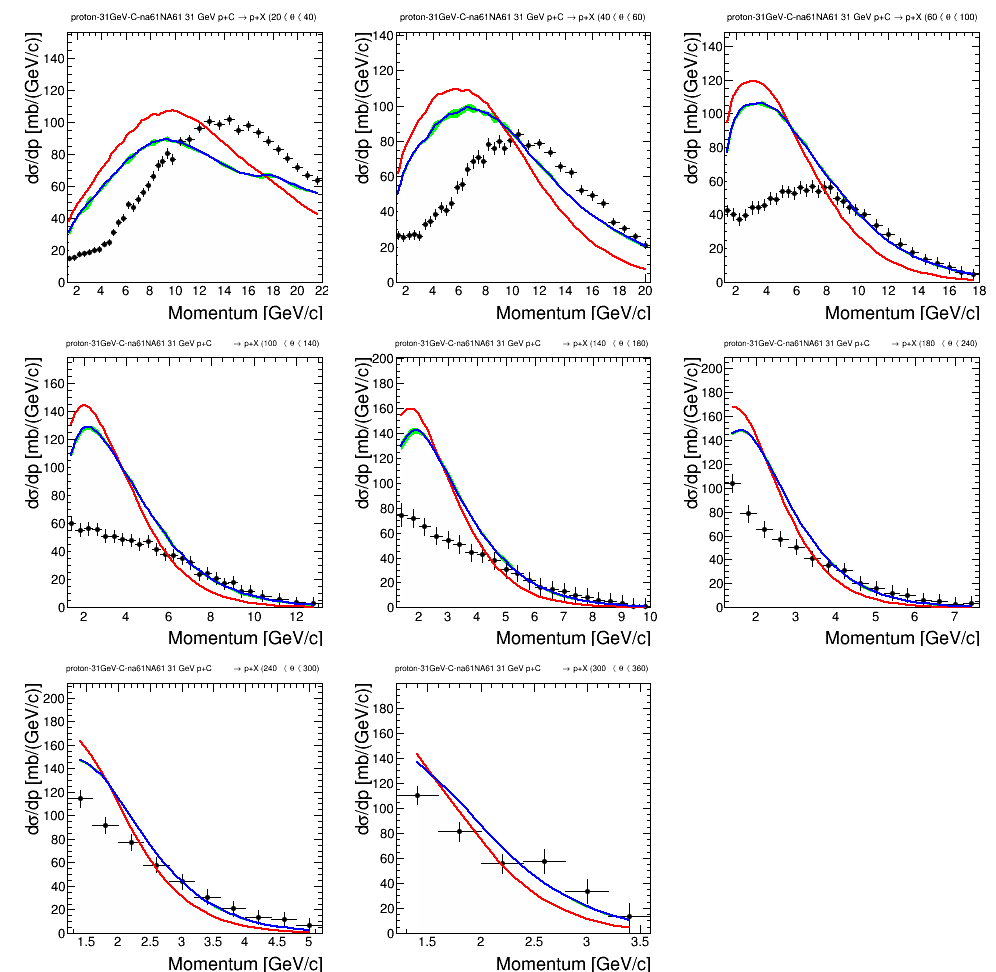}
\caption{\label{fig:fit_ftf4} Results of the global FTF parameter fit, compared to NA61 31 GeV $pC\rightarrow \pi^+X$ and $pC\rightarrow pX$ data in bins of final state hadron angle.  Data points are shown in black; default Geant4 is red and Geant4 with best fit parameters in black; the green band shows uncertainties propagated from parameter uncertainties returned by the fit.   }
\end{figure}

\begin{figure}[htbp]
\centering 
\includegraphics[width=15cm, height=10.5cm]{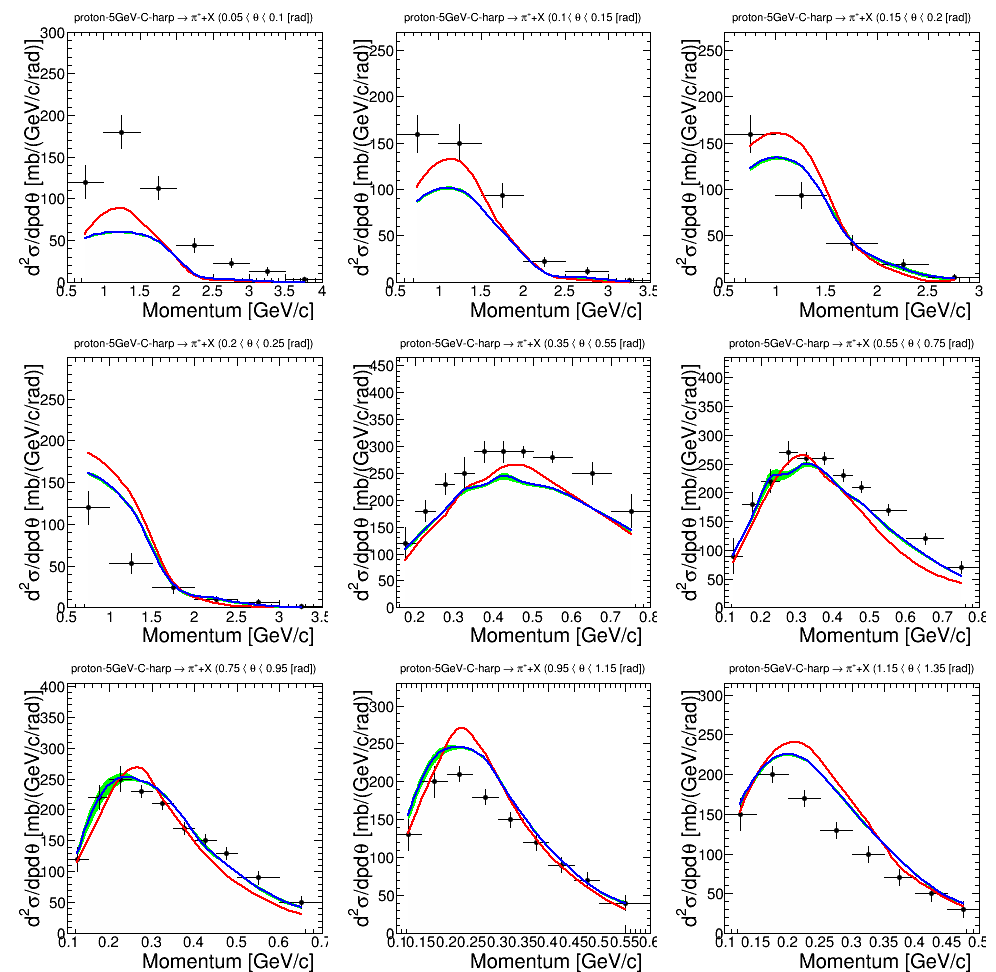}
\includegraphics[width=15cm, height=10.5cm]{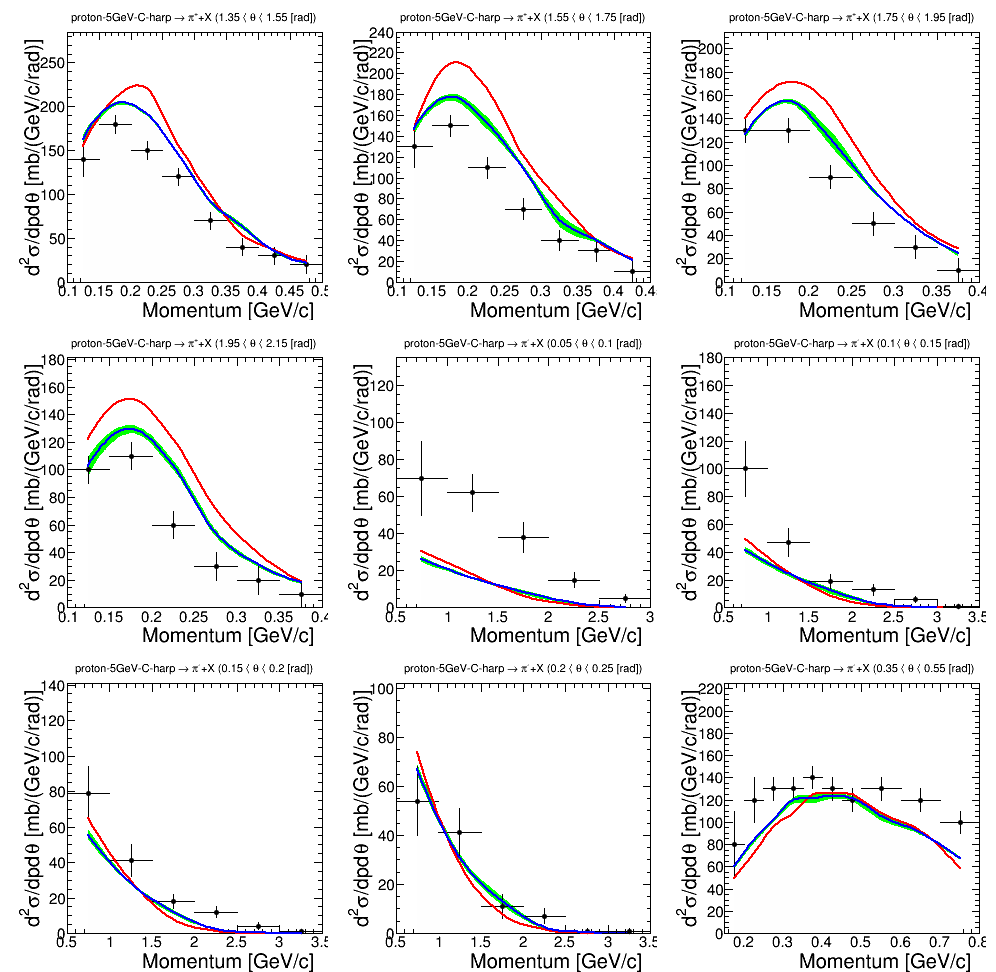}

\caption{\label{fig:fit_ftf5} Results of the global FTF parameter fit, compared to HARP 5 GeV $pC\rightarrow \pi^+X$ and $pC\rightarrow \pi^-X$ data in bins of final state hadron angle.  Data points are shown in black; default Geant4 is red and the global fit result in blue; the green band shows uncertainties propagated from parameter uncertainties returned by the fit.    }
\end{figure}

\begin{figure}[htbp]
\centering 
\includegraphics[width=15cm, height=10.5cm]{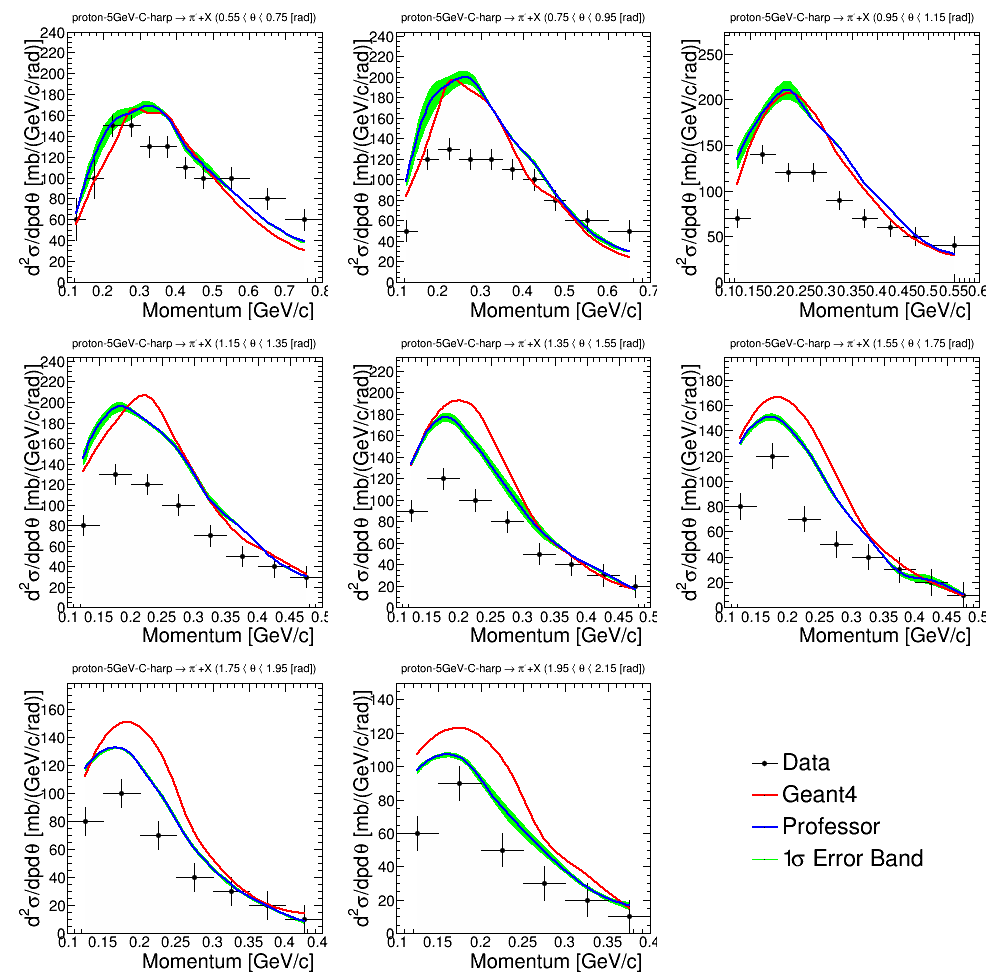}
\includegraphics[width=15cm, height=10.5cm]{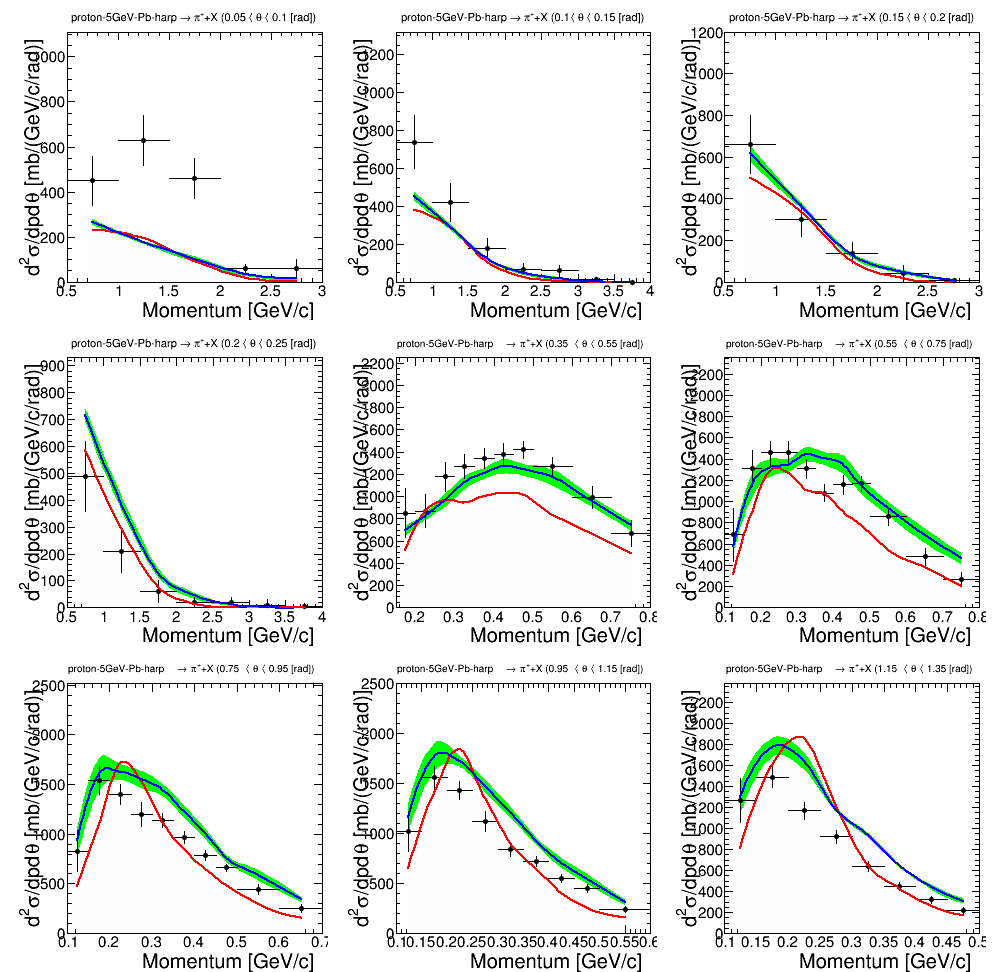}
\caption{\label{fig:fit_ftf6} Results of the global FTF parameter fit, compared to HARP 5 GeV $pC\rightarrow \pi^-X$ and $pPb\rightarrow \pi^+X$ data in bins of final state hadron angle.  Data points are shown in black; default Geant4 is red and the global fit result in blue; the green band shows uncertainties propagated from parameter uncertainties returned by the fit.    }
\end{figure}

\begin{figure}[htbp]
\centering 
\includegraphics[width=15cm, height=10.5cm]{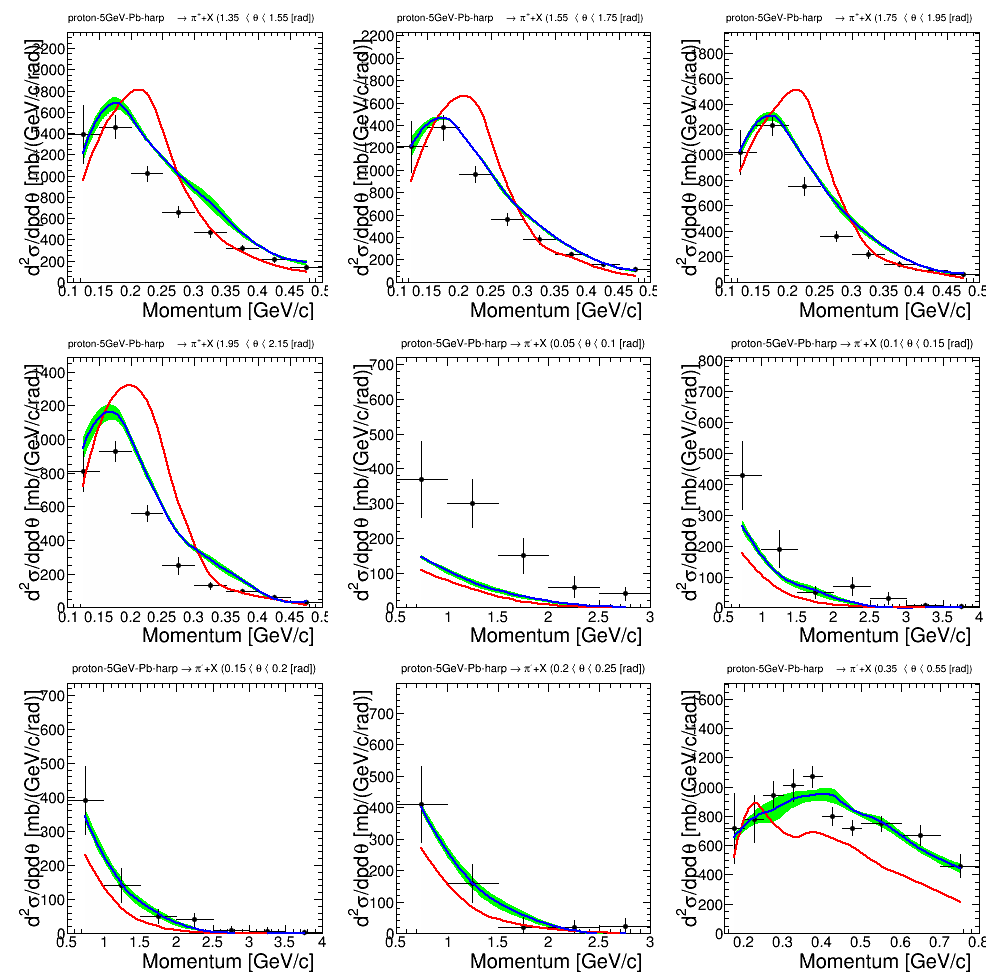}
\includegraphics[width=15cm, height=10.5cm]{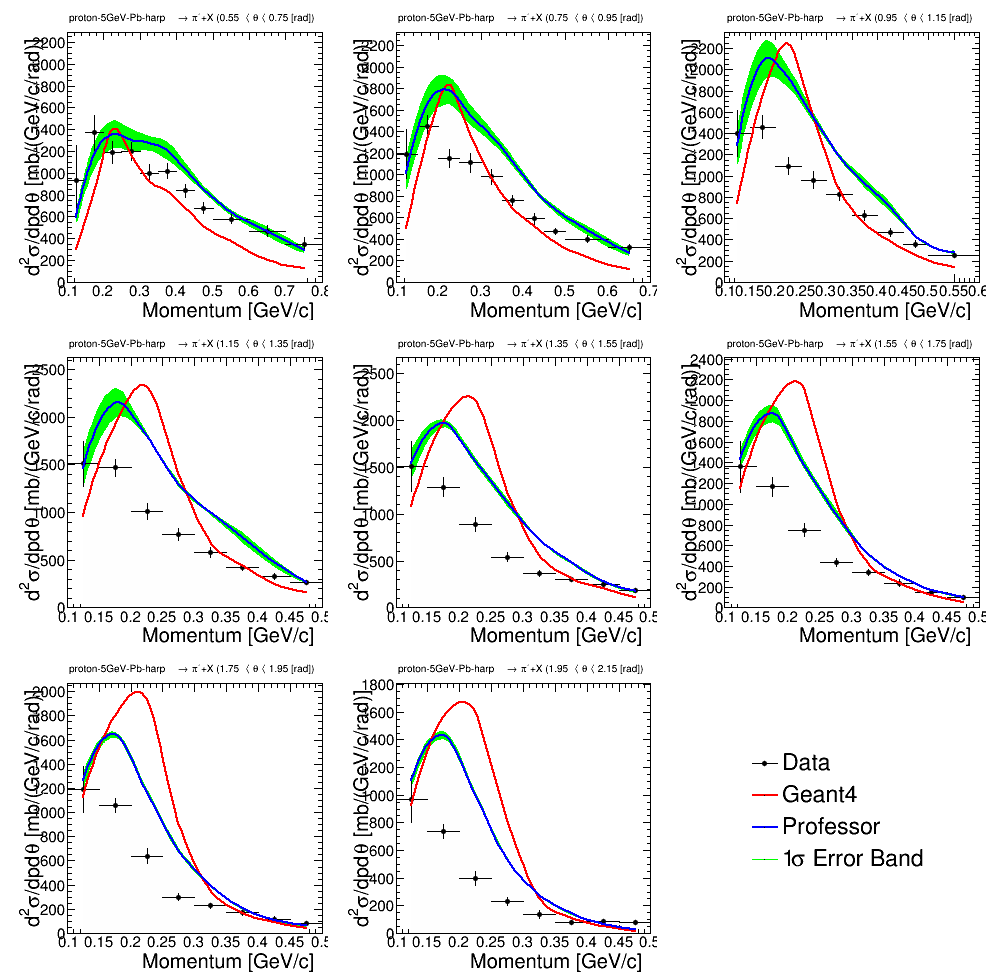}
\caption{\label{fig:fit_ftf7} Results of the global FTF parameter fit, compared to HARP 5 GeV $pPb      \rightarrow \pi^+X$ and $pPb\rightarrow \pi^-X$ data in bins of final state hadron angle.  Data points are shown in black; default Geant4 is red and the global fit result in blue; the green band shows uncertainties propagated from parameter uncertainties returned by the fit.    }
\end{figure}

\begin{figure}[htbp]
\centering 
\includegraphics[width=15cm, height=10.5cm]{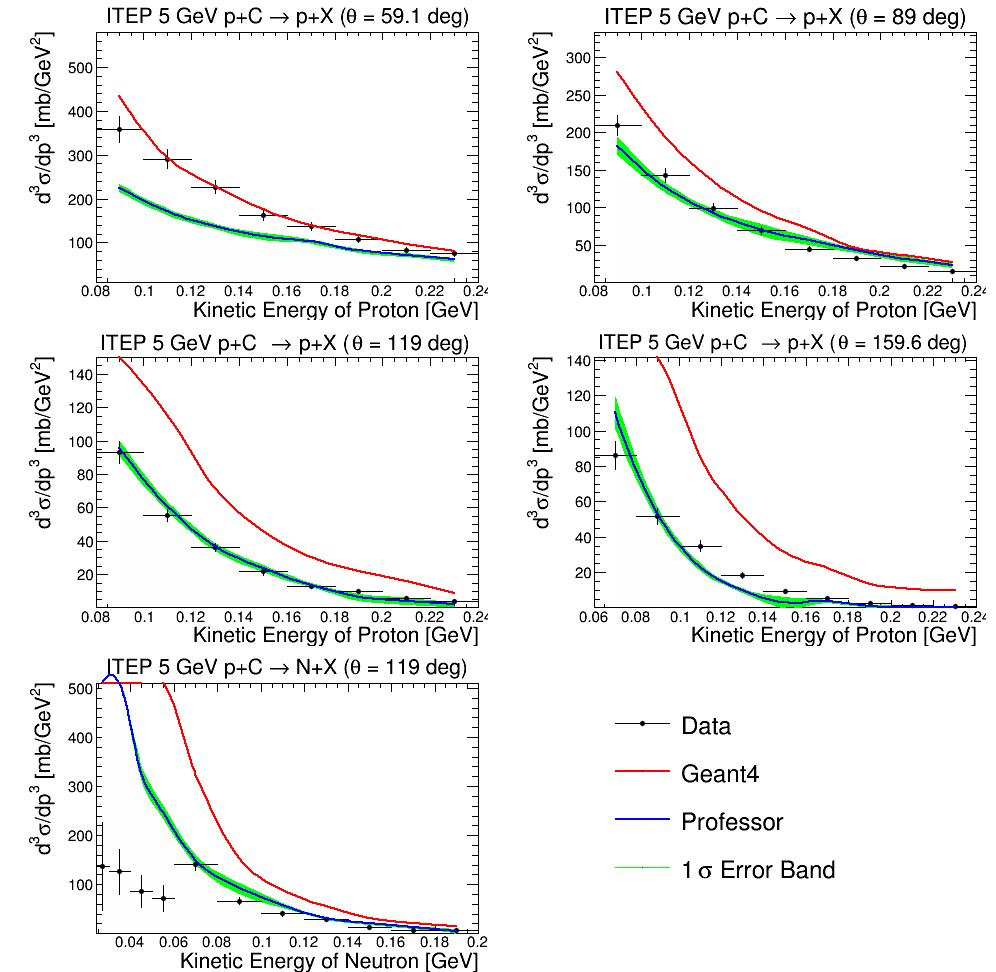}
\includegraphics[width=15cm, height=10.5cm]{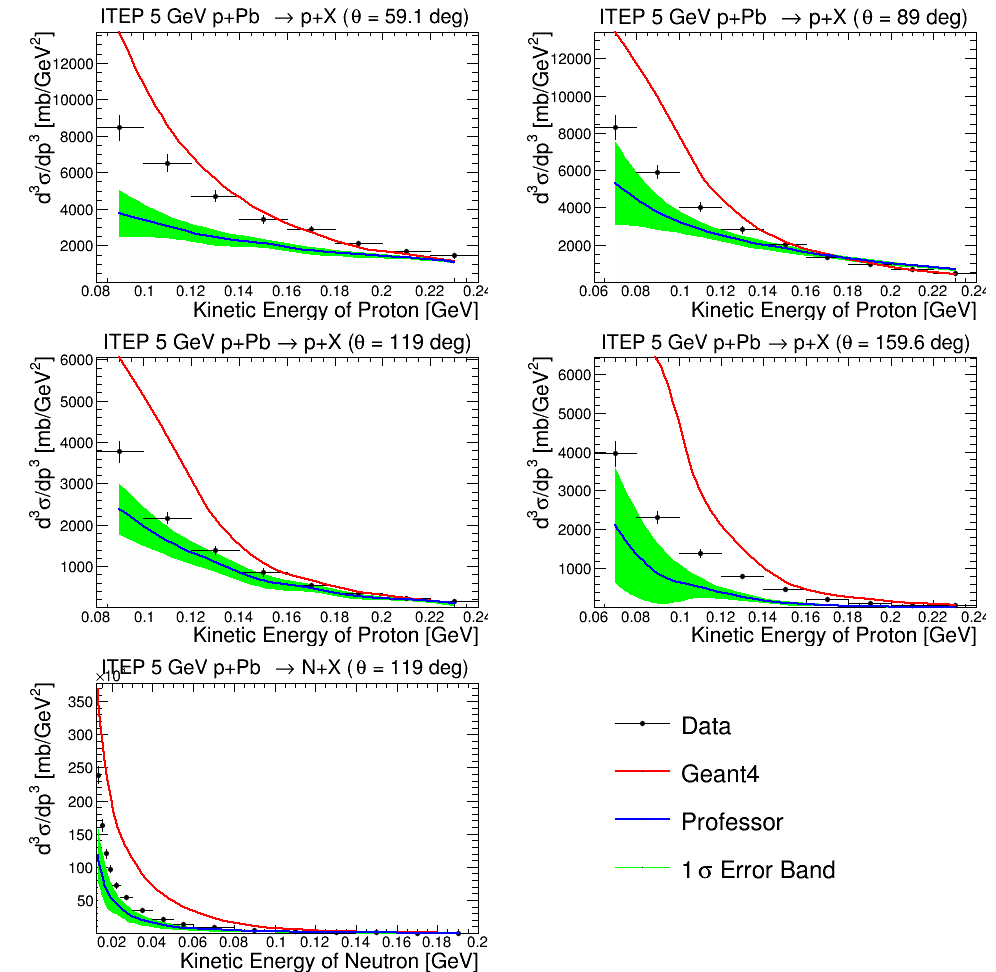}
\caption{\label{fig:fit_ftf8} Results of the global FTF parameter fit, compared to ITEP771 5 GeV $pC\rightarrow pX$,
$pC\rightarrow nX$, $pPb\rightarrow pX$,
and $pPb\rightarrow nX$ data in bins of final state hadron angle.  Data points are shown in black; default Geant4 is red and the global fit result in blue; the green band shows uncertainties propagated from parameter uncertainties returned by the fit.    }
\end{figure}

\begin{figure}[htbp]
\centering 
\includegraphics[width=15cm, height=10.5cm]{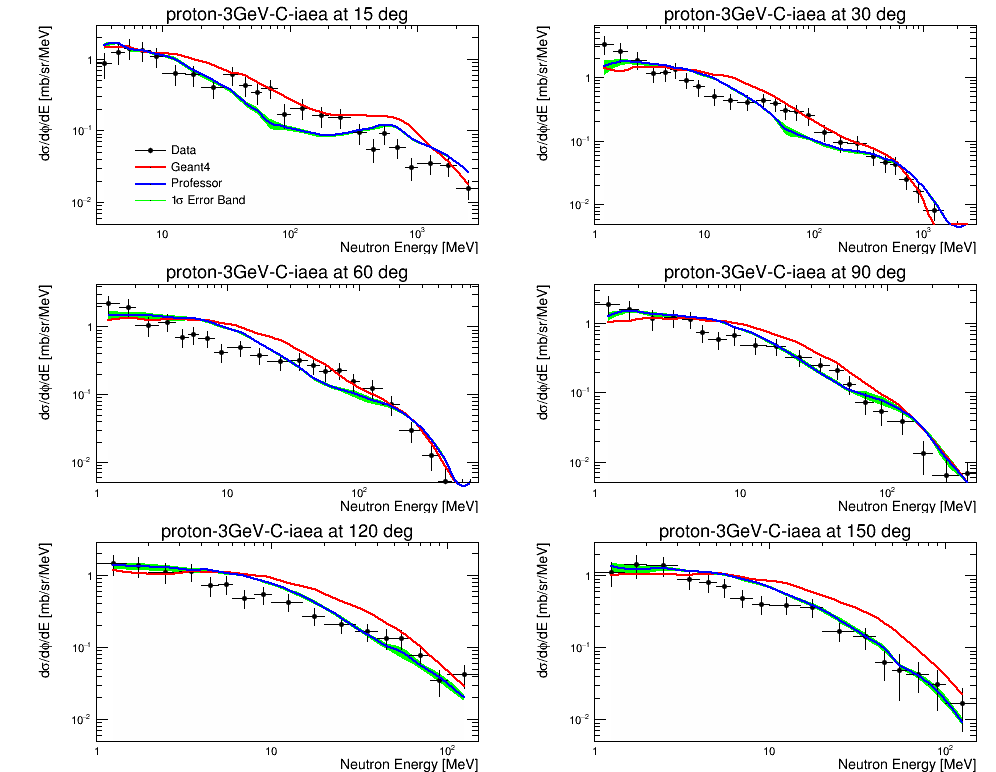}
\includegraphics[width=15cm, height=10.5cm]{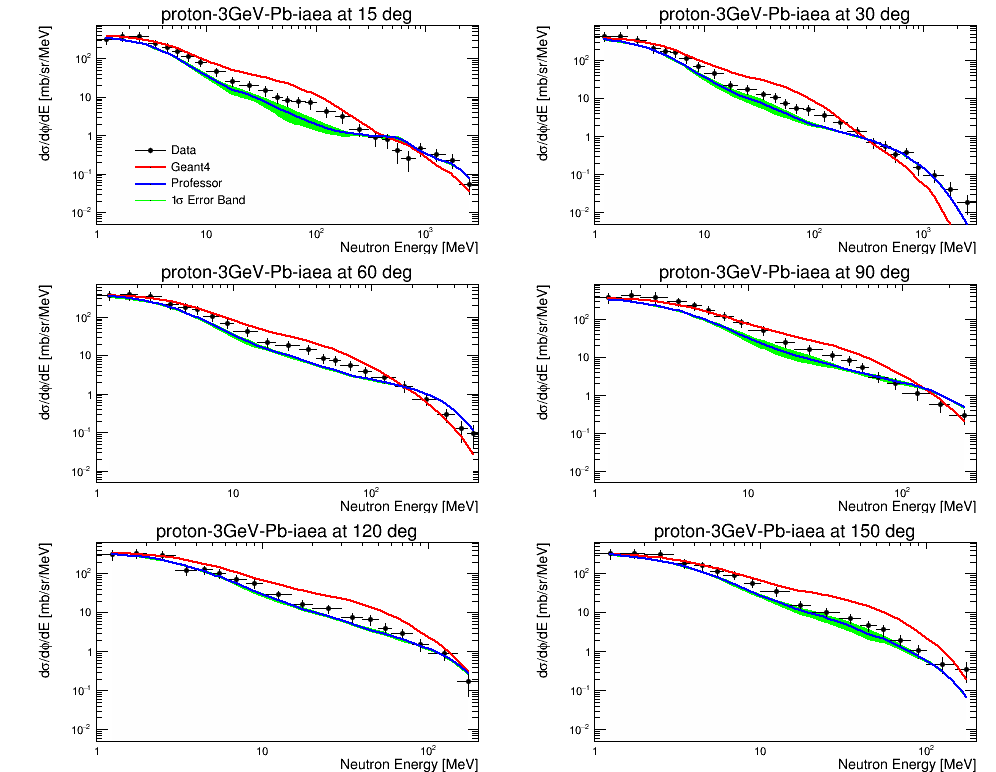}
\caption{\label{fig:fit_ftf9} Results of the global FTF parameter fit, compared to IAEA 3 GeV $pC\rightarrow nX$ and
$pPb\rightarrow nX$ data in bins of final state neutron angle.  Data points are shown in black; default Geant4 is red and the global fit result in blue; the green band shows uncertainties propagated from parameter uncertainties returned by the fit.    }
\end{figure}

\end{document}